\documentclass[american,aps,pra,reprint,floatfix,superscriptaddress,longbibliography,nofootinbib]{revtex4-2}
\usepackage[unicode=true,pdfusetitle, bookmarks=true,bookmarksnumbered=false,bookmarksopen=false, breaklinks=false,pdfborder={0 0 0},backref=false,colorlinks=false,linktocpage=true]{hyperref}
\hypersetup{colorlinks,linkcolor=blue,citecolor=red,urlcolor=cyan}
\usepackage{graphics,epstopdf,amsthm,amsmath,amssymb,mathptmx,colortbl,color,bm,framed,mathrsfs}
\usepackage{comment}
\usepackage{csquotes}
\usepackage[T1]{fontenc}
\usepackage[up]{subfigure}
\usepackage[braket, qm]{qcircuit}
\usepackage{physics}
\usepackage{commath}
\usepackage{mathtools}
\usepackage{float}
\usepackage{dsfont}
\usepackage[capitalise]{cleveref}

\makeatother

\newcommand{\RGA}[1]{\text{RGA}_{#1}}
\newcommand{\UR}[1]{\text{UR}_{#1}}
\newcommand{\CDD}[1]{\text{CDD}_{#1}}
\newcommand{\UDD}[1]{\text{UDD}_{#1}}
\newcommand{\UDDx}[1]{\text{UDDx}_{#1}}
\newcommand{\UDDy}[1]{\text{UDDy}_{#1}}
\newcommand{\QDD}[2]{\text{QDD}_{#1,#2}}

\newcommand{\XY}{\text{XY4}}
\newcommand{\CPMG}{\text{CPMG}}
\newcommand{\PX}{\text{PX}}

\newcommand{\armonk}{ibmq\_armonk}
\newcommand{\bogota}{ibmq\_bogota}
\newcommand{\jakarta}{ibmq\_jakarta}
\newcommand{\cnot}{\text{CNOT}}
\newcommand{\mc}{\mathcal}

\newcommand{\freea}{f_{\tau}}

\newcommand{\bes} {\begin{subequations}}
\newcommand{\ees} {\end{subequations}}
\newcommand{\bea} {\begin{eqnarray}}
\newcommand{\eea} {\end{eqnarray}}
\newcommand{\beq}{\begin{equation}}
\newcommand{\eeq}{\end{equation}}

\def\err{\text{err}}

\def\tDD{t_\text{DD}}

\def\DD{\text{DD}}

\usepackage[demo]{graphicx}
\usepackage{centernot}

\usepackage[normalem]{ulem}

\usepackage{csquotes}

\usepackage{xcolor}

\newtheorem*{assumption*}{\assumptionnumber}
\providecommand{\assumptionnumber}{}
\makeatletter
\newenvironment{assumption}[1]
 {%
  \renewcommand{\assumptionnumber}{Assumption #1}%
  \begin{assumption*}%
  \protected@edef\@currentlabel{#1}%
 }
 {%
  \end{assumption*}
 }
\makeatother

\newtheorem*{aresult*}{\aresultnumber}
\providecommand{\aresultnumber}{}
\makeatletter
\newenvironment{aresult}[1]
 {%
  \renewcommand{\aresultnumber}{Result #1}%
  \begin{aresult*}%
  \protected@edef\@currentlabel{#1}%
 }
 {%
  \end{aresult*}
 }

\newtheorem*{test*}{\testnumber}
\providecommand{\testnumber}{}
\makeatletter

\newenvironment{dataAndAvail} {\begin{acknowledgements}} {\end{acknowledgements}}

\begin{document}

\title{Dynamical decoupling for superconducting qubits: a performance survey}

\author{Nic Ezzell}
\email{nezzell@usc.edu}
\affiliation{Department of Physics and Astronomy, University of Southern California, Los Angeles, CA, 90089, USA}
\affiliation{Center for Quantum Information Science \& Technology, University of Southern California, Los Angeles, CA, 90089, USA}

\author{Bibek Pokharel}
\email{pokharel@usc.edu}
\affiliation{Department of Physics and Astronomy, University of Southern California, Los Angeles, CA, 90089, USA}
\affiliation{Center for Quantum Information Science \& Technology, University of Southern California, Los Angeles, CA 90089, USA}

\author{Lina Tewala}
\email{lina.tewala@emory.edu}
\affiliation{Johns Hopkins University Applied Physics Laboratory, Laurel, MD, 20723, USA}
\affiliation{Thomas C. Jenkins Department of Biophysics,
Johns Hopkins University Baltimore, MD, 21218, USA}

\author{Gregory Quiroz}
\email{gregory.quiroz@jhuapl.edu}
\affiliation{Johns Hopkins University Applied Physics Laboratory, Laurel, MD, 20723, USA}
\affiliation{William H. Miller III Department of
Physics $\&$ Astronomy, Johns Hopkins University, Baltimore, Maryland 21218, USA}

\author{Daniel A. Lidar}
\email{lidar@usc.edu}

\affiliation{Department of Physics and Astronomy, University of Southern California, Los Angeles, CA, 90089, USA}
\affiliation{Center for Quantum Information Science \& Technology, University of Southern California, Los Angeles, CA 90089, USA}
\affiliation{Department of Electrical Engineering, University of Southern California, Los Angeles, CA, 90089, USA}
\affiliation{Department of Chemistry, University of Southern California, Los Angeles, CA, 90089, USA}

\date{\today}

\begin{abstract}
Dynamical Decoupling (DD) is perhaps the simplest and least resource-intensive error suppression strategy for improving quantum computer performance. Here we report on a large-scale survey of the performance of  
$60$ different DD sequences from $10$ families, including basic as well as advanced sequences with high order error cancellation properties and built-in robustness.
The survey is performed using three different superconducting-qubit IBMQ devices, with the goal of assessing the relative performance of the different sequences in the setting of arbitrary quantum state preservation. 
We find that the high-order universally robust (UR) and quadratic DD (QDD) sequences generally outperform all other sequences across devices and pulse interval settings. Surprisingly, we find that DD performance for basic sequences such as \CPMG\ and \XY\ can be made to nearly match that of UR and QDD by optimizing the pulse interval, with the optimal interval being substantially larger than the minimum interval possible on each device. 
\end{abstract}
\maketitle


\section{Introduction}
In the pre-fault-tolerance era, quantum computing research has two main near-term goals: to examine the promise of quantum computers via demonstrations of quantum algorithms~\cite{montanaroQuantumAlgorithmsOverview2016,Preskill2018,Daley:2022vu} and to understand how quantum error correction and other noise mitigation methods can pave a path towards fault-tolerant quantum computers~\cite{Campbell:2017aa,Gottesman:FT-review-2022}. The last decade has seen the rise of multiple cloud-based quantum computing platforms that allow a community of researchers to test error suppression and correction techniques~\cite{PAL:13,Devitt:2016ti,vinci2015nested,Wootton:2018aa,Vuillot2018,Roffe:2018aa,Willsch:2018wu,Pearson:2019aa,Harper:2019aa,ryan2021realization,Chen:2021up}. Error suppression using dynamical decoupling (DD)~\cite{Viola:98, Zanardi:1999fk, Vitali:99, Duan:98e, Viola:99} is among the earliest methods to have been experimentally demonstrated, using experimental platforms such as trapped ions~\cite{Biercuk:09,biercuk-2009}, photonic qubits~\cite{Damodarakurup:08}, electron paramagnetic resonance~\cite{Du:09}, nuclear magnetic resonance (NMR)~\cite{Jenista2009Optimized,Alvarez:2010ve,Peng:2011ly}, trapped atoms~\cite{Sagi:2010vq} and nitrogen vacancies in diamond~\cite{Wang:2012ut}. It is  known that DD can be used to improve the fidelity of quantum computation both without~\cite{Viola:1999:4888,Viola:01a,KhodjastehLidar:02,Viola:2005:060502,KhodjastehLidar:08,PhysRevLett.100.160506,West:10,Sar:2012km} and with quantum error correction~\cite{Ng:2011dn, Paz-Silva:2013tt}. Several recent cloud-based demonstrations have shown that DD can unequivocally improve the performance of superconducting-qubit based devices~\cite{Pokharel:2018aa,souza2020process,jurcevic2020demonstration,tripathi2021suppression,Pokharel:better-than-classical-Grover,zhou2022quantum}, even leading to algorithmic quantum advantage~\cite{pokharel2022demonstration}.  

In this work, we systematically compare a suite of known and increasingly elaborate DD sequences developed over the past two decades (see \cref{tab:dd-seq-sum} for a complete list and Ref.~\cite{souza_robust_2012} for a detailed review). These DD sequences reflect a growing understanding of how to build features that suppress noise to increasingly higher order and with greater robustness to pulse imperfections. Our goal is to study the efficacy of the older and the more recent advanced sequences on currently available quantum computers. To this end, we implement these sequences on three different IBM Quantum Experience (IBMQE) transmon qubit-based platforms: ibmq\_armonk (Armonk), ibmq\_bogota (Bogota), and ibmq\_jakarta (Jakarta). We rely on the open-pulse functionality~\cite{openpulse} of IBMQE, which enables us to precisely control the pulses and their timing. The circuit-level implementation of the various sequences can be suboptimal, as we detail in the Appendix. 

We assess these DD sequences for their ability to preserve an arbitrary single-qubit state. Previous work, focused on the \XY\ sequence, has studied the use of DD to improve two-qubit entanglement~\cite{Pokharel:2018aa} and the fidelity of two-qubit gates~\cite{tripathi2021suppression}, and we leave a systematic survey of the multi-qubit problem for a future publication, given that the single-qubit case is already a rich and intricate topic, as we discuss below. By and large, we find that all DD sequences outperform the ``unprotected'' evolution (without DD). The higher-order DD sequences, like concatenated DD (CDD~\cite{Khodjasteh:2005xu}), Uhrig DD (UDD~\cite{Uhrig:2007qf}), quadratic DD (QDD~\cite{west_near-optimal_2010}), nested UDD (NUDD~\cite{Wang:10}) and universally robust (UR~\cite{Genov:2017aa}), perform consistently well across devices and pulse placement settings.  While these more elaborate sequences are statistically better than the traditional sequences such as Hahn echo \cite{hahn_spin_1950}, Carr-Purcell-Meiboom-Gill (CPMG), and XY4~\cite{maudsley_modified_1986} for short pulse intervals, their advantage diminishes with sparser pulse placement. As both systematic and random errors, e.g., due to finite pulse-width and limited control, are reduced, advanced sequences will likely provide further performance improvements. Overall, our study indicates that the robust DD sequences can be viewed as the preferred choice over their traditional counterparts.

The structure of this paper is as follows. In \cref{sec:DD-background} we review the pertinent DD background and describe the various pulse sequences we tested. In \cref{sec:methods}, we detail the cloud-based demonstration setup, the nuances of DD sequence implementation, and the chosen success metrics. We describe the results and what we learned about the sequences and devices in \cref{sec:results}. A summary of results and possible future research directions are provided in \cref{sec:Conclusion}. Additional details are provided in the Appendix.

\section{Dynamical decoupling background}
\label{sec:DD-background}

For completeness, we first provide a brief review of DD. In this section we focus on a small subset of all sequences studied in this work, primarily to introduce key concepts and notation. The details of all the other sequences are provided in \cref{app:DD-summary}. The reader who is already an expert in the theory may wish to skim this section to become familiar with our notation.

\subsection{DD with perfect pulses}
\label{sec:DD-perf}

Consider a time-independent Hamiltonian 
\beq
H = H_{S} + H_B + H_{SB},
\eeq 
where \(H_{S}\) and \(H_{B}\) contains terms that act, respectively, only on the system or the bath, and \(H_{SB}\) contains the system-bath interactions. We write $H_S = H_S^0+H_S^1$, where \(H_{S}^1\) represents an undesired, always on term (e.g., due to crosstalk), so that 
\beq
H_\err = H_{S}^1 + H_{SB}
\label{eq:Herr}
\eeq 
represents the ``error Hamiltonian'' we wish to remove using DD. $H_S^0$ contains all the terms we wish to keep. The corresponding \emph{free} unitary evolution for duration \(\tau\) is given by
\beq
f_{\tau} \equiv U(\tau)=\exp(-i\tau H). 
\eeq
DD is generated by an additional, time-dependent control Hamiltonian $H_c(t)$ acting purely on the system, so that the total Hamiltonian is
\beq
H(t) = H_S^0 + H_\err + H_B + H_c(t).
\eeq
An ``ideal'', or ``perfect'' pulse sequence is generated by a control Hamiltonian that is a sum of error-free, instantaneous Hamiltonians $\{\Omega_0 H_{P_k}\}_{k=1}^n$ that generate the pulses at corresponding intervals 
$\{\tau_k\}_{k=1}^n$:
\beq
\label{eq:hatHc}
\hat{H}_c(t) = \Omega_0 \sum_{k=1}^n \delta(t-t_k) H_{P_k} , \quad t_k = \sum_{j=1}^{k}\tau_j ,
\eeq 
where we use the hat notation to denote ideal conditions and let $\Omega_0$ have units of energy. Choosing $\Omega_0$ such that $\Omega_0\Delta = \pi/2$, where $\Delta$ is the ``width'' of the Dirac-delta function (this is made rigorous when we account for pulse width in \cref{sec:width} below), this gives rise to instantaneous unitaries or pulses 
\beq
\hat{P}_{k} = e^{-i\frac{\pi}{2}H_{P_k}} ,
\eeq
so that the total evolution is:

\beq
\label{eq:ideal-DD}
\tilde{U}(T) = f_{\tau_n}  \hat{P}_{n} \cdots f_{\tau_2} \hat{P}_{2} f_{\tau_1}  \hat{P}_{1} ,
\eeq
where $T \equiv t_n = \sum_{j=1}^{n}\tau_j$ is the total sequence time.
The unitary \(\tilde{U}(T) = U_{0}(T) B(T)\) can be decomposed into the desired error-free evolution $U_{0}(T) = \exp(-i T H_S^0)\otimes I_B$ and the unitary error \(B(T)\). Ideally,
\(B(T) = I_S \otimes e^{-iT\tilde B}\), where \(\tilde B\) is an arbitrary Hermitian bath operator. Hence, by applying $N$ repetitions of an ideal DD sequence of duration $T$, the system stroboscopically decouples from the bath at uniform intervals $T_j = j T$ for $j = 1, \ldots, N$. In reality, we only achieve approximate decoupling, so that $B(T) = I_S \otimes e^{-iT\tilde B} + \text{err}$, and the history of DD design is motivated by making the error term as small as possible under different and increasingly more realistic physical scenarios. 

\subsubsection{First order protection}

Historically, the first observation of stroboscopic decoupling came from nuclear magnetic resonance (NMR) spin echoes observed by Erwin Hahn in 1950~\cite{hahn_spin_1950} with a single $X$ pulse.\footnote{We use $X=\sigma^x$ interchangeably, and likewise for $Y$ and $Z$, where $\sigma^\alpha$ denotes the $\alpha$'th Pauli matrix, with $\alpha\in\{x,y,z\}$. See \cref{app:DD-summary} for a precise definition of all sequences.} Several years later, Carr \& Purcell~\cite{Carr:54} and Meiboom \& Gill~\cite{meiboom_modified_1958} independently proposed the improved \CPMG\ sequence with two $X$ pulses. In theory, both sequences are only capable of partial decoupling in the ideal pulse limit. In particular, $B(T) \approx I_S \otimes e^{-iT\tilde B}$ only for states near $\ket{\pm} = (\ket{0}\pm\ket{1})/\sqrt{2}$ (where $\ket{0}$ and $\ket{1}$ are the $+1$ and $-1$ eigenstates of $\sigma^z$, respectively), as we explain below. Nearly four decades after Hahn's work, Maudsley proposed the \XY\ sequence~\cite{maudsley_modified_1986},
which is \emph{universal} since $B(T) \approx I_S \otimes e^{-iT \tilde{B}}$ on the full Hilbert space, which means all states are equally protected. Equivalently, universality means that arbitrary single-qubit interactions with the bath are decoupled to first order in $\tau$.

To make this discussion more precise, we first write $H_B + H_{SB}$ in a generic way for a single qubit:
\begin{equation}
    \label{eq:err-H}
    \overline{H} \equiv H_B + H_{SB} = \sum_{\alpha = 0}^{3} \gamma_{\alpha} \sigma^{\alpha} \otimes B^{\alpha}, 
\end{equation}
where $B^\alpha$ are bath terms and $\sigma^{(0)} = I$.
 
Since distinct Pauli operators anti-commute, i.e., $\{\sigma_i, \sigma_j\} = 2I \delta_{i j}$, then for $k \neq 0$, 
\begin{equation}
    \label{eq:anticommute}
    \sigma_k \overline{H} \sigma_k = -\sum_{\alpha \neq k} \gamma_{\alpha} \sigma^{\alpha} \otimes B^{\alpha} + \gamma_k \sigma_k \otimes B_k .
\end{equation}
The minus sign is an effective time-reversal of the terms that anticommute with $\sigma_k$.
In the ideal pulse limit, this is enough to show that \enquote{pure-X} defined as
    \begin{align}
        \label{eq:cpmg-def1}
        \PX &\equiv X - \freea - X - \freea\ ,
    \end{align}
induces an effective error Hamiltonian 
\beq
\label{eq:CPMG}
H^{\text{eff}}_{\PX} = \gamma_x \sigma^x \otimes B^x + I_S \otimes \tilde{B} + \mathcal{O}(\tau^2)
\eeq 
every $2 \tau$. Note that \CPMG\ is defined similarly:
\beq
\CPMG \equiv f_{\tau/2} - X - f_\tau - X - f_{\tau/2}\ ,
\label{eq:CPMG2}
\eeq
which is just a symmetrized placement of the pulse intervals; see \cref{subsec:opt-free-evo-periods}. PX and \CPMG\ have the same properties in the ideal pulse limit, but we choose to begin with PX for simplicity of presentation. Intuitively, the middle $X - \freea - X$ is a time-reversed evolution of the $\sigma^{(y,z)}$ terms, followed by a forward evolution, which cancel to first order in $\tau$ using the Zassenhaus formula~\cite{Suzuki:1977fk}, $\exp{\tau(A + B)} = e^{\tau A}e^{\tau B} + \mathcal{O}(\tau^2)$, an expansion that is closely related to the familiar Baker-Campbell-Hausdorff (BCH) formula. 
The undesired noise term $\gamma_x \sigma^x \otimes B^x$ does not decohere $\ket{\pm}$, but all other states are subject to bit-flip noise in the absence of suppression. By adding a second rotation around $y$, the \XY\ sequence, 
\begin{equation}
    \label{eq:xy4-def1}
    \XY \equiv Y - \freea - X - \freea - Y - \freea - X - \freea
\end{equation}
cancels the remaining $\sigma^x$ term and achieves universal (first order) decoupling at time $4 \tau$: 
\beq
\label{eq:XY}
H^{\text{eff}}_{\XY} = I_S \otimes \tilde{B} + \mathcal{O}(\tau^2) .
\eeq 
Practically, this means that all single-qubit states are equally protected to first order. These results can be generalized by viewing DD as a symmetrization procedure~\cite{zanardi_symmetrizing_1999}, with an intuitive geometrical interpretation wherein the pulses replace the original error Hamiltonian by a sequence of Hamiltonians that are arranged symmetrically so that their average cancels out~\cite{ByrdLidar:01}.

\subsubsection{Higher order protection}
\label{sec:HOP}

While the \XY\ sequence is universal for qubits, it only provides first-order protection. A great deal of effort has been invested in developing DD sequences that provide higher order protection.
We start with concatenated dynamical decoupling, or $\CDD{n}$~\cite{Khodjasteh:2005xu}. $\CDD{n}$ is an $n^{\text{th}}$-order recursion of \XY.\footnote{The construction works for any base sequence, but we specify \XY\ here for ease of presentation since our results labeled with $\CDD{n}$ always assume \XY\ is the base sequence.} 
For example, $\CDD{1} \equiv \XY$ is the base case, and
 \begin{subequations}
 \label{eq:CDD-def}
    \begin{align}
        \text{CDD}_n &\equiv{} \XY([\text{CDD}_{n-1}]) \\
        \begin{split}
        &= Y - [\text{CDD}_{n-1}] - X - [\text{CDD}_{n-1}] \\
        {} &\qquad -Y - [\text{CDD}_{n-1}] - X - [\text{CDD}_{n-1}],
        \end{split}
    \end{align}
\end{subequations}
 which is just the definition of \XY\ in \cref{eq:xy4-def1} with every $\freea$ replaced by $\CDD{n-1}$. This recursive structure leads to an improved error term $\mathcal{O}(\tau^{n+1})$ provided $\tau$ is \enquote{small enough.} To make this point precise, we must define a measure of error under DD. Following Ref.~\cite{Ng:2011dn}, one useful way to do this is to separate the \enquote{good} and \enquote{bad} parts of the joint system-bath evolution, i.e., to split \(\tilde{U}(T)\) [\cref{eq:ideal-DD}] as
 \beq
 \tilde{U}(T) = \mc{G}+\mc{B},
 \eeq
 where $\mc{G} = U_0(T)\otimes B'(T)$, and where -- as above -- $U_0(T)$ is the ideal operation that would be applied to the system in the absence of noise, and $B'(T)$ is a unitary transformation acting purely on the bath. The operator $\mc{B}$ is the \enquote{bad}
 part, i.e., the deviation of $\tilde{U}(T)$ from the ideal operation.
 The error measure is then\footnote{We use the sup operator norm (the largest singular value of $A$): $\|A\| \equiv \sup_{\{ \ket{v} \}} \frac{\|A \ket{v} \|}{\norm{\ket{v}}} = \sup_{\{ \ket{v} \text{ s.t. }  \norm{\ket{v}}=1 \}} \|A \ket{v} \|$.}
 \begin{equation}
     \eta_{\DD} = \| \mc{B} \|
 \end{equation}
 Put simply, $\eta_{\DD} $ measures how far  
 the DD-protected evolution $\tilde{U}(T)$ is from 
 the ideal evolution $\mc{G}$. 
 With this error measure established, we can bound the performance of various DD sequences in terms of the relevant energy scales:
 \begin{align}
      \beta \equiv \|H_B\|, \ \ J \equiv \|H_{SB}\|, \ \ \epsilon \equiv  \beta + J.
 \end{align}
 
Using these definitions, we can replace the coarse $\mathcal{O}$ estimates with rigorous upper bounds on $\eta_{\DD}$.
In particular, as shown in Ref.~\cite{Ng:2011dn}, 
 \begin{subequations}
     \begin{align}
     \label{eq:etaXY4-ideal}
         \eta_{\XY} &= (4 J \tau) \left[ \frac12 (4 \epsilon \tau) + \frac29 (4 \epsilon \tau)^2 \right] + \mathcal{O}(\tau^3) \\
         \eta_{\CDD{n}} &= 4^{n(n+3)/2}(c \epsilon \tau)^n (J \tau) + \mathcal{O}(\tau^{n+2}),
     \end{align}
 \end{subequations}
 where $c$ is a constant of order $1$. This more careful analysis implies that (1) 
  $\epsilon \tau \lesssim 1$ is sufficient for \XY\ to provide error suppression,
 and (2) $\CDD{n}$ has an optimal concatenation level induced by the competition between taking longer (the bad $\sim 4^{n^2}$ scaling) and more error suppression [the good $(c \epsilon \tau)^n$ scaling]. The corresponding optimal concatenation level is
 \begin{equation}
    \label{eq:opt-cdd-order}
     n_{\text{opt}} = \lfloor \log_4 (1 / \overline{c} \epsilon \tau) - 1 \rfloor,
 \end{equation}
 where $\lfloor \cdot \rfloor$ is the floor function and $\bar{c}$ is another constant of order $1$ (defined in Eq.~(165) of Ref.~\cite{Ng:2011dn}). That such a saturation in performance should occur is fairly intuitive. By adding more layers of recursion, we suppress noise that was unsuppressed before. However, at the same time, we introduce more periods of free evolution $\freea$ which cumulatively add up to more noise. At some point, the noise wins since there is no active noise removal in an open loop procedure such as DD.
 
 Though $\CDD{n}$ derived from recursive symmetrization allows for $\mathcal{O}(\tau^{n+1})$ order suppression, it employs $\sim 4^n$ pulses. One may ask whether a shorter sequence could achieve the same goal. The answer is provided by the Uhrig DD (UDD) sequence~\cite{Uhrig:2007qf}. The idea is to find which DD sequence acts as an optimal filter function on the noise-spectral density of the bath while relaxing the constraint of uniform pulse intervals~\cite{Uhrig:2007qf, cywi-filter-func, biercuk_optimized_2009, alvarez_measuring_2011}. For a brief overview, we first assume that a qubit state decoheres as $e^{-\chi (t)}$. For a given noise spectral density $S(\omega)$,
 \begin{equation}
    \label{eq:coherence}
     \chi (t) = \frac{2}{\pi} \int_0^{\infty} \frac{S(\omega)}{\omega^2} F(\omega t) d \omega, 
 \end{equation}
where the frequency response of the system to DD is captured by the filter function $F(\omega t)$. For example, for $n$ ideal $\pi$ pulses executed at times $\{t_j\}$~\cite{Uys:09}, 
\begin{equation}
    F_n(\omega \tau) = \left|1 + (-1)^{n+1} e^{i \omega \tau} + 2 \sum_{j=1}^n (-1)^j e^{i \omega t_j} \right|^2,
\end{equation}
which can be substituted into \cref{eq:coherence} and optimized for $\{t_j\}$; the result is UDD~\cite{Uhrig:2007qf}. For a desired total evolution $T$, the solution (and definition of UDD) is simply to place $\pi$ pulses with \emph{nonuniform} pulse intervals,
\begin{equation}
    \label{eq:udd-tj-times}
    t_j = T \sin \left( \frac{j \pi}{2 (n + 1)} \right)^2.
\end{equation}
When we use $n$ $X$-type pulses in particular, we obtain $\UDDx{n}$. It turns out that $\UDDx{n}$ achieves $\mathcal{O}(\tau^n)$ suppression for states near $\ket{\pm}$ using only $n$ pulses, and this is the minimum number of $\pi$ pulses needed~\cite{Uhrig:2007qf, QL:11}. It is in this sense that UDD is provably optimal. However, it is important to note that this assumes that the total sequence time is fixed; only in this case can the optimal sequence be used to make the distance between the protected and unperturbed qubit states arbitrarily small in the number of applied pulses. On the other hand, if the minimum pulse interval is fixed and the total sequence time is allowed to scale with the number of pulses, then -- as in CDD -- longer sequences need not always be advantageous~\cite{UL:10}.

UDD can be improved from a single axis sequence to the universal \emph{quadratic} DD sequence (QDD)~\cite{west_near-optimal_2010, QL:11,KL:11} using recursive design principles similar to those that lead to \XY\ and eventually $\CDD{n}$ from \PX. Namely, to achieve universal decoupling, we use a recursive embedding of a $\UDDy{m}$ sequence into a $\UDDx{n}$ sequence. Each $X$ pulse in $\UDDx{n}$ is separated by a free evolution period $f_{t_{j+1} - t_{j}}$ which can be filled with a $\UDDy{m}$ sequence. Hence, we can achieve $\min\{\tau^n, \tau^m\}$ universal decoupling, and when $m = n$, we obtain universal order $\tau^n$ decoupling using only $n^2$ pulses instead of the $\sim 4^n$ in $\CDD{n}$. This is nearly optimal~\cite{west_near-optimal_2010}, and an exponential improvement over $\CDD{n}$. When $m \neq n$, the exact decoupling properties are more complicated~\cite{QL:11}. Similar comments as for $\UDDx{n}$ regarding the difference between a fixed total sequence time $T$ \textit{vs} a fixed minimum pulse interval apply for QDD as well~\cite{Xia:2011uq}.

While QDD is universal and near-optimal for single-qubit decoherence, the ultimate recursive generalization of UDD is nested UDD (NUDD)~\cite{Wang:10}, which applies for general multi-qubit decoherence, and whose universality and suppression properties have been proven and analyzed in a number of works~\cite{Jiang:2011ta,Xia:2011uq,Kuo:2012rf}. In the simplest setting, suppression to $N$'th order of general decoherence afflicting an $m$-qubit system requires $(N+1)^{2m}$ pulses under NUDD.

\begin{table}[t]
    \centering
    \begin{tabular}{|c|c|c|c|}
\hline
sequence & uniform interval & universal & Needs OpenPulse\\
\hline
Hahn Echo~\cite{hahn_spin_1950}& Y & N & N \\
\hline
PX/ CPMG~\cite{Carr:54, Meiboom:58} & Y & N & N \\
\hline
XY4~\cite{maudsley_modified_1986} & Y & Y & Y \\ 
\hline 
$\CDD{n}$~\cite{Khodjasteh:2005xu} & Y & Y & Y \\
\hline
EDD~\cite{viola-knill-robustDD} & Y & Y \& N & Y \\ 
\hline
$\RGA{n}$~\cite{Quiroz:2013fv} & Y & Y \& N & Y\\
\hline
KDD~\cite{souza_robust_2011} & Y & Y & Y \\
\hline
$\UR{n}$~\cite{Genov:2017aa} & Y & Y & Y \\
\hline
$\UDDx{n}$~\cite{Uhrig:2007qf} & N & N & N \\                                      
\hline
$\QDD{n}{m}$~\cite{west_near-optimal_2010} & N & Y & Y \\                               
\hline
    \end{tabular}
    \caption{Summary of the DD sequences surveyed in this work, along with the original references. A sequence has a uniform (pulse) interval provided $\tau_i = \tau_j \ \forall i, j$ [see \cref{eq:ideal-DD}] and is nonuniform otherwise. A sequence is universal (in theory) if it cancels an arbitrary $H_\err$ [\cref{eq:Herr}] to first order in the pulse interval, and practically this means it protects all states equally well. Otherwise, it only protects a subset of states (e.g.,  CPMG, which only protects $\ket{\pm}$). In practice, this distinction is more subtle due to rotating frame effects, as discussed in \cref{subsec:sc-physics}. For those listed as both Y \& N, such as $\RGA{n}$, we mean that not all sequences in the family are universal. For example, $\RGA{2}$ is not universal but $\RGA{n}$ for $n \geq 4$ is universal. The last column lists whether our eventual implementation requires OpenPulse~\cite{alexander_qiskit_2020} or can be implemented faithfully just with the traditional circuit API~\cite{Qiskit, mckay_qiskit_2018} (see \cref{app:CvsP}).} 
    \label{tab:dd-seq-sum}
\end{table}

\subsection{DD with imperfect pulses}
\label{sec:DD-imperf}

So far, we have reviewed DD theory with ideal pulses. An ideal pulse is instantaneous and error-free, but in reality, finite bandwidth constraints and control errors matter. Much of the work since \CPMG\ has been concerned with (1) accounting for finite pulse width, (2) mitigating errors induced by finite width, and (3) mitigating systematic errors such as over- or under-rotations. We shall address these concerns in order. 

\subsubsection{Accounting for finite pulse width}
\label{sec:width}

During a finite width pulse, $P_j$, the effect of $H_\err+H_B$ cannot be ignored, so the analysis of \cref{sec:DD-perf} needs to be modified correspondingly. Nevertheless, both the symmetrization and filter function approaches can be augmented to account for finite pulse width.

We may write a realistic DD sequence with $\Delta$-width pulses just as in \cref{eq:ideal-DD}, but with the ideal control Hamiltonian replaced by
\beq
\hat{H}_c(t) = \sum_k \Omega(t-\tau_k) H_{P_k} ,\quad \int_{-\Delta/2}^{\Delta/2}  \Omega(t) dt = \Omega_0 \Delta = \frac{\pi}{2},
\eeq 
where $\Omega(t)$ is sharply (but not infinitely) peaked at $t=0$ and vanishes for $|t|>\Delta/2$. The corresponding DD pulses are of the form
    \begin{align}
        \label{eq:arb-dd-seq}
        P_k &\equiv \exp{-{i}\int_{\tau_k-\Delta/2}^{\tau_k+\Delta/2} dt [\Omega(t-\tau_k) H_{P_k} + H_\err + H_B]}.
    \end{align}
Note that the pulse intervals remain $\tau_k$ as before, now denoting the peak-to-peak interval; the total sequence time therefore remains $T =\sum_{j=k}^{n}\tau_k$.
 The ideal pulse limit of \cref{eq:ideal-DD} is obtained by taking the pulse width to zero, so that $H_\err + H_B$ can be ignored:
\beq
\hat{P}_k = \lim_{\Delta\to 0}P_k= e^{-i\frac{\pi}{2}H_{P_k}} , \quad \lim_{\Delta\to 0}\Omega(t) = \Omega_0  \delta(t) .
\eeq
We can then recover a result similar to \cref{eq:anticommute} by entering the toggling frame with respect to the control Hamiltonian $H_c(t)$ (see \cref{app:toggling-frame}),
and computing $\eta_{\DD} $ with a Magnus expansion or Dyson series~\cite{Ng:2011dn}. Though the analysis is involved, the final result is straightforward: $\eta_{\DD} $ picks up an additional dependence on the pulse width $\Delta$. For example, \cref{eq:etaXY4-ideal} is modified to
     \begin{align}
         \eta_{\XY}^{(\Delta)} &= 4 J \Delta +  \eta_{\XY} ,
     \end{align}
 which now has a linear dependence on $\Delta$. This new dependence is fairly generic, i.e., the previously discussed \PX, \XY, $\CDD{n}$, $\UDD{n}$, and $\QDD{n}{m}$ all have an error $\eta$ with an additive $\mathcal{O}(\Delta)$ dependence. Nevertheless, (1) DD is still effective provided $J \Delta \ll 1$, and (2) concatenation to order $n$ is still  effective provided the $J\Delta$ dependence does not dominate the $\epsilon \tau^n$ dependence. For $\CDD{n}$ this amounts to an effective noise strength floor~\cite{Ng:2011dn},
 \begin{equation}
    \eta_{\CDD{n}} \geq 16 \Delta J,
 \end{equation}
which modifies the optimal concatenation level 
$n_{\text{opt}}$.

\subsubsection{Mitigating errors induced by finite width}
 
A natural question is to what extent we can suppress this first order $\mathcal{O}(\Delta)$ dependence. One solution is Eulerian symmetrization,\footnote{The terminology arises from Euler cycles traversed by the control unitary in the Cayley graph of the group generated by the DD pulses.} which exhibits robustness to pulse-width errors~\cite{viola-knill-robustDD, viola-advanced-dd,smith_degree_nodate}. 
For example, the palindromic sequence
 \begin{equation}
     \text{EDD} \equiv X\freea Y\freea X\freea Y\freea Y\freea X\freea Y\freea X\freea,
 \end{equation}
 which is an example of Eulerian DD (EDD),
 has error term~\cite{Ng:2011dn}
 \begin{equation}
     \eta_\text{EDD} = (8 J \tau)\left[ \frac12 (8 \epsilon \tau) + \frac29 (8 \epsilon \tau)^2 + \mathcal{O}(\tau^3) \right],
 \end{equation}
 which contains no first order $\mathcal{O}(\Delta)$ term. Nevertheless, the constant factors are twice as large compared to \XY, 
 and it turns out that EDD 
 outperforms \XY\ when $\Delta / \tau 
 \gtrsim 8\epsilon \tau$ (see Fig.~9 in Ref.~\cite{Ng:2011dn}).\footnote{To clarify their relationship, suppose we take the ideal $\Delta \rightarrow 0$ limit. Here, \XY\ is strictly better since EDD uses twice as many pulses (and therefore free periods) to accomplish the same $\mathcal{O}(\tau)$ decoupling.} The same Eulerian approach can be used to derive the pulse-width robust version of the Hahn echo and CPMG, which we refer to with a ``super'' prefix (derived from the ``Eulerian supercycle'' terminology of Ref.~\cite{smith_degree_nodate}): 
 \begin{subequations}
     \begin{align}
         \text{super}-\text{Hahn} &\equiv X\freea \overline{X}\freea \\
         \text{super}-\text{CPMG} &\equiv X\freea X\freea \overline{X}\freea \overline{X}\freea,
     \end{align}
 \end{subequations}
 where 
 \begin{equation}
     \label{eq:overline-def}
     \overline{P}_k \equiv \exp{-i\int_{\tau_k-\Delta/2}^{\tau_k+\Delta/2} dt [-\Omega(t-\tau_k) H_{P_k} + H_\err + H_B]}
 \end{equation}
[compare to \cref{eq:arb-dd-seq}]. Intuitively, if $X$ is a finite pulse that generates a rotation about the $x$ axis of the Bloch sphere, then $\overline{X}$ is (approximately) a rotation about the $-x$ axis, i.e., with opposite orientation.
 
These robust sequences, coupled with concatenation, suggest that we can eliminate the effect of pulse width to arbitrary order $\mathcal{O}(\Delta^n)$; up to certain caveats this indeed holds with concatenated dynamically corrected gates (CDCG)~\cite{Khodjasteh:2010qd} (also see Ref.~\cite{Cai:2012aa}).\footnote{Caveats include: 
(1) The analytical CDCG constructions given in Ref.~\cite{Khodjasteh:2010qd} do not accommodate for the setting where always-on terms in the system's Hamiltonian are needed for universal control (so that they cannot just be included in $H_\err$); (2) Control-induced (often multiplicative) noise can be simultaneously present along with bath-induced noise; multiplicative noise requires modifications to the formalism of Ref.~\cite{Khodjasteh:2010qd}.}
However, this approach deviates significantly from the sequences consisting of only $\pi$ rotations we have considered so far. To our  knowledge, no strategy better than EDD exists for sequences consisting of only $\pi$ pulses~\cite{viola-advanced-dd}. 
 
\subsubsection{Mitigating systematic errors}
\label{sec:MSE}

In addition to finite width errors, real pulses are also subject to systematic errors. For example, $\Omega(t)$ might be slightly miscalibrated, leading to a systematic over- or under-rotation, and any aforementioned gain might be lost due to the accumulation of these errors. A useful model of pulses subject to systematic errors is 
\begin{equation}
    \label{eq:square-flip-angle-pulses}
    P^r_j = \exp{\pm i \frac{\pi}{2} ( 1 + \epsilon_r) \sigma^{\alpha})}
\end{equation}
for $\alpha\in\{x,y,z\}$. This represents instantaneous $X,Y,Z$ and $\overline{X}, \overline{Y}, \overline{Z}$ pulses subject to systematic over- or under-rotation by $\epsilon_r$, also known as a \emph{flip-angle error}. Another type of systematic control error is \emph{axis-misspecification}, where instead of the intended $\sigma^{\alpha}$ in \cref{eq:square-flip-angle-pulses} a linear combination of the form $\sigma^{\alpha}+\epsilon_\beta\sigma^\beta+\epsilon_\gamma\sigma^\gamma$ is implemented, with $\epsilon_\beta,\epsilon_\gamma \ll 1$ and $\alpha\ne\beta\ne \gamma$ denoting orthogonal axes \cite{Wang:2012ut}.

Fortunately, even simple $\pi$ pulses can mitigate systematic errors if rotation axes other than $+x$ and $+y$ are used. We consider three types of sequences: robust genetic algorithm (RGA) DD~\cite{Quiroz:2013fv}, Knill DD (KDD)~\cite{composite-knill,souza_robust_2011} and universally robust (UR) DD~\cite{Genov:2017aa}.

\paragraph{RGA DD.---}
The basic idea of RGA is as follows. A universal DD sequence should satisfy $\prod_{k=1}^n P_k = {I}$ up to a global phase, but there is a great deal of freedom in what combination of pulses are used that satisfy this constraint. In Ref.~\cite{Quiroz:2013fv}, this freedom was exploited to find, by numerical optimization with genetic algorithms, a class of sequences robust to over- or under-rotations.

Subject to a generic single-qubit error Hamiltonian as in \cref{eq:err-H}, optimal DD sequences were then found for a given number of pulses under different parameter regimes (i.e., the relative magnitude of $J$, $ \beta$, etc.). This numerical optimization \enquote{rediscovered} CPMG, \XY, and Eulerian DD as base sequences with $\mathcal{O}(\tau^2)$ errors. Higher-order sequences were then found to be concatenations of the latter. For example, 
\begin{subequations}
\begin{align}
    \RGA{8c} &\equiv \text{EDD} \\
    \RGA{64c} &\equiv \RGA{8c}[\RGA{8c}] .
\end{align}
\end{subequations}
A total of 12 $\text{RGA}$ sequences were found in total; more details are given in \cref{app:DD-summary}.

\paragraph{KDD.---}
The KDD sequence is similar in its goal to RGA because it mitigates systematic over or under-rotations. In design, however, it uses the principle of composite pulses~\cite{Freeman:book,Brown:04,Brown:2005vp}. The idea is to take a universal sequence such as \XY\ and replace each pulse $P_j$ with a composite series of pulses $(CP)_j$ that have the same effect as $P_j$ but remove pulse imperfections by self-averaging; for details, see \cref{app:DD-summary}.

The KDD sequence is robust to flip-angle errors [\cref{eq:square-flip-angle-pulses}]. For example,  suppose $\epsilon_r = \pi / 20$, and we apply idealized $\text{KDD}$ 10 times [which we denote by $(\text{KDD})^{10}$]. Then by direct calculation, $\|(\text{KDD})^{10} - I\| \approx 7\times 10^{-7}$ whereas  $\|(\XY)^{10} - I\| \approx 3\times 10^{-2}$, and in fact, KDD is robust up to $\mathcal{O}(\epsilon_r^5)$~\cite{souza_robust_2012, Genov:2017aa}. This robustness to over-rotations comes at the cost of 20 free evolution periods instead of 4, so we only expect KDD to work well in an $\epsilon_r$ dominated parameter regime. As a preview of the results we report in \cref{sec:results}, KDD is not among the top-performing sequences. Hence it appears reasonable to conclude that our demonstrations are conducted in a regime which is \emph{not} $\epsilon_r$ dominated.

\paragraph{UR DD.---}
An alternative approach to devise robust DD sequences is the $\UR{n}$ DD family~\cite{Genov:2017aa}, developed for a semiclassical noise model. In particular, the system Hamiltonian is modified by a random term instead of including an explicit quantum bath and system-bath interaction as for the other DD sequences we consider in this work. The model is expressed using an arbitrary unitary pulse that includes a fixed systematic error $\epsilon_r$ as in \cref{eq:square-flip-angle-pulses}, and reduces to a $\pi$ pulse in the ideal case. As detailed in \cref{app:DD-summary}, this leads to a family of sequences that give rise to an error scaling as $\epsilon_n \sim \epsilon_r^{n/2}$ using $n$ pulses. 

These sequences recover some known results at low order: $\UR{2}$ is CPMG, and $\UR{4}$ is \XY. In other words, \CPMG\ and \XY\ are also robust to some pulse imperfections as they cancel flip-angle errors to a certain order. Moreover, by the same recursive arguments, $\CDD{n}$ can achieve arbitrary flip-angle error suppression while achieving arbitrary $\mathcal{O}(\tau^n)$ (up to saturation) protection. Still,  $\CDD{n}$  requires exponentially more pulses than $\UR{n}$, since $\UR{n}$ is by design a semiclassical $\mathcal{O}(\tau^2)$ sequence. Whether $\UR{n}$ is also an $\mathcal{O}(\tau^2)$ sequence for a fully quantum bath is an interesting open problem.

\subsection{Optimizing the pulse interval}
\label{subsec:opt-free-evo-periods} 

Nonuniform pulse interval sequences such as UDD and QDD already demonstrate that introducing pulse intervals longer than the minimum possible ($\tau_{\min}$) can be advantageous. In particular, such alterations can reduce spectral overlap between the filter function and bath spectral density. A longer pulse interval also results in pulses closer to the ideal limit of a small $\Delta/\tau$ ratio when $\Delta$ is fixed. Empirical studies have also found evidence for better DD performance under longer pulse intervals~\cite{alvarez_measuring_2011, souza_robust_2011,souza_robust_2012}.

We may distinguish two ways to optimize the pulse interval: an asymmetric or symmetric placement of additional delays. For example, the asymmetric and symmetric forms of \XY\ we optimize take the form
\begin{subequations}
    \begin{align}
     \label{eq:xy4-asym}
    \XY_a(d) &\equiv Y f_d X f_d Y f_d X f_d  \\
    \label{eq:xy4-sym}
    \XY_s(d) &\equiv f_{d / 2} Y f_d X f_d Y f_d X f_{d / 2},
\end{align}
\end{subequations}
where $d$ sets the duration of the pulse interval. The asymmetric form here is consistent with how we defined \XY\ in \cref{eq:xy4-def1}, and except \CPMG\ we have tacitly defined every sequence so far in its asymmetric form for simplicity. The symmetric form is a simple alteration that makes the placement of pulse intervals symmetric about the mid-point of the sequence. For a generic sequence whose definition requires nonuniform pulse intervals like UDD, we can write the two forms as
\begin{subequations}
    \begin{align}
     \label{eq:seq-asym}
    \text{DD}_a(d) &\equiv P_1 f_{\tau_1 + d} P_2 f_{\tau_2 + d} \ldots P_n f_{\tau_n + d} \\
    \label{eq:seq-sym}
    \text{DD}_s(d) &\equiv f_{d / 2} P_1 f_{\tau_1 + d} P_2 f_{\tau_2 + d} \ldots P_n f_{\tau_n + d / 2}.
\end{align}
\end{subequations}
Here, the symmetric nature of $\text{DD}_s$ is harder to interpret. The key is to define $\Tilde{P}_j = P_1 f_{\tau_1}$ as the \enquote{effective pulse}. In this case, the delay-pulse motif is $f_{d / 2} \Tilde{P}_j f_{d / 2}$ for every pulse in the sequence exhibiting a reflection symmetry of the pulse interval about the center of the pulse. In the asymmetric version, it is $\Tilde{P}_j f_d$ instead. Note that $\PX_s(\tau) = \CPMG$, i.e., the symmetric form of the \PX\ sequence is \CPMG. As \CPMG\ is a well-known sequence, hereafter we refer to the \PX\ sequence as \CPMG\ throughout our data and analysis regardless of symmetry.

\subsection{Superconducting hardware physics relevant to DD}
\label{subsec:sc-physics} 

So far, our account has been abstract and hardware agnostic. Since our demonstrations involve IBMQ superconducting hardware, we now provide a brief background relevant to the operation of DD in such devices. We closely follow Ref.~\cite{tripathi2021suppression}, which derived the effective system Hamiltonian of transmon superconducting qubits from first principles and identified the importance of modeling DD performance within a frame rotating at the qubit drive frequency. It was found that DD is still effective with this added complication both in theory and cloud-based demonstrations and in practice, DD performance can be modeled reasonably well by ideal pulses with a minimal pulse spacing of $\tau_{\text{min}} = \Delta$. We now address these points in more detail.

The effective system Hamiltonian for two qubits (the generalization to $n > 2$ is straightforward) is 
\begin{equation}
    H_S = -\frac{\omega_{q_1}}{2} Z_1 - \frac{\omega_{q2}}{2}Z_2 +  J Z Z,
\end{equation}
where the $\omega_{q_i}$'s are qubit frequencies and $J \neq 0$ is an undesired, always-on $ZZ$ crosstalk. The appropriate frame for describing the superconducting qubit dynamics is the one co-rotating with the number operator $\hat{N} = \sum_{k,l \in \{0, 1\}} (k + l) \ket{kl}\!\!\bra{kl} = I - \frac12 (Z_1 + Z_2)$. The unitary transformation into this frame is $U(t) = e^{-i \omega_d \hat{N} t}$, where $\omega_d$ is the drive frequency used to apply gates.\footnote{This frame choice is motivated by the observation that in the IBMQ devices, the drive frequency is calibrated to be resonant with a particular eigenfrequency of the devices' qubits, which depends on the state of the neighboring qubits. Transforming into a frame that rotates with one of these frequencies gets one as close as possible to describing the gates and pulses as static in the rotating frame.}  In this frame, the effective dynamics are given by the Hamiltonian
\begin{equation}
    \label{eq:sc-ham}
    \tilde{H}(t) = \sum_{i=1}^2 \left( \frac{\omega_d - \omega_{q_i}}{2} \right)   Z_i + J ZZ + \tilde{H}_{SB}(t) + H_B, 
\end{equation}
where $\tilde{H}_{SB} = U^{\dagger}(t) H_{SB} U(t)$. To eliminate unwanted interactions, DD must symmetrize $JZZ$ and $\tilde{H}_{SB}$. The $JZZ$ term is removed by applying an $X$-type DD to the first qubit (``the DD qubit''): $X_1Z_1 Z_2 X_1 = -Z_1 Z_2$, so symmetrization still works as intended. However, the $\tilde{H}_{SB}$ term is time-dependent in the rotating frame $U(t)$, which changes the analysis. First, the sign-flipping captured by \cref{eq:anticommute} no longer holds due to the time-dependence of $\tilde{H}_{SB}$. Second, some terms in $\tilde{H}_{\text{err}}$ self-average and nearly cancel even without applying DD~\footnote{Intuitively, the $Z_1 Z_2$ rotation at frequency $\omega_d$ is already self-averaging the effects of noise around the $z$-axis, so in a sense, is acting like a DD sequence.}: 

\begin{equation}
    \label{eq:approx-cancelled-terms}
    \{\sigma^x_1, \sigma^x_2, \sigma^y_1, \sigma^y_2, \sigma^x \sigma^z, \sigma^z \sigma^x, \sigma^y \sigma^z, \sigma^z \sigma^y\}
\end{equation}
The remaining terms, 
\begin{equation}
    \label{eq:not-cancelled-at-all-terms}
    \{ \sigma^z_2, \sigma^x \sigma^x, \sigma^x \sigma^y, \sigma^y \sigma^x, \sigma^y \sigma^y\},
\end{equation}
are not canceled at all.
This differs from expectation in that the two terms containing $\sigma_1^y$ in \cref{eq:not-cancelled-at-all-terms} are not canceled, whereas in the $\omega_d \rightarrow 0$ limit, all terms containing $\sigma_1^y$ are fully canceled.

Somewhat surprisingly, a nominally universal sequence such as \XY, is no longer universal in the rotating frame, again due to the time-dependence acquired by $\tilde{H}_{SB}$. In particular, the only terms that perfectly cancel to $\mathcal{O}(\tau)$ are the same $Z_1$ and $ZZ$ as with \CPMG. However, the list of terms that approximately cancels grows to include
\begin{equation}
    \label{eq:xy4-partial-cancel-added}
    \{\sigma^x \sigma^x, \sigma^x \sigma^y, \sigma^y \sigma^x, \sigma^y \sigma^y \},
\end{equation}
and when $\tau$ is fine-tuned to an integer multiple of $2\pi / \omega_d$, then \XY\ cancels all terms except, of course, terms involving $I_1$, which commute with the DD sequence.

Consequently, without fine-tuning $\tau$, we should expect \CPMG\ and \XY\ to behave similarly when the terms in \cref{eq:xy4-partial-cancel-added} are not significant. Practically, this occurs when $T_1 \gg T_2$ for the qubits coupled to the DD qubit~\cite{tripathi2021suppression}. However, when instead $T_1 \lesssim T_2$ for coupled qubits, \XY\ should prevail. In addition, the analysis in Ref.~\cite{tripathi2021suppression} was carried out under the assumption of ideal $\pi$ pulses with $\tau_{\min} = \Delta$, and yet, the specific qualitative and quantitative predictions bore out in the cloud-based demonstrations. Hence, it is reasonable to model DD sequences on superconducting transmon qubits as
\begin{subequations}
    \begin{align}
    \label{eq:dd-approx-on-sc}
    \text{DD}_{\text{sc}} &\equiv \hat{P}_1 f_{\tau_1 + \Delta_1} \ldots \hat{P}_n f_{\tau_n + \Delta_n},
    \end{align}
\end{subequations}
where $\hat{P}_j$ is once again an ideal pulse with zero width, and the free evolution periods have been incremented by $\Delta_j$ -- the width of the actual pulse $P_j$.  

\subsection{What this theory means in practice}
 \label{subsec:theory-means-in-practice}
 
We conclude our discussion of the background theory by discussing its practical implications for actual DD memory cloud-based demonstrations.  This section motivates many of our design choices and our interpretation of the results.

At a high level, our primary demonstration -- whose full details are fleshed out in \cref{sec:methods} below -- is to prepare an initial state and apply DD for a duration $T$.  We estimate the fidelity overlap of the initial state and the final prepared state at time $T$ by an appropriate measurement at the end.  By adjusting $T$,  we map out a \emph{fidelity decay curve} (see \cref{subfig:long-fid-decay} for examples), which serves as the basis of our performance assessment of the DD sequence.

The first important point from the theory is the presence of $ZZ$ crosstalk, a spurious interaction that is always present in fixed-frequency transmon processors, even when the intentional coupling between qubits is turned off. 
Without DD, the always-on $ZZ$ term induces coherent oscillations in the fidelity decay curve, such as the Free curve in \cref{subfig:long-fid-decay}, depending on the transmon drive frequency, the dressed transmon qubit eigenfrequency, and calibration details~\cite{tripathi2021suppression}. Applying DD via pulses that anti-commute with the $ZZ$ term makes it possible to cancel the corresponding crosstalk term to first order~\cite{tripathi2021suppression,zhou2022quantum}.  For some sequences -- such as $\UR{20}$ in \cref{subfig:long-fid-decay} -- this first order cancellation almost entirely dampens the oscillation.  One of our goals in this work is to rank DD sequence performance, and $ZZ$ crosstalk cancellation is an important feature of the most performant sequences.  
The simplest way to cancel $ZZ$ crosstalk is to apply DD to a single qubit (the one we measure) and leave the remaining qubits idle.  We choose this simplest strategy in our work since it accomplishes our goal without adding complications to the analysis.  

Second, we must choose which qubit to perform our demonstrations on. As we mentioned in our discussion in the previous subsection, we know that the \XY\ sequence only approximately cancels certain terms.  Naively, we expect these remaining terms to be important when $T_1 \lesssim T_2$ and negligible when $T_1 \gg T_2$.  Put differently,  a universal sequence such as \XY\ should behave similarly to CPMG when these terms are negligible anyway ($T_1 \gg T_2$) and beat CPMG when the terms matter ($T_1 \lesssim T_2$).  To test this prediction, we pick three qubits where $T_1 > T_2$,  $T_2 > T_1$ and $T_1 \lesssim T_2)$, as summarized in \cref{tab:3proc}.

Third, we must contend with the presence of systematic gate errors.  When applying DD,  these systematic errors can manifest as coherent oscillations in the fidelity decay profile~\cite{souza_robust_2011, souza2020process}.  For example, suppose the error is a systematic over-rotation by $\epsilon_r$ within each cycle of DD with $N$ pulses. In that case, we over-rotate by an accumulated error $N \epsilon_r$, which manifests as fidelity oscillations.  This explains the oscillations for CPMG and XY4 in \cref{subfig:long-fid-decay}. To stop this accumulation for a given fixed sequence, one strategy is to apply DD sparsely by increasing the pulse interval $\tau$. However, increasing $\tau$ increases the strength of the leading error term obtained by symmetrization, which scales as $\mathcal{O}(\tau^k)$ (where $k$ is a sequence-dependent power).  

Thus, there is a tension between packing pulses together to reduce the leading error term and applying pulses sparsely to reduce the build-up of coherent errors.  Here, we probe both regimes (dense/sparse pulses) and discuss how this trade-off affects our results. Sequences that are designed to be robust to pulse imperfections play an important role in this study.  As the $\UR{20}$ sequence demonstrates in \cref{subfig:long-fid-decay}, this design strategy can be effective at suppressing oscillations due to both $ZZ$ crosstalk and the accumulation of coherent gate errors.  Our work attempts to empirically disentangle the various trade-offs in DD sequence design.

\section{Methods}
\label{sec:methods}

We performed our cloud-based demonstrations on three IBM Quantum superconducting processors~\cite{ibmq_armonk} with OpenPulse access~\cite{mckay_qiskit_2018, alexander_qiskit_2020}: ibmq\_armonk, ibmq\_bogota, and ibmq\_jakarta. Key properties of these processors are summarized in \cref{tab:3proc}.

\begin{table}[t]
    \begin{tabular}{|l|l|l|l|}
        \hline
            Device & ibmq\_armonk & ibmq\_bogota & ibmq\_jakarta \\ \hline
            \# qubits & 1 & 5 & 7 \\ \hline
            qubit used & q0 & q2 & q1 \\ \hline
            T1 ($\mu$s) & 140 $\pm$ 41 & 105 $\pm$ 41 & 149 $\pm$ 61 \\ \hline
            T2 ($\mu$s) & 227 $\pm$ 71 & 145 $\pm$ 63 & 21 $\pm$ 3 \\ \hline
            pulse duration (ns) & $71.1\overline{1}$  & $35.5\overline{5}$ & $35.5\overline{5}$  \\ \hline
        \end{tabular}
                    \caption{The three processors used in our cloud-based demonstrations. The total number of qubits $n$ varies from $1$ to $7$, but in all cases, we applied DD to just one qubit: number 2 for Bogota and 1 for Jakarta (see the insets of \cref{fig:pauliPlot} for the connectivity graph of each device). The choice of the qubit used is motivated by the prediction that $T_1 \gg T_2$ and $T_1 \lesssim T_2$ lead to different DD sequence behavior (see \cref{subsec:theory-means-in-practice} and Ref.~\cite{tripathi2021suppression}).  The qubits we have chosen have the highest connectivity on their respective devices and therefore are subject to maximal $ZZ$ crosstalk.  The $T_1$ and $T_2$ times are the averages of all reported values during data collection for the specified qubit, along with the $2 \sigma$ sample standard deviation. Data was collected over roughly $20$ different calibrations, mostly between Aug.~9-25, 2021, for the Pauli demonstration and Jan.~11-19, 2022, for the Haar-interval demonstration. }
    \label{tab:3proc}
    \end{table}

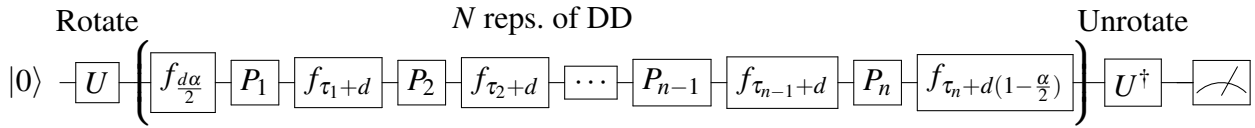
\begin{figure*}[htpb]
    \hspace{1cm}
    \large
        \Qcircuit @C=0.5em @R=1.0em {
        & \mbox{Rotate} & & & & & & \hspace{1.cm} \mbox{$N$ reps.\ of DD} & & & & & & & \mbox{Unrotate}  \\
        \lstick{\ket{0}} & \gate{U} & \qw & \gate{f_{\frac{d \alpha}{2}}}  &  \gate{P_1} & \gate{f_{\tau_1 + d}} & \gate{P_2} & \gate{f_{\tau_2 + d}} & \gate{\cdots} & \gate{P_{n-1}} & \gate{f_{\tau_{n-1} + d}} & \gate{P_n} & \gate{f_{\tau_n + d(1 - \frac{\alpha}{2})}} &  \qw & \gate{U^{\dagger}}  & \qw & \meter  \qw \gategroup{2}{4}{2}{7}{.7em}{(} \gdef\gatesup{N}\gategroup{2}{4}{2}{13}{.7em}{)}
        }
    \caption{The \enquote{quantum memory} circuit that underlies all of our demonstrations. By sampling the circuit using 8192 shots, we estimate the Uhlmann fidelity, $f_e(t)$ in \cref{eq:emp-fid}, between the prepared initial state and the final state under the DD sequence $P_1 f_{\tau_1} P_2 f_{\tau_2} \cdots P_n f_{\tau_n}$. We have included an additional adjustable pulse interval $f_d$, where we set $d=0$ for the Pauli demonstration~(\cref{subsec:pauliExperiment}) and systematically vary $d$ for the Haar demonstration~(\cref{subsec:haarIntervalExperiments}). The choice of $\alpha = 0$ ($1$) corresponds to an asymmetric (symmetric) placement of the additional delays.}
    \label{subfig:dd-circuit}
    \end{figure*}

All our demonstrations follow the same basic structure, summarized in \cref{subfig:dd-circuit}. Namely, we prepare a single-qubit initial state $\ket{\psi} = U\ket{0}$, apply $N$ repetitions of a given DD sequence $S$ lasting total time $t$, undo $U$ by applying $U^\dagger$, and finally measure in the computational basis. Note that this is a single-qubit protocol. Even on multi-qubit devices, we intentionally only apply DD to the single qubit we intend to measure to avoid unwanted $ZZ$ crosstalk effects as discussed in \cref{subsec:theory-means-in-practice}.  The qubit used on each device is listed in \cref{tab:3proc}.  

We empirically estimate the Uhlmann fidelity
\begin{equation}
    \label{eq:emp-fid}
    f_e(t) = |\bra{\psi}\rho_{\text{final}}(t) \ket{\psi}|^2
\end{equation}
with 95\% confidence intervals by counting the number of $0$'s returned out of $8192$ shots and bootstrapping. The results we report below were obtained using OpenPulse~\cite{alexander_qiskit_2020}, which allows for refined control over the exact waveforms sent to the chip instead of the coarser control that the standard Qiskit circuit API gives~\cite{Qiskit, mckay_qiskit_2018}. \cref{app:CvsP} provides a detailed comparison between the two APIs, highlighting the significant advantage of using OpenPulse. 
  
We utilize this simple procedure to perform two types of demonstrations which we refer to as Pauli and Haar, and explain in detail below. Briefly, the Pauli demonstration probes the ability of DD to maintain the fidelity of the six Pauli states (the eigenstates of $\{\sigma^x,\sigma^y,\sigma^z\}$) over long times, while the Haar demonstrations address this question for Haar-random states and short times. 
 
\def\gatesup{{}}

\begin{figure}
    \subfigure[]{
    \centering
    \label{subfig:long-fid-decay}
    \includegraphics[width=0.35\textwidth]
    {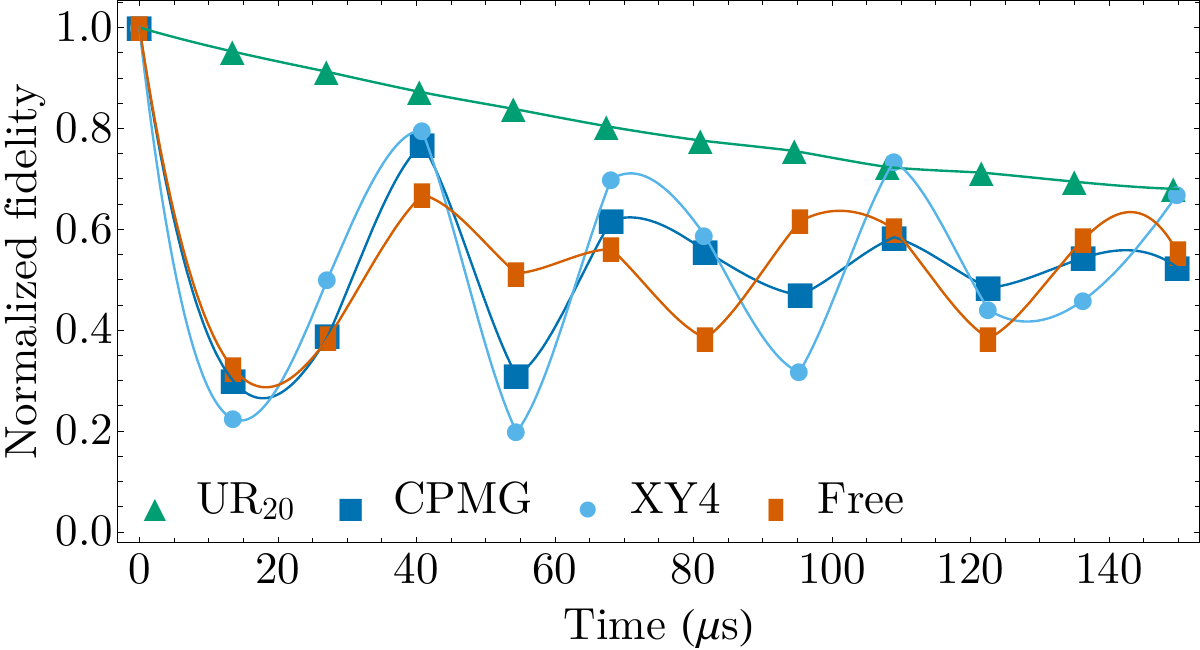}
    \includegraphics[width=0.12\textwidth]
    {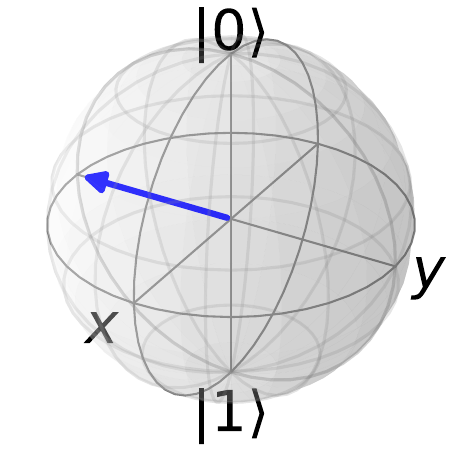}
    }
    \hfill
    \subfigure[]{
    \centering
    \label{subfig:box-plot-summary}
    \includegraphics[trim =0 0 0 25, clip,width=0.24\textwidth]{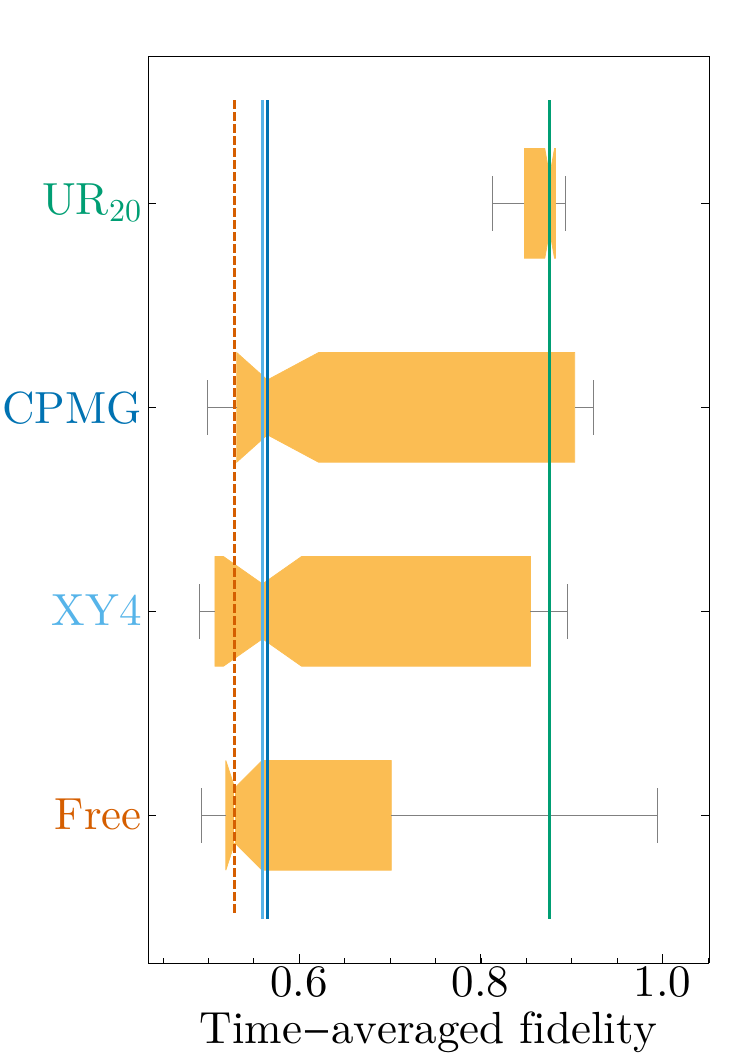}
    \includegraphics[width=0.12\textwidth]
    {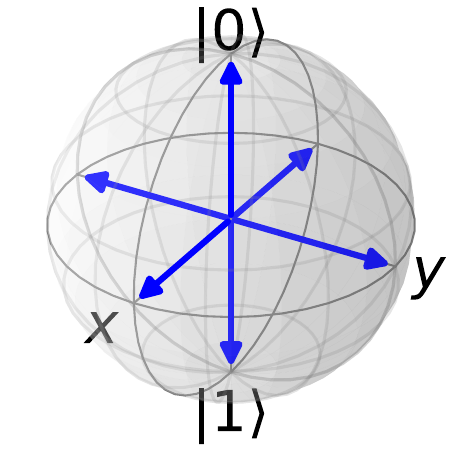}}
    \hfill
    \subfigure[]{
    \centering
    \label{fig:armonkRollingAvgPlot}
    \includegraphics[width=0.35\textwidth]{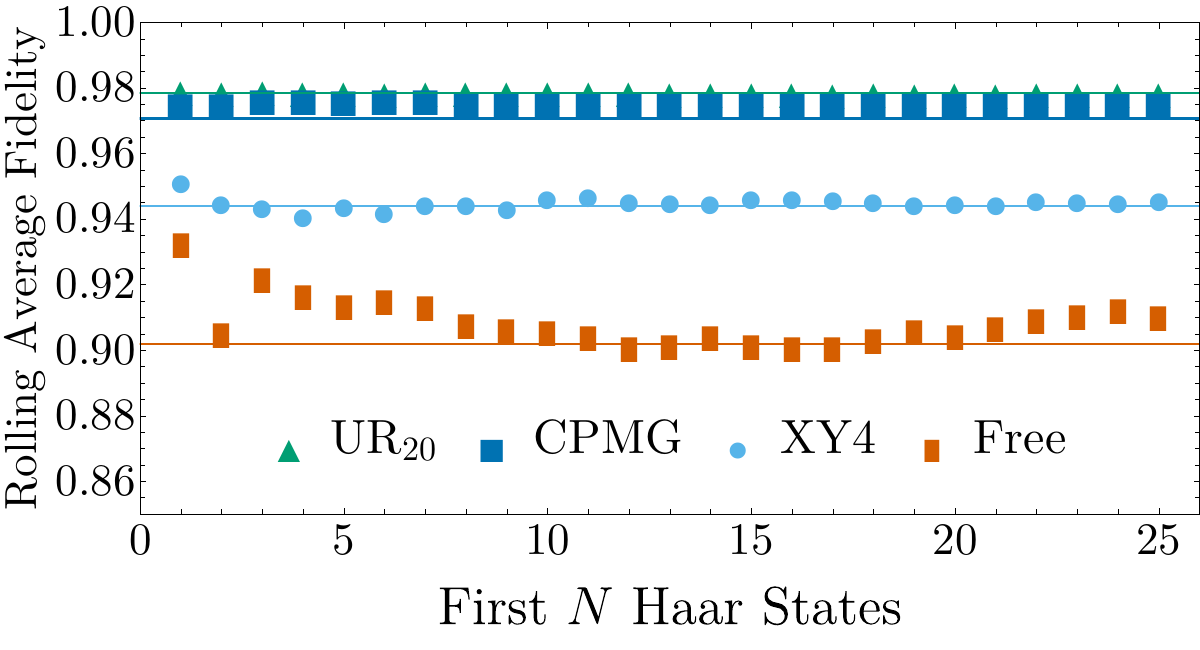}
    \includegraphics[width=0.12\textwidth]
    {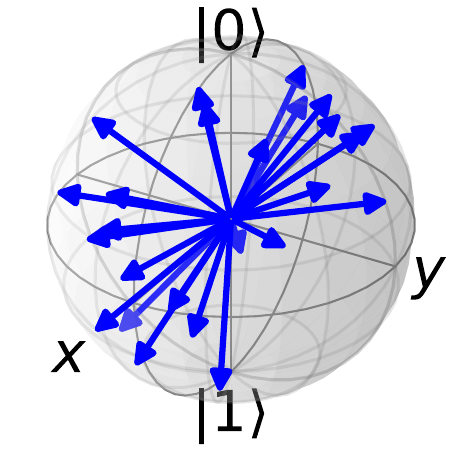}}
    \hfill
    \caption{Representative samples of our results for three DD sequences and free evolution. The Bloch sphere representation of the quantum states used for each plot is shown on the bottom right. (a) Normalized fidelity $\frac{f_e(t)}{f_e(0)}$ under four DD sequences for the initial state $\ket{-i}$ and a fixed calibration cycle on Bogota. (b) We summarize the result of many such fidelity decay curves using a box-plot. Each box shows the max (right-most vertical black lines), inner-quartile range (the orange box), median (the skinny portion of each orange box), and minimum (left-most vertical black lines) time-averaged fidelity, $F(T = 75\mu s)$ in \cref{eq:time-avg-fidelity}, across the $6$ Pauli states and $10$ calibration cycles (for a total of $60$ data points each) on Bogota. The vertical lines denote the performance of each sequence by its median, colored with the same color as the corresponding sequence. We use this type of box plot to summarize the Pauli demonstration results. (c) We show average fidelity convergence as a function of the number of Haar-random states. In particular, the horizontal lines represent $\mathbb{E}_{100-\text{Haar}}[f_e(T = 3.27\mu s)]$ whereas each point represents $\mathbb{E}_{N-\text{Haar}}[f_e(T = 3.27\mu s)]$ for increasing $N$, i.e., the rolling average fidelity. In all cases, we find that $25$ states are sufficient for reasonable convergence as $\mathbb{E}_{25-\text{Haar}}[f_e(T)]$ is within $1\%$ of $\mathbb{E}_{100-\text{Haar}}[f_e(T)]$. The data shown is for Jakarta, but a similar result holds across all devices tested.}
    \label{fig:methods-summary}
\end{figure}

 \subsection{Pauli demonstration for long times}
 \label{subsec:pauliExperiment}
 
 For the Pauli demonstration, we keep the pulse interval $\tau$ fixed to the smallest possible value allowed by the relevant device, $\tau_{\min} = \Delta \approx 36ns$ (or $\approx 71ns$), the width of the $X$ and $Y$ pulses (see \cref{tab:3proc} for specific $\Delta$ values for each device). Practically, this corresponds to placing all the pulses as close to each other as possible, i.e., the peak-to-peak spacing between pulses is just $\Delta$ (except for nonuniformly spaced sequences such as QDD). For ideal pulses and uniformly spaced sequences, this is expected to give the best performance~\cite{Viola:99,Khodjasteh:2007zr}, so unless otherwise stated, all DD is implemented with minimal pulse spacing, $\tau = \Delta$. 

Within this setting, we survey the capacity of each sequence to preserve the six Pauli eigenstates $\{\ket{0},\ket{1},\ket{+},\ket{-},\ket{+i},\ket{-i}\}$ for long times, which we define as $T \geq 75\mu$s. In particular, we generate {fidelity decay} curves like those shown in \cref{subfig:long-fid-decay} by incrementing the number of repetitions of the sequence, $N$, and thereby sampling $f_e(t)$ [\cref{eq:emp-fid}] for increasingly longer times $t$. Using \XY\ as an example, we apply $P_1 P_2 \cdots P_n = (XYXY)^N$ for different values of $N$ while keeping the pulse interval fixed. After generating fidelity decay curves over $10$ or more different calibration cycles across a week, we summarize their performance using a box-plot like that shown in \cref{subfig:box-plot-summary}. For the Pauli demonstrations, the box-plot bins the average normalized fidelity, 
\begin{equation}
    \label{eq:time-avg-fidelity}
    F(T) \equiv \frac{1}{f_e(0)}\expval{f_e(t)}_T = \frac{1}{T} \int_0^T dt \frac{f_e(t)}{f_e(0)},
\end{equation}
at time $T$ computed using numerical integration with Hermite polynomial interpolation. Note that no DD is applied at $f_e(0)$, so we account for state preparation and measurement errors by normalizing. (A sense of the value of $f_e(0)$ and its variation can be gleaned from \cref{fig:sample-fid-decay-curves}. ) The same holds for \cref{subfig:long-fid-decay}.

We can estimate the best-performing sequences for a given device by ranking them by the median performance from this data. In \cref{subfig:box-plot-summary}, for example, this leads to the fidelity ordering $\UR{20} > \CPMG > \XY >$ free evolution on Bogota, which agrees with the impression left by the decay profiles in \cref{subfig:long-fid-decay} generated in a single run. We use $F(T)$ because fidelity profiles $f_e(t)$ are generally oscillatory and noisy, so fitting $f_e(t)$ to extract a decay constant (as was done in Ref.~\cite{Pokharel:2018aa}) does not return reliable results across the many sequences and different devices we tested. We provide a detailed discussion of these two methods in App.~\ref{app:fits}.

\subsection{Haar interval demonstrations}
\label{subsec:haarIntervalExperiments}
The Pauli demonstration estimates how well a sequence preserves quantum memory for long times without requiring excessive data, but it leaves several open questions. Namely, (1) Does DD preserve quantum memory for an \emph{arbitrary} state? (2) Is $\tau = \tau_{\min}$ the best choice empirically? And (3) how effective is DD for short times? In the Haar interval demonstration, we address all of these questions. This setting -- of short times and arbitrary states -- is particularly relevant to improving quantum computations with DD~\cite{Viola:1999:4888,KhodjastehLidar:03,KhodjastehLidar:08,Ng:2011dn}. For example, DD pulses have been inserted into idle spaces to improve circuit performance~\cite{jurcevic2020demonstration,raviVAQEMVariationalApproach2021,Pokharel:better-than-classical-Grover,pokharel2022demonstration,amico2023defining}. 

In contrast to the Pauli demonstration, where we fixed the pulse delay $d = 0$ and the symmetry $ \zeta = a$ (see \cref{eq:xy4-asym} and~\eqref{eq:xy4-sym}) and varied $t$, here we fix $t = T$ and vary $d$ and $ \zeta$, writing $f_e(d,  \zeta; t)$. 
Further, we now sample over a fixed set of $25$ Haar random states instead of the $6$ Pauli states. Note that we theoretically expect the empirical Haar-average over $n$ states $\mathbb{E}_{n-\text{Haar}}[f_e] \equiv \frac{1}{n}\sum_{i=1}^n f_e(\psi_i)$ for $\ket{\psi_i} \sim \text{Haar}$ to converge to the true Haar-average $\expval{f_e}_{\text{Haar}}$ for sufficiently large $n$. As shown by \cref{fig:armonkRollingAvgPlot}, $25$ states are enough for a reasonable empirical convergence while keeping the total number of circuits to submit and run on IBMQ devices manageable in practice. 

\begin{figure*}[ht]
    \centering
    \subfigure{\centering
    \label{subfig:armonkPauli}
    \includegraphics[width=0.32\textwidth]{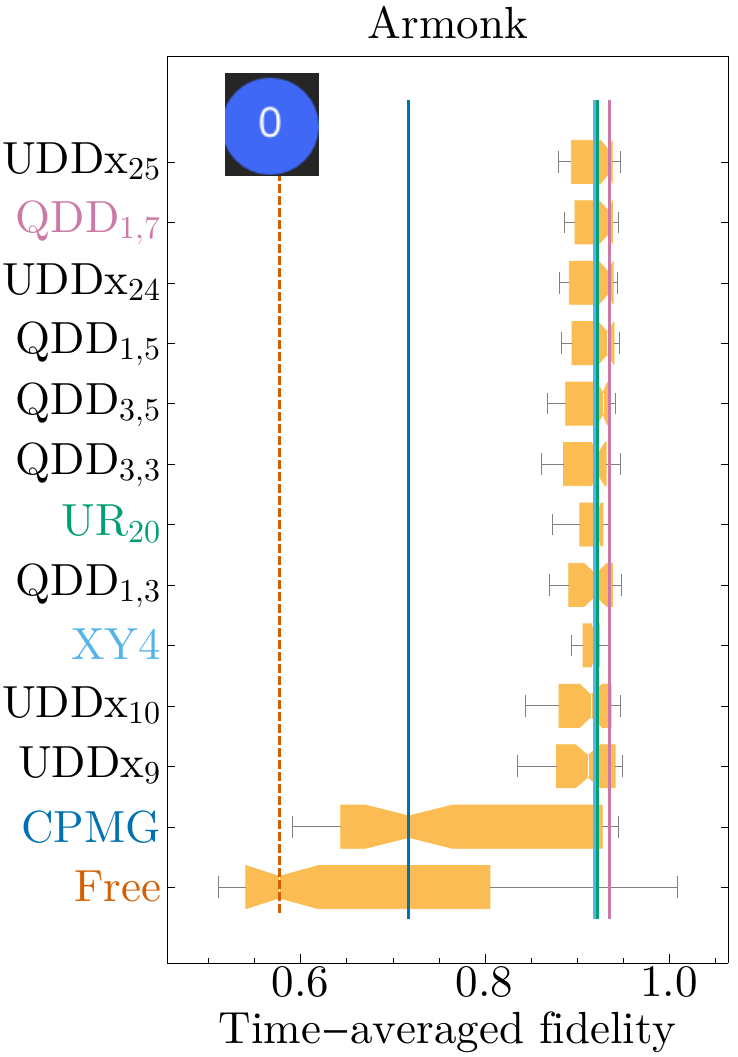}}
    \subfigure{\centering
    \label{subfig:bogotaPauli}
    \includegraphics[width=0.32\textwidth]{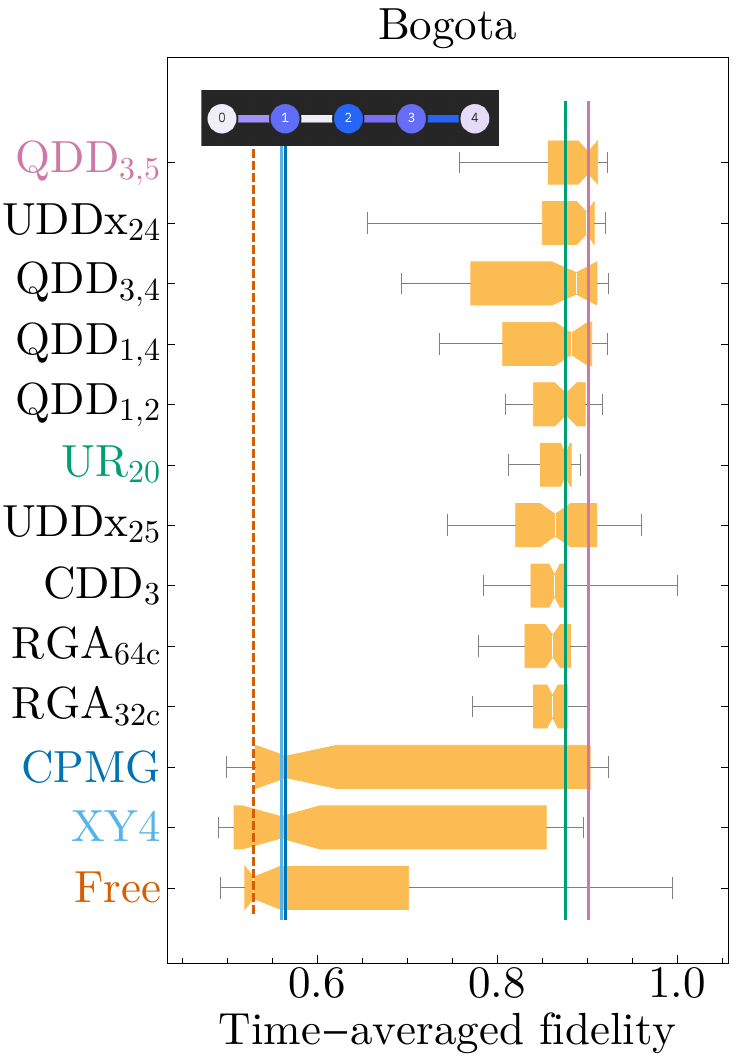}}
    \subfigure{\centering
    \label{subfig:jakartaPauli}
    \includegraphics[width=0.32\textwidth]{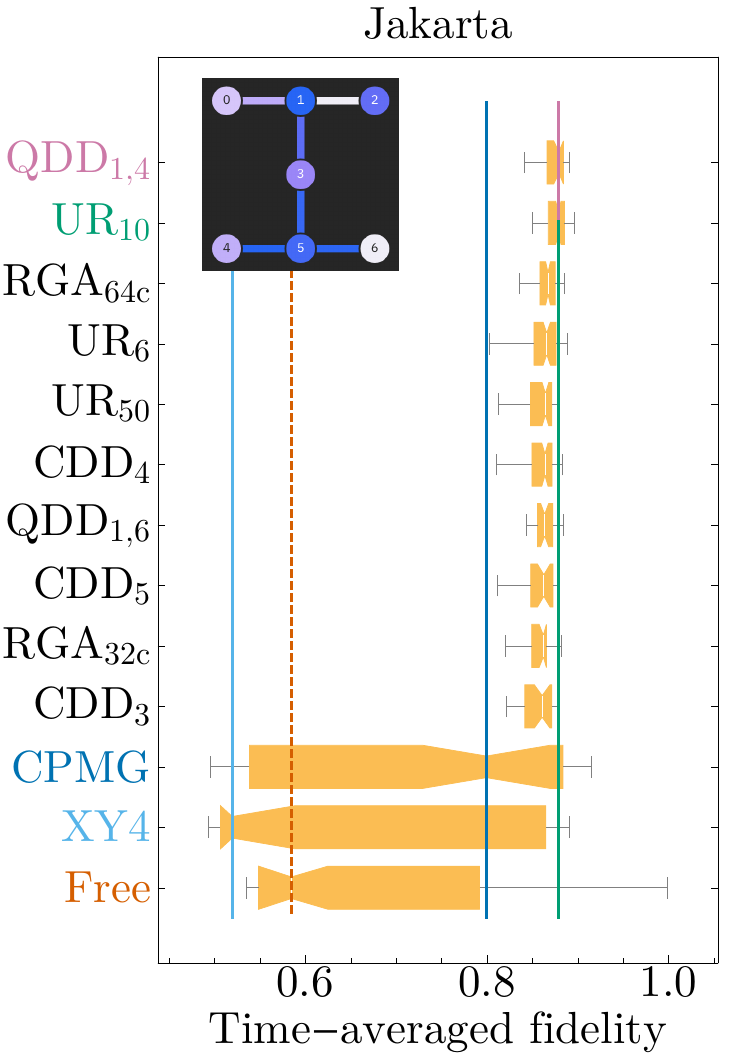}
    }
 
    \caption{A summary of the Pauli demonstration results for all three devices. The top ten sequences are ranked from top to bottom by median average fidelity [\cref{eq:time-avg-fidelity}] for the listed time $T = 75 \mu s$. Also displayed in all cases are \CPMG\ and \XY, along with free evolution (Free). Colored vertical lines indicate the median fidelity of the correspondingly colored sequence (best UR, best QDD, CPMG, \XY, and free evolution). Thin white lines through the orange boxes indicate the median fidelities in all other cases. Otherwise, the conventions of \cref{subfig:box-plot-summary} apply. Two main observations emerge: (1) DD systematically outperforms free evolution as indicated by \enquote{Free} being at the bottom. The corresponding dot-dashed red vertical line denotes the median average fidelity of Free, $F_{\text{Free}}(75\mu s)$, which is below $0.6$ on every device. 
    (2) Advanced DD sequences, especially the UDD and QDD families, provide a substantial improvement over both \CPMG\ and \XY. The best median performance of the top sequence is indicated by a solid green line, $F_{\XY}(75\mu s)$ by a cyan line, and $F_{\text{CPMG}}(75\mu s)$ by a dark blue line.}
    \label{fig:pauliPlot}
\end{figure*}

The Haar interval demonstration procedure is now as follows. For a given DD sequence and time $T$, we sample $f_e(d,  \zeta; t)$ for $ \zeta \in \{a,s\}$ from $d = 0$ to $d = d_{\max}$ for $8$ equally spaced values across $25$ fixed Haar random states and $10$ calibration cycles ($250$ data points for each $d$ value). Here $d = 0$ and $d=d_{\max}$ correspond to the tightest and sparsest pulse placements, respectively. At $d_{\max}$, we consider only a single repetition of the sequence during the time window $T$. To make contact with DD in algorithms, we first consider a short-time limit $T = T_{5\cnot} \approx 4 \mu$s, which is the amount of time it takes to apply 5 CNOTs on the DD-qubit. As shown by the example fidelity decay curves of \cref{subfig:long-fid-decay}, we expect similar results for $T \lesssim 15 \mu$s before fidelity oscillations begin. To make contact with the Pauli demonstration, we also consider a long-time limit of $T = 75\mu$s. Finally, to keep the number of demonstrations manageable, we only optimize the interval of (i) the best-performing UR sequence, (ii) the best-performing QDD sequence, and CPMG, \XY, and Free as constant references. 


\section{Results}
\label{sec:results}

We split our results into three subsections. The first two summarize the results of demonstrations aimed at preserving Pauli eigenstates (\cref{subsec:pauliExperiment}) and Haar-random states (\cref{subsec:haarIntervalExperiments}). In the third subsection, we discuss how theoretical expectations about the saturation of $\CDD{n}$~\cite{Khodjasteh:2007zr} and $\UR{n}$~\cite{Genov:2017aa} compare to the demonstration results.

\subsection{The Pauli demonstration result: DD works and advanced DD works even better}
\label{subsec:pauliResult}

\begin{figure*}[ht]
    \renewcommand{\thesubfigure}{(a1)}
    \centering
    \subfigure[Armonk Short Time, $T_{5\text{CNOT}} = 5\mu s$]{\centering
    \label{subfig:armonkShortIntHaar}
    \includegraphics[trim = 0 40 0 0, clip, width=0.45\textwidth]{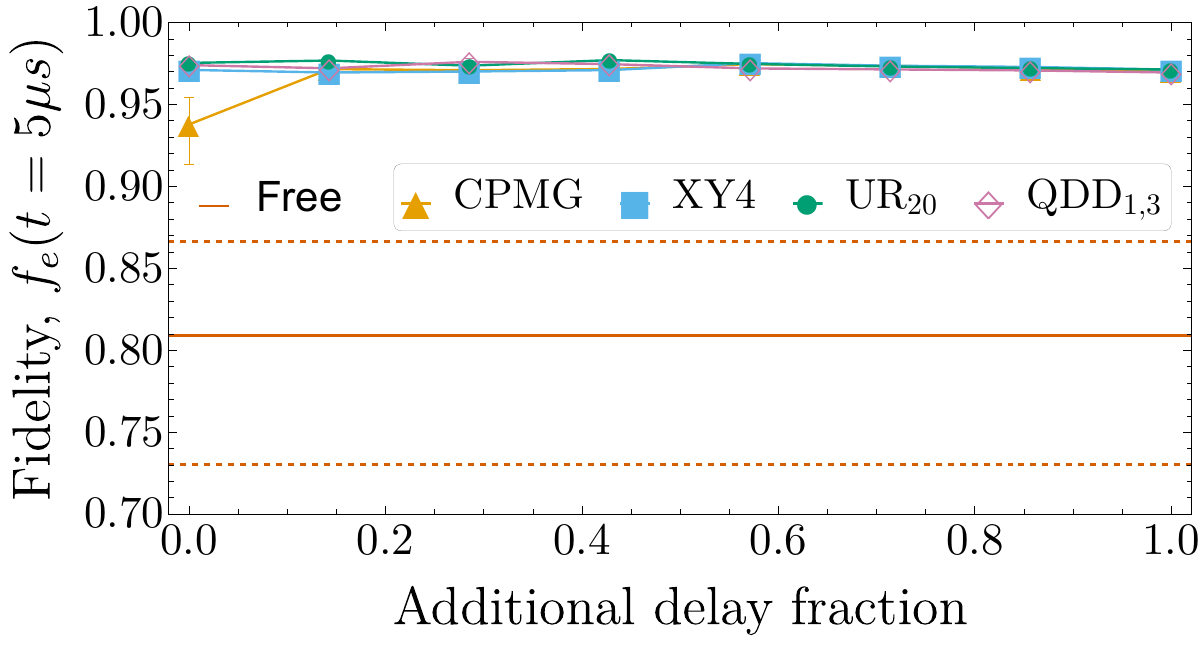}}
    \renewcommand{\thesubfigure}{(a2)}
    \centering
    \subfigure[Armonk Long Time, $T = 75 \mu s$]{\centering
    \label{subfig:armonkLongIntHaar}
    \includegraphics[trim = 0 40 0 0, clip, width=0.45\textwidth]{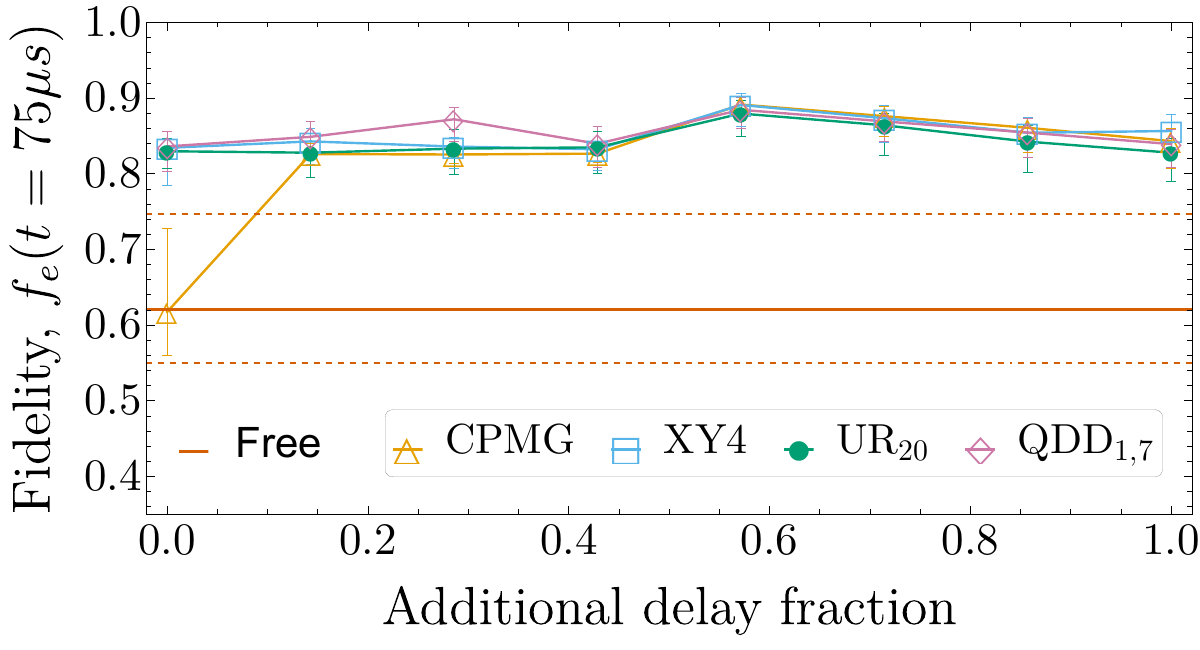}}
    \renewcommand{\thesubfigure}{(b1)}
    \subfigure[Bogota Short Time, $T_{5\text{CNOT}} = 4.65\mu s$]{\centering
    \label{subfig:bogotaShortIntHaar}
    \includegraphics[trim = 0 40 0 0, clip, width=0.45\textwidth]{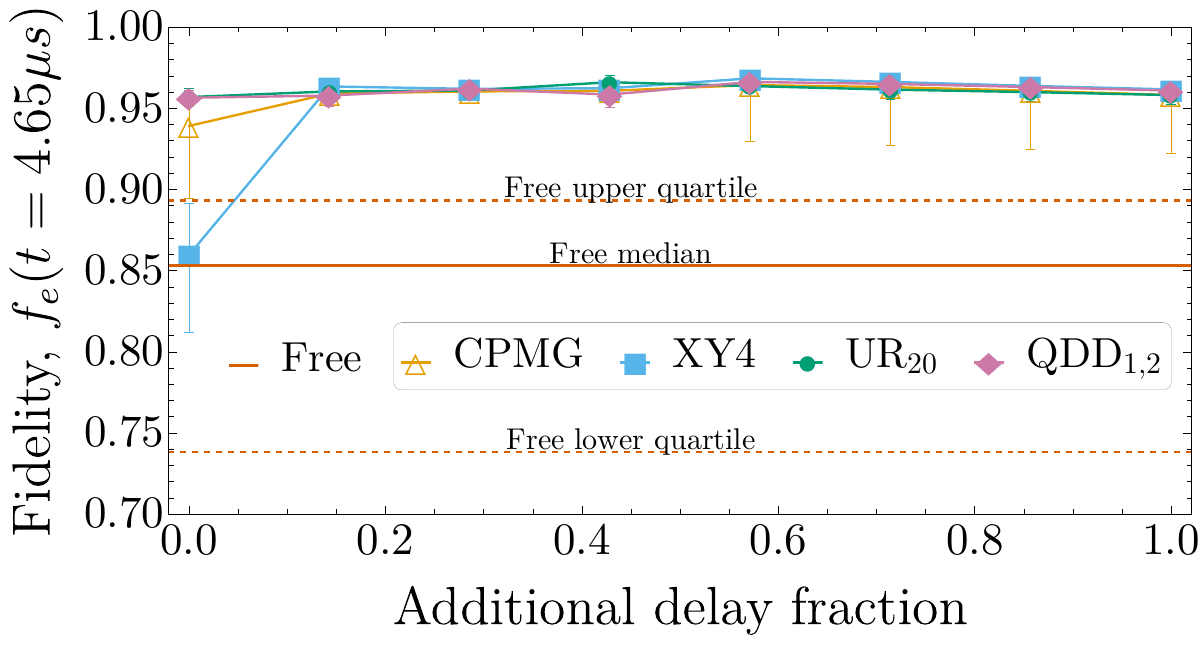}}
    \renewcommand{\thesubfigure}{(b2)}
    \subfigure[Bogota Long Time, $T = 75 \mu s$]{\centering
    \label{subfig:bogotaLongIntHaar}
    \includegraphics[trim = 0 40 0 0, clip, width=0.45\textwidth]{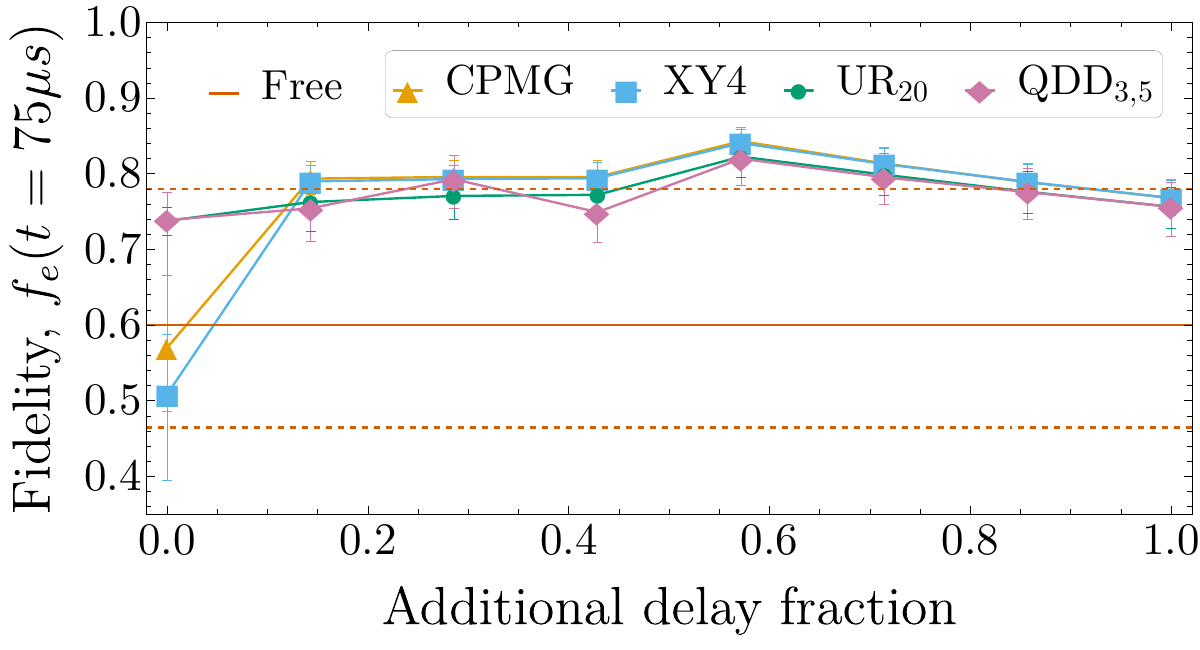}}
    \renewcommand{\thesubfigure}{(c1)}
    \subfigure[Jakarta Short Time, $T_{5\text{CNOT}} = 3.27\mu s$]{\centering
    \label{subfig:jakartaShortIntHaar}
    \includegraphics[width=0.45\textwidth]{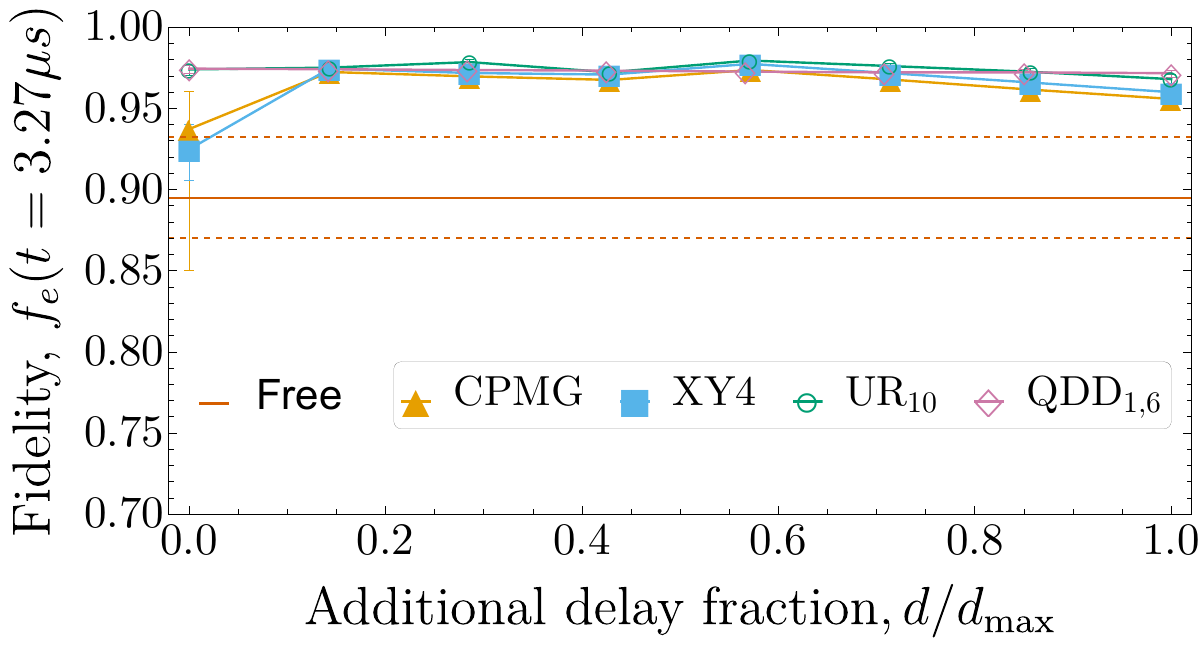}}
    \renewcommand{\thesubfigure}{(c2)}
    \subfigure[Jakarta Long Time, $T = 75 \mu s$]{\centering
    \label{subfig:jakartaLongIntHaar}
    \includegraphics[width=0.45\textwidth]{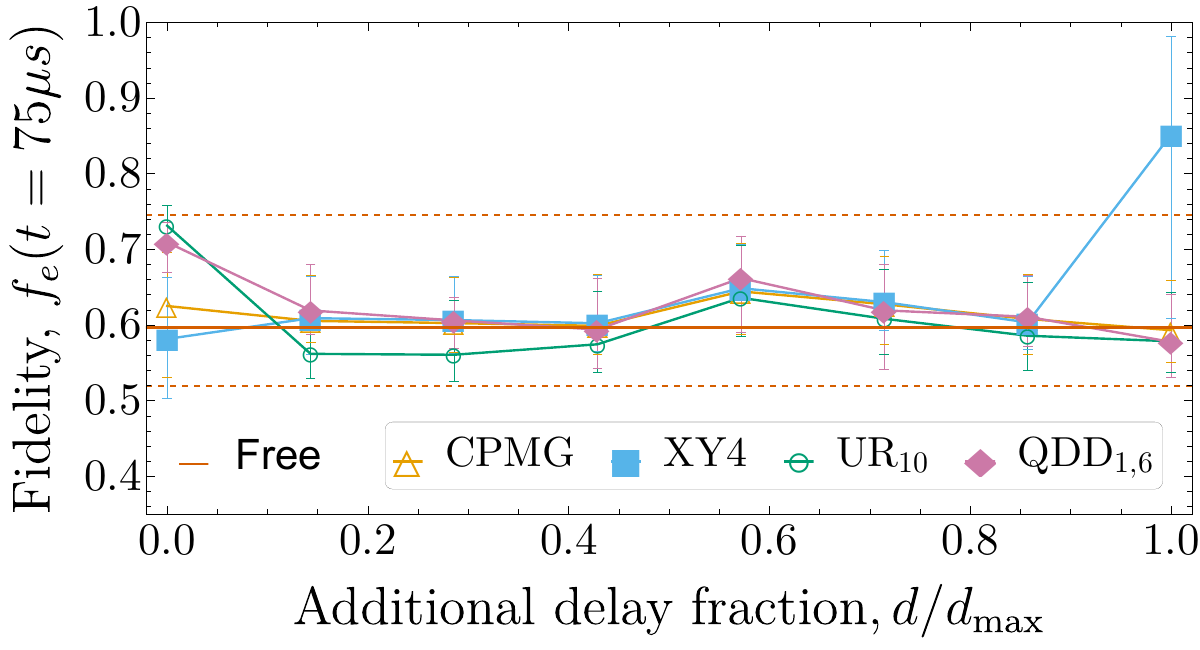}}
    \caption{Summary of the Haar interval demonstration across the three devices. We explore the relationship between the median fidelity $f_e(d,  \zeta^*; t)$ across Haar random states as a function of the relative spacing of the pulse intervals, $d / d_{\max}$. The value $d_{\max}$ corresponds to the largest possible pulse spacing where only a single repetition of a sequence fits. While $d_{\max}$ is a sequence-dependent quantity, we sample $d / d_{\max}$ evenly regardless of the sequence. For a given device and $T$, we plot the best robust sequence ($\UR{n}$) and the best nonuniform sequence $\QDD{n}{m}$ from \cref{fig:pauliPlot}, as well as \CPMG, \XY, and Free evolution as reference sequences. The Free curve has a solid line for its median and a dashed line above and below, representing its upper and lower quartiles.  The parameters $n,m$ in $\QDD{n}{m}$ are chosen so that the sequence fits the time window.  The left and right columns correspond to short ($T = T_{5\text{CNOT}}$) and long times ($T = 75\mu$s), respectively. The confidence intervals are upper and lower quartiles -- 75\% and 25\% of all fidelities lie below them, respectively. When the interval is not visible, this is because it is smaller than the marker for the median.  Both asymmetric (open symbols) and symmetric (closed symbols) sequences were considered, but we display only the better of the two (for some empirically optimal $d^*$).}
    \label{fig:haarIntervalPlot}
\end{figure*}
      
In \cref{fig:pauliPlot}, we summarize the results of the Pauli demonstration. We rank each device's top $10$ sequences by median performance across the six Pauli states and $10$ or more calibration cycles, followed by \CPMG, \XY, and free evolution as standard references. As discussed in \cref{subsec:pauliExperiment}, the figure of merit is the normalized, time-averaged fidelity at $75\mu s$ [see \cref{eq:time-avg-fidelity}] which is a long-time average. 

The first significant observation is that DD is better than free evolution, consistent with numerous previous DD studies. This is evidenced by free evolution (Free) being close to the bottom of the ranking for every device.

Secondly, advanced DD sequences outperform Free, CPMG, and \XY\ (shown as dark-blue and light-blue vertical lines in \cref{fig:pauliPlot}). In particular, 29/30 top sequences across all three devices are advanced -- the exception being \XY\ on \armonk. These sequences perform so well that there is a 50\% improvement in the median fidelity of these sequences (0.85-0.95) over Free (0.45-0.55). The best sequences also have a much smaller variance in performance, as evidenced by their smaller interquartile range in $F$. For example, on \armonk\, 75\% of all demonstration outcomes for $F_{\UDDx{25}}(75\mu s)$ fall between 0.9 and 0.95, whereas for Free, the same range is between 0.55 and 0.8. Similar comparisons show that advanced DD beats \CPMG\ and \XY\ for every device to varying degrees.

Among the top advanced sequences shown, $16/29 \sim 55\%$ are UDD or QDD, which use nonuniform pulse intervals. On the one hand,  the dominance of advanced DD strategies, especially UDD and QDD, is not surprising. After all, these sequences were designed to beat the simple sequences. On the other hand, as reviewed above, many confounding factors affect how well these sequences perform, such as finite pulse width errors and the effect of the rotating frame. It is remarkable that despite these factors, predictions made based on ideal pulses apply to actual noisy quantum devices.

Finally, we comment more specifically on \CPMG\ and \XY, as these are widely used and well-known sequences. Generally, they do better than free evolution, which is consistent with previous results. On \armonk\, \XY\  outperforms CPMG, which outperforms free evolution. On \bogota, both \XY\ and \CPMG\ perform comparably and marginally better than free evolution. Finally, On \jakarta\ \XY\ is worse than \CPMG\ -- and even free evolution -- but the median performance of \CPMG\ is substantially better than that of free evolution. It is tempting to relate these results to the relative values of $T_1$ and $T_2$, as per \cref{tab:3proc}, in the sense that \CPMG\ is a single-axis (``pure-X'') type sequence [\cref{eq:cpmg-def1}] which does not suppress the system-bath $\sigma^x$ coupling term responsible for $T_1$ relaxation, while \XY\ does. Nevertheless, a closer look at \cref{fig:pauliPlot} shows such an explanation would be inconsistent with the fact that both single-axis and multi-axis sequences are among the top $10$ performers for \armonk\ and \bogota. 

The exception is \jakarta, for which there are no single-axis sequences in the top $10$. This processor has a much smaller $T_2$ than $T_1$ (for \armonk\ and \bogota, $T_1>T_2$), so one might expect that a single-axis sequence such as UDD or \CPMG\ would be among the top performers, but this is not the case. In the same vein, the top performing asymmetric $\QDD{n}{m}$ sequences all have $n<m$, despite this opposite ordering of $T_1$ and $T_2$. These results show that the backend values of $T_1$ than $T_2$ are not predictive of DD sequence performance.

\subsection{Haar Interval demonstration Results: DD works on arbitrary states, and increasing the pulse interval can help substantially}

We summarize the Haar interval demonstration results in \cref{fig:haarIntervalPlot}. 
Each plot corresponds to $f_e(d,  \zeta^*; t)$ as a function of $d$, the additional pulse interval spacing. We plot the spacing in relative units of $d / d_{\text{max}}$, i.e., the additional delay fraction, for each sequence. The value $d_{\max}$ corresponds to the largest possible pulse spacing where only a single repetition of a sequence fits within a given unit of time. Hence, $d_{\max}$ depends both on the demonstration time $T$ and the sequence tested, and plotting $d \in [0, d_{\max}]$ directly leads to sequences with different relative scales. Normalizing with respect to $d_{\max}$ therefore makes the comparison between sequences easier to visualize in a single plot. 

 For each device, we compare \CPMG, \XY, the best robust sequence from the Pauli demonstration, the best nonuniform sequence from the Pauli demonstration, and Free. The best robust and nonuniform sequences correspond to $\UR{n}$ and $\QDD{n}{m}$ for each device. 
We only display the choice of $ \zeta$ with the better optimum, and the error bars correspond to the inner-quartile range across the $250$ data points. These error bars are similar to the ones reported in the Pauli demonstration. 

\subsubsection{$d = 0$: DD continues to outperform Free evolution also over Haar random states}
The $d = 0$ (i.e., additional delay fraction $d / d_{\text{max}} = 0$) limit is identical to those in the Pauli demonstration, i.e., with the minimum possible pulse spacing. The advanced sequences, $\UR{n}$ and $\QDD{n}{m}$, outperform Free by a large margin for short times and by a moderate margin for long times. In particular, they have higher median fidelity, a smaller inter-quartile range than Free, and are consistently above the upper quartile in the short-time limit.  But, up to error bars, $\UR{n}$ and $\QDD{n}{m}$ are statistically indistinguishable in terms of their performance, except that $\UR{n}$ has a better median than $\QDD{n}{m}$ for \jakarta.

Focusing on CPMG, for short times, it does slightly worse than the other sequences, yet still much better than Free, but for long times, it does about as poorly as Free. On \bogota\ and \jakarta, \XY\ performs significantly worse than $\UR{n}$, and $\QDD{n}{m}$ and is always worse than CPMG. The main exception to this rule is \armonk\ where \XY is comparable to $\UR{n}$ and $\QDD{n}{m}$ in both the short and long time limits.

Overall, while all sequences lose effectiveness in the long-time limit, the advanced sequences perform well when the pulse spacing is $d=0$. 

\subsubsection{$d>0$: Increasing the pulse interval can improve DD performance}
It is clear from \cref{fig:haarIntervalPlot} (most notably from \cref{subfig:bogotaLongIntHaar}) that increasing the pulse interval can sometimes significantly improve DD performance. For example, considering \cref{subfig:armonkShortIntHaar}, at $d = 0$ \CPMG\ is worse than the other sequences, but by increasing $d$, \CPMG\ matches the performance of the sequences. The same qualitative result occurs even more dramatically for long times (\cref{subfig:armonkLongIntHaar}). Here, \CPMG\ goes from a median fidelity around 0.6 -- as poor as Free -- to around 0.9 at $d / d_{\max} \approx 0.55$. The significant improvement of \CPMG\ with increasing $d$ is fairly generic across devices and times, with the only exception being \jakarta\ and long times. Thus, even a simple sequence such as \CPMG\ (or \XY, which behaves similarly) can compete with the highest-ranking advanced sequences if the pulse interval is optimized. Unsurprisingly, optimizing pulse-interval can help, but the degree of improvement is surprising, particularly the ability of simple sequences to match the performance of advanced ones. 

\subsubsection{$d = d_{\max}$: Performance at the single cycle limit}
In a similar vein, it is notable how well sequences do in the $d = d_{\max}$ limit (at the right of each plot in \cref{fig:haarIntervalPlot} where $d / d_{\max} = 1$). For \CPMG\ at $T = 75\mu s$, this corresponds to applying a pulse on average every $37.5 \mu s$; this certainly does not obey the principle of making $\tau$ small, implied from error terms scaling as $\mathcal{O}(\tau^n)$ as in the standard theory. There is always a tradeoff between combating noise and packing too many pulses on real devices with finite width and implementation errors. Incoherent errors result in a well-documented degradation that reduces the cancelation order for most of the sequences we have considered; see \cref{sec:DD-imperf}. In addition, coherent errors build up as oscillations in the fidelity; see \cref{subsec:theory-means-in-practice}. Low-order sequences such as CPMG and \XY\ are particularly susceptible to both incoherent and coherent errors, and are therefore expected to exhibit an optimal pulse interval. Moreover, the circuit structure will restrict the pulse interval when implementing DD on top of a quantum circuit. The general rule gleaned from \cref{fig:haarIntervalPlot} is to err toward the latter and apply DD sparsely. In particular, $d = d_{\max}$ results in a comparable performance to $d = 0$ or a potentially significant improvement.

Nevertheless, whether dense DD is better than sparse DD can depend on the specific device characteristics, desired timescale, and relevant metrics. As a case in point, note the Haar demonstration results on \jakarta\ in the long-time limit. Here, most sequences -- aside from \XY\ -- deteriorate with increasing $d$. Surprisingly, the best strategy in median performance is \XY, which does come at the cost of a sizeable inner-quartile range.  

$\UR{n}$ for $d = 0$ does substantially better than Free for all six panels, which is an empirical confirmation of its claimed robustness. Even for $d >0$, $\UR{n}$ remains a high fidelity and low variance sequence. Since other sequences only roughly match $\UR{n}$ upon interval optimization, using a robust sequence is a safe default choice.

\subsection{Saturation of CDD and UDD, and an optimum for UR}
\label{subsec:sat-of-cdd-and-ur}

\begin{figure*}[ht]
    \centering
    \subfigure[Comparison of CDD sequences on Bogota]{\centering
    \label{subfig:bogotaCDDComp}
    \includegraphics[width=0.3\textwidth]{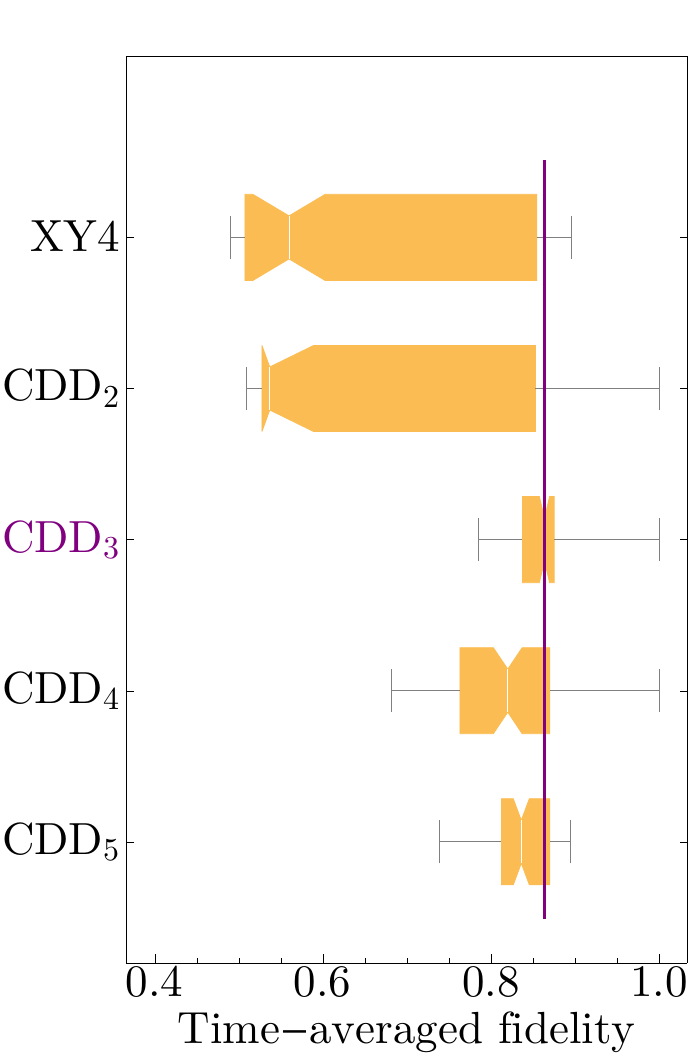}}
    \subfigure[Comparison of UDD sequences on Bogota]{\centering
    \label{subfig:bogotaUDDComp}
    \includegraphics[width=0.318\textwidth]{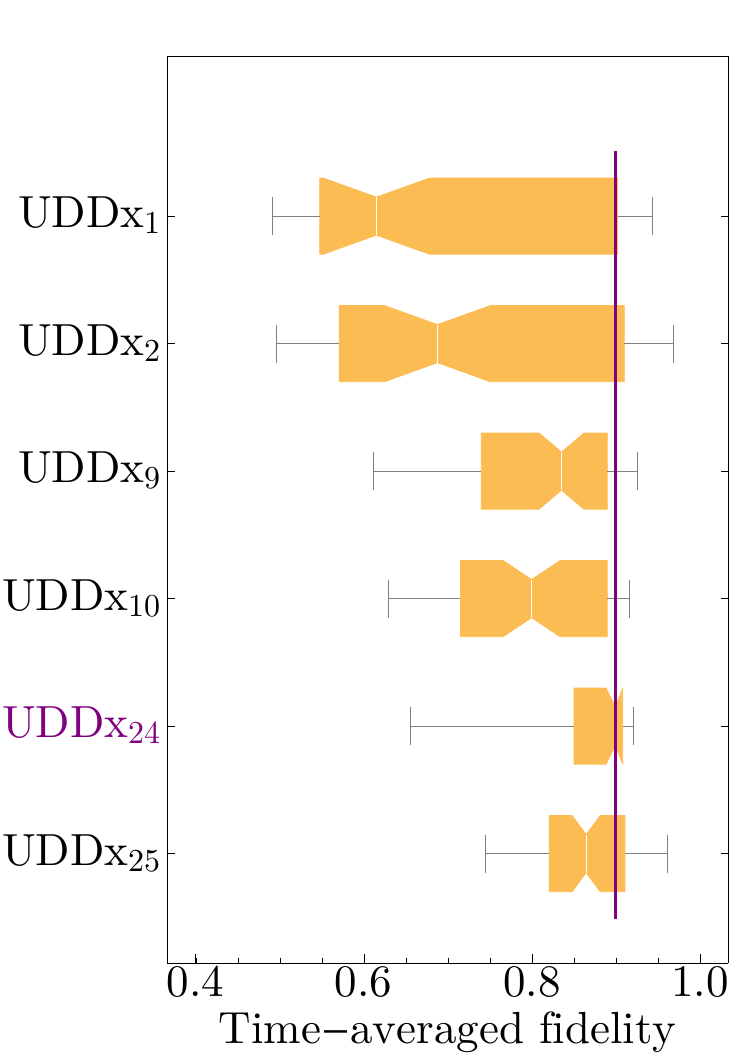}}
    \subfigure[Comparison of UR sequences on Bogota]{\centering
    \label{subfig:bogotaURComp}
    \includegraphics[width=0.3\textwidth]{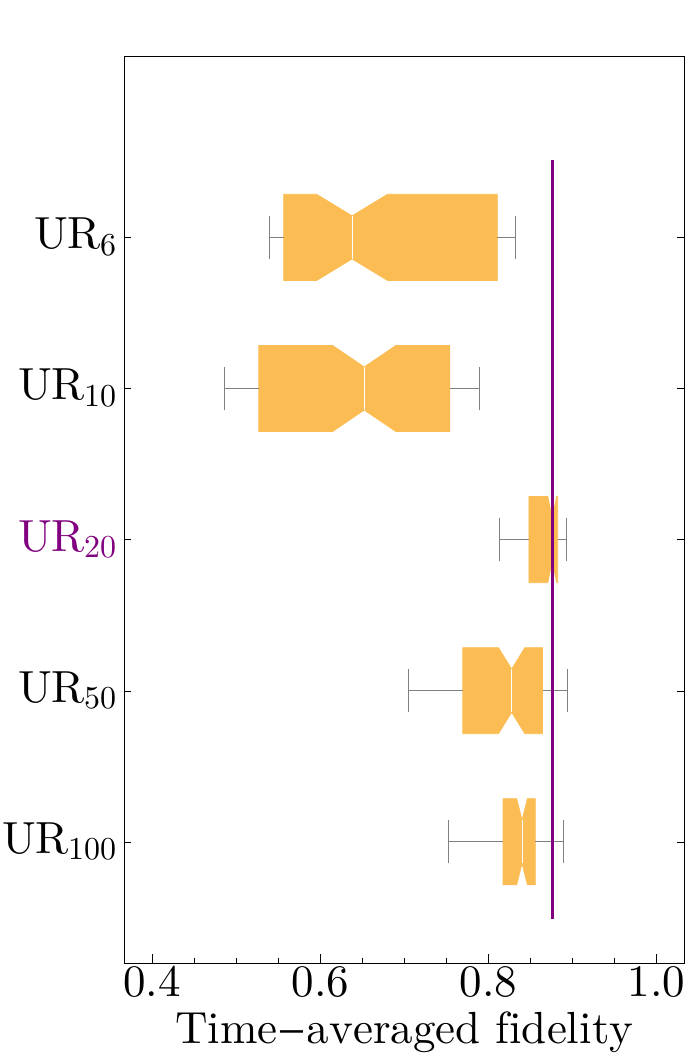}}
    \caption{$f_e(t)$ Average fidelity at $75\mu$s for (a) $\CDD{n}$ for $n\in\{1,\dots,5\}$, (b) $\text{UDD}_{x_{n}}$ for $x\in \{1,2,9,24,25\} $, (c) $\UR{n}$ for $n\in\{6,10,20,50,100\}$. All three sequence families exhibit saturation on \bogota\ as we increase $n$, as expected from theory~\cite{Ng:2011dn,UL:10,Genov:2017aa}. The vertical purple line denotes the median performance of the correspondingly colored sequence in each panel, which is also the top-performing sequence by this metric.}
    \label{fig:CDDandURsaturation}
\end{figure*}

In \cref{fig:CDDandURsaturation}, we display the time-averaged fidelity from \bogota\ for $\text{CDD}_n$, $\text{UR}_n$ and $\text{UDD}_{x_{n}}$ as a function of $n$.  Related results for the other DD sequence families are discussed in  \cref{app:all-pauli-data}. As discussed in \cref{sec:HOP}, $\CDD{n}$ performance is expected to saturate at some $n_{\text{opt}}$ according to \cref{eq:opt-cdd-order}. In \cref{subfig:bogotaCDDComp}, we observe evidence of this saturation at $n_{\text{opt}} = 3$ on \bogota. We can use this to provide an estimate of $\epsilon = \|H_B\| + \|H_{SB}\|$. 
Substituting $n_{\text{opt}}$ into \cref{eq:opt-cdd-order} we find:
\begin{equation}
    \overline{c} \epsilon \Delta \in [4^{-5}, 4^{-4}] = [9.77 \times 10^{-4}, 3.91 \times 10^{-3}].
\end{equation}
This means that $\epsilon \Delta \ll 1$ (we set $\overline{c} \approx 1$), which confirms the assumption we needed to make for DD to give a reasonable suppression given that \XY\ yields $\mathcal{O}(\epsilon \tau^2)$ suppression. This provides a level of empirical support for the validity of our assumptions. In addition, $\Delta \approx 51$ns on \bogota, so we conclude that $\epsilon\approx 0.5$MHz. Since qubit frequencies are roughly $\omega_q \approx 4.5 - 5$ GHz on IBM devices, this also confirms that $\omega_q \gg \epsilon$, as required for a DD pulse.  We observe a similar saturation in $\text{CDD}_{n}$ on Jakarta and Armonk as well (\cref{app:all-pauli-data}). 

Likewise, for an ideal demonstration with fixed time $T$, the performance of $\UDDx{n}$ should scale as $\mathcal{O}(\tau^n)$, and hence we expect a performance that increases monotonically $n$. In practice, this performance should saturate once the finite-pulse width error $\mathcal{O}(\Delta)$ is the dominant noise contribution~\cite{UL:10}. Once again, the $\UDD{n}$ sequence performance on \bogota\ is consistent with theory. In particular, we expect (and observe) a consistent increase in performance with increasing $n$ until performance saturates. While this saturation is also seen on Armonk, on Jakarta $\UDDx{n}$'s performance differs significantly from theoretical expectations (see \cref{app:all-pauli-data}).

$\UR{n}$ also experiences a tradeoff between error suppression and noise introduced by pulses. After a certain optimal $n$, the performance for $\UR{n}$ is expected to drop~\cite{Genov:2017aa}.  In particular, while $\UR{n}$ yields $\mathcal{O}(\epsilon_r^{n/2})$ suppression with respect to flip angle errors (\cref{sec:MSE}), all $\UR{n}$ provide $\mathcal{O}(\tau^2)$ decoupling, i.e., adding more free evolution periods also means adding more noise. Thus, we expect $\UR{n}$ to improve with increasing $n$ until performance saturates.  On \bogota, by increasing up to $\UR{20}$, we gain a large improvement over $\UR{10}$, but increasing further to $\UR{50}$ or $\UR{100}$ results in a small degradation in performance; see \cref{fig:CDDandURsaturation}. A similar saturation occurs with \jakarta\ and \armonk\ (see \cref{app:all-pauli-data}).

\section{Summary and Conclusions}
\label{sec:Conclusion} 
We performed an extensive survey of $10$ DD families (a total of $60$ sequences) across three superconducting IBM devices. In the first set of demonstrations (the Pauli demonstration, \cref{subsec:pauliExperiment}), we tested how well these 60 sequences preserve the six Pauli eigenstates over relatively long times ($25 - 75\mu $s). Motivated by theory, we used the smallest possible pulse interval for all sequences. We then chose the top-performing QDD and UR sequences from the Pauli demonstration for each device, along with CPMG, \XY, and free evolution as baselines, and studied them extensively. In this second set of demonstrations (the Haar demonstration \cref{subsec:haarIntervalExperiments}), we considered $25$ (fixed) Haar-random states for a wide range of pulse intervals, $\tau$. 

In the Pauli demonstration (\cref{subsec:pauliExperiment}), we ranked sequence performance by the median time-averaged fidelity at $T = 75\mu s$. This ranking is consistent with DD theory. The best-performing sequence on each device substantially outperforms free evolution. Moreover, the expected deviation, quantified using the inner-quartile range of the average fidelity, was much smaller for DD than for free evolution. Finally, $29$ out of the $30$ best performing sequences were ``advanced'' DD sequences, explicitly designed to achieve high-order cancellation or robustness to control errors.

We reported point-wise fidelity rather than the coarse-grained time-averaged fidelity in the Haar-interval demonstration (\cref{subsec:haarIntervalExperiments}). At $\tau = 0$, the Haar-interval demonstration is identical to the Pauli demonstration except for the expanded set of states. Indeed, we found the same hierarchy of sequence performance between the two demonstrations. For example, on \jakarta, we found \XY\ < \CPMG\ < $\QDD{1}{6}$ < $\UR{10}$ for both \cref{fig:pauliPlot} and \cref{fig:haarIntervalPlot}. This suggests that a test over the Pauli states is a good proxy to Haar-random states for our metric.

However, once we allowed the pulse interval ($\tau$) to vary, we found two unexpected results. First, contrary to expectations, advanced sequences, which theoretically provide a better performance, do not retain their performance edge. Second, for most devices and times probed, DD sequence performance improves or stays roughly constant with increasing pulse intervals before decreasing slightly for very long pulse intervals. This effect is particularly significant for the basic \CPMG\ and \XY\ sequences. Relating these two results, we found that with pulse-interval optimization, the basic sequences' performance is statistically indistinguishable from that of the advanced UR and QDD sequences. In stark contrast to the theoretical prediction favoring short intervals, choosing the longest possible pulse interval, with one sequence repetition within a given time, is generally better than the minimum interval. The one exception to these observations is the \jakarta\ processor for $T = 75\mu\text{s}$, which is larger than its mean $T_2$ of $20.7 \mu$s. Here, the advanced sequences significantly outperform the basic sequences at their respective optimal interval values ($\tau=0$ for the advanced sequences), and DD performance degrades with sufficiently large pulse intervals. The short $T_2$ for \jakarta\ is notable, since in contrast, $T = 75\mu\text{s} < \expval{T_2}$ for both \armonk\ ($\expval{T_2}=230\mu$s) and \bogota\ ($\expval{T_2}=146\mu$s). We may thus conclude that, overall, sparse DD is preferred to tightly-packed DD, provided decoherence in the free evolution periods between pulses is not too strong.

The UR sequence either matched or nearly matched the best performance of any other tested sequence at any $\tau$ for each device. It also achieved near-optimal performance at $\tau = 0$ in four of the six cases shown in \cref{fig:haarIntervalPlot}. This is a testament to its robustness and suggests the $\UR{n}$ family is a good generic choice, provided an optimization to find a suitable $n$ for a given device is performed. In our case, this meant choosing the top performing $\UR{n}$ member from the Pauli demonstration. Alternatively, our results suggest that as long as $T < T_2$, one can choose a basic sequence and likely achieve comparable performance by optimizing the pulse interval. In other words, \emph{optimizing the pulse interval for a given basic DD sequence or optimizing the order of an advanced DD sequence at zero pulse interval are equally effective strategies for using DD in practice}. However, the preferred strategy depends on hardware constraints. For example, if OpenPulse access (or a comparable control option) is not available so that a faithful $\UR{n}$ implementation is not possible, one would be constrained to optimizing \CPMG\ or \XY\ pulse intervals. Under such circumstances, where the number of DD optimization parameters is restricted, using a variational approach to identify the remaining parameters (e.g., pulse intervals) can be an effective approach \cite{raviVAQEMVariationalApproach2021}.

Overall, theoretically promising, advanced DD sequences work well in practice. However, one must fine-tune the sequence to obtain the best DD performance for a given device. A natural and timely extension of our work would be developing a rigorous theoretical understanding of our observations, which do not always conform to previous theoretical expectations. Developing DD sequences for specific hardware derived using the physical model of the system instead of trial-and-error optimization, or using machine learning methods~\cite{alex2020deep}, are other interesting directions. A thorough understanding of how to tailor DD sequences to today's programmable quantum computers could be vital in using DD to accelerate the attainment of fault-tolerant quantum computing~\cite{Ng:2011dn}.

\begin{dataAndAvail}
    The code and data that support the findings of this study are openly available at the following URL/DOI: \href{https://doi.org/10.5281/zenodo.7884641}{https://doi.org/10.5281/zenodo.7884641}. In more detail, this citable Zenodo repository~\cite{nicholas_ezzell_2023_7884641} contains (i) all raw and formatted data in this work (including machine calibration specifications), (ii) python code to submit identical demonstrations, (iii) the Mathematica code used to analyze the data, and (iv) a brief tutorial on installing and using the code as part of the GitHub ReadMe.
\end{dataAndAvail}

\acknowledgments

NE was supported by the U.S. Department of Energy (DOE) Computational Science Graduate Fellowship under Award Number DE-SC0020347. This material is based upon work supported by
the National Science Foundation the Quantum Leap Big
Idea under Grant No. OMA-1936388. LT and GQ acknowledge funding from the U.S. Department of Energy, Office of Science, Office of Advanced Scientific Computing Research, Accelerated Research in Quantum Computing under Award Number DE-SC0020316. This material is based upon work supported by the U.S. Department of Energy, Office of Science, SC-1
U.S. Department of Energy
1000 Independence Avenue, S.W.
Washington DC 20585,
under Award Number(s) DE-SC0021661. We acknowledge
the use of IBM Quantum services for this work. The
views expressed are those of the authors, and do not reflect the official policy or position of IBM or the IBM
Quantum team. We acknowledge the access to advanced
services provided by the IBM Quantum Researchers Program. We thank Vinay Tripathi for insightful discussions, particularly regarding device-specific effects. D.L. is grateful to Prof. Lorenza Viola for insightful discussions and comments.


\appendix

\section{Summary of the DD sequences benchmarked in this work}
\label{app:DD-summary}

Here we provide definitions for all the DD sequences we tested. To clarify what free evolution periods belong between pulses, we treat uniform and nonuniform pulse interval sequences separately. When possible, we define a pulse sequence in terms of another sequence (or entire DD sequence) using the notation $[ \cdot ]$. In addition, several sequences are recursively built from simpler sequences. When this happens, we use the notation $S = s_1([s_2])$, whose meaning is illustrated by the example of $\text{CDD}_n$ [\cref{eq:CDD-def}].

\subsection{Uniform pulse interval sequences}

All the uniform pulse interval sequences are of the form
    \begin{equation}
        f_{\tau} - P_1 - f_{2 \tau} - P_2 - f_{2 \tau} - \ldots - f_{2 \tau} - P_n - f_{\tau}.
    \end{equation}
For brevity, we omit the free evolution periods in the following definitions. 

We distinguish between single and multi-axis sequences, by which we mean the number of orthogonal interactions in the system-bath interaction (e.g., pure dephasing is a single-axis case), not the number of axes used in the pulse sequences.

First, we list the single-axis DD sequences:
\begin{subequations}
    \begin{align}
        \text{Hahn} &\equiv X \\
        \text{super-Hahn} / \text{RGA}_{2x} &\equiv X - \overline{X} \\
        \text{RGA}_{2y} &\equiv Y - \overline{Y} \\
        \text{CPMG} &\equiv X - X  \\
        \text{super-CPMG} &\equiv X - X - \overline{X} - \overline{X} .
    \end{align}
\end{subequations}

Second, the $\text{UR}_n$ sequence for $n \geq 4$ and $n$ even is defined as
\begin{subequations}
    \begin{align}
        \text{UR}_n &= (\pi)_{\phi_1} - (\pi)_{\phi_2} - \ldots - (\pi)_{\phi_n} \\
        \phi_k &= \frac{(k-1)(k-2)}{2} \Phi^{(n)} + (k - 1) \phi_2 \\
        \Phi^{(4m)} &= \frac{\pi}{m} \ \ \ \Phi^{(4m + 2)} = \frac{2 m \pi}{2m + 1},
    \end{align}
\end{subequations}
where $(\pi)_{\phi}$ is a $\pi$ rotation about an axis which makes an angle $\phi$ with the $x$-axis, $\phi_1$ is a free parameter usually set to $0$ by convention, and $\phi_2 = \pi / 2$ is a standard choice we use. This is done so that $\text{UR}_4 = \XY$ as discussed in Ref.~\cite{Genov:2017aa} (note that despite this $\text{UR}_n$ was designed for single-axis decoherence).

Next, we list the multi-axis sequences.  We start with \XY\ and all its variations:
\begin{subequations}
    \begin{align}
        \text{XY4}\text{\textbackslash}\text{CDD}_1 &\equiv Y-X-Y-X \\
        \text{CDD}_n &\equiv \text{XY4}([\text{CDD}_{n-1}]) \\
        \text{RGA}_4 &\equiv \overline{Y}-X-\overline{Y}-X \\
        \text{RGA}_{4p} &\equiv \overline{Y}-\overline{X}-\overline{Y}-\overline{X} \\
        \text{RGA}_{8c}/\text{XY8} &\equiv X - Y - X - Y - Y - X - Y - X  \\
        \text{RGA}_{8a} &\equiv X - \overline{Y} - X - \overline{Y} - Y - \overline{X} - Y - \overline{X} \\
        \begin{split}
        \text{super-Euler} &\equiv 
        X - Y - X - Y - Y - X - Y - X \\
        {}-& \overline{X} - \overline{Y} - \overline{X} - \overline{Y} - \overline{Y} - \overline{X} - \overline{Y} - \overline{X}
        \end{split} \\
        \text{KDD} &\equiv [{K}_{\pi / 2}] - [{K}_{0}] - [{K}_{\pi / 2}] - [{K}_{0}],
        \label{eq:KDD}
    \end{align}
\end{subequations}
where $K_\phi$ is a composite of 5 pulses:
    \begin{equation}
        K_\phi \equiv (\pi)_{\pi / 6 + \phi} - (\pi)_{\phi} - (\pi)_{\pi / 2 + \phi} - (\pi)_{\phi} - (\pi)_{\pi / 6 + \phi} .
    \end{equation}
For example, $(\pi)_0 = X$ and $(\pi)_{\pi / 2} = Y$. The series $K(\phi)$ is itself a $\pi$ rotation about the $\phi$ axis followed by a $-\pi/3$ rotation about the $z$-axis. To see this, note that a $\pi$ rotation about the $\phi$ axis can be written $(\pi)_\phi = R_Z(-\phi) R_Y(-\pi / 2) R_Z(\pi) R_Y(\pi / 2) R_Z(\phi)$, and one can verify the claim by direct matrix multiplication. KDD~\eqref{eq:KDD} is the Knill-composite version of \XY\ with a total of 20 pulses~\cite{souza_robust_2011, souza_robust_2012}. Note that the alternation of $\phi$ between $0$ and $\pi/2$ means that successive pairs give rise to a $-\pi/3 + \pi/3 = 0$ $z$-rotation at the end.

Next, we list the remaining multi-axis RGA sequences: 
\begin{subequations}
    \begin{align}
        \text{RGA}_{16b} &\equiv \text{RGA}_{4p}([\text{RGA}_{4p}]) \\
        \text{RGA}_{32a} &\equiv \text{RGA}_{4}( [\text{RGA}_{8a}] ) \\
        \text{RGA}_{32c} &\equiv \text{RGA}_{8c}([\text{RGA}_{4}]) \\
        \text{RGA}_{64a} &\equiv \text{RGA}_{8a}([\text{RGA}_{8a}]) \\
        \text{RGA}_{64c} &\equiv \text{RGA}_{8c}([\text{RGA}_{8c}]) \\
        \text{RGA}_{256a} &\equiv \text{RGA}_{4}([\text{RGA}_{64a}])
    \end{align}
\end{subequations}

\subsection{Nonuniform pulse interval sequences}
The nonuniform sequences are described by a general DD sequence of the form
    \begin{equation}
        f_{\tau_1} - P_1 - f_{\tau_2} - \ldots - f_{\tau_{m}} -  P_m - f_{\tau_{m+1}} 
    \end{equation}
for pulses $P_j$ applied at times $t_j$ for $1 \leq j \leq m$. Thus for ideal, zero-width pulses, the interval times are $\tau_j = t_j - t_{j-1}$ with $t_0 \equiv 0$ and $t_{m+1} \equiv T$, the total desired duration of the sequence.

\subsubsection{Ideal UDD}
For ideal pulses, $\UDDx{n}$ is defined as follows. For a desired evolution time $T$,  apply $X$ pulses at times $t_j$ given by
\begin{equation}
    \label{eq:uhrig-times}
    t_j = T \sin^2 \left( \frac{j \pi}{2n + 2} \right),
\end{equation}
for $j = 1, 2, \ldots, n$ if $n$ is even and $j = 1, 2, \ldots, n + 1$ if $n$ is odd. Hence, $\UDDx{n}$ always uses an even number of pulses -- $n$ when $n$ is even and $n+1$ when $n$ is odd -- so that when no noise or errors are present $\UDDx{n}$ faithfully implements a net identity operator. 

\subsubsection{Ideal QDD}
To define QDD it is useful to instead define $\UDDx{n}$ in terms of the pulse intervals, $\tau_j = t_j - t_{j-1}$. By defining the normalized pulse interval, 
\begin{equation}
    \label{eq:norm-uhrig-times}
    s_j = \frac{t_j - t_{j-1}}{t_1} = \sin\left(\frac{(2j-1)\pi}{2n + 2}\right) \csc\left(\frac{\pi}{2n + 2}\right),
\end{equation}
for $j = 1, 2, \ldots, n+1$, we can define $\UDDx{n}$ over a total time $T$,
\begin{equation}
    \label{eq:uddx-def}
    \UDDx{n}(T) \equiv f_{s_1 T} - X - \ldots - f_{s_n T} - X - f_{s_{n+1} T} - X^n,
\end{equation}
where the notation $X^n$ means that the sequence ends with $X$ ($I$) for odd (even) $n$.
From this, $\QDD{n}{m}$ has the recursive definition
\begin{align}
    \QDD{n}{m} &\equiv \UDDx{m}(s_{1} T) - Y - \UDDx{m}(s_{n} T) - \ldots \notag \\
    &-\UDDx{m}(s_{n} T) - Y - \UDDx{m}(s_{n+1} T) - Y^n
    \label{eq:qddnm-def1} 
\end{align}
This means that we implement $\UDDy{n}$ (the outer $Y$ pulses) and embed an $m^{\text{th}}$ order $\UDDx{m}$ sequence within the free evolution periods of this sequence. The inner $\UDDx{m}$ sequences have a rescaled total evolution time $s_k T$; since the decoupling properties only depend on $\tau_j$ (and not the total time), we still obtain the expected inner cancellation. Written in this way, the total evolution time of $\QDD{n}{m}$ is $S_n^2 T$ where $S_n = \sum_{j=1}^{n+1} s_j$. 

To match the convention of all other sequences presented, we connect this definition to one in which the total evolution time of $\QDD{n}{m}$ is itself $T$. First, we implement the outer $\UDDy{n}$ sequence with $Y$ pulses placed at times $t_j$ according to \cref{eq:uhrig-times}. The inner $X$ pulses must now be applied at times 
\begin{equation}
    t_{j,k} = \tau_j \sin^2\left(\frac{k \pi}{2m + 2}\right) + t_{j-1},
\end{equation}
where $j = 1, 2, \ldots, n$ if $n$ is even (or $j = 1, 2, \ldots, n + 1$ if $n$ is odd), with a similar condition for $k$ up to $m$ if $m$ is even (or $m+1$ if $m$ is odd). Note that when $m$ is odd, we end each inner sequence with an $X$, and then the outer sequence starts where a $Y$ must be placed simultaneously. In these cases, we must apply a $Z = X Y$ pulse (ignoring the global phase, which does not affect DD performance). Hence, we must apply rotations about $X$ and $Z$ when $m$ is odd and $X$ and $Y$ when $m$ is even. This can be avoided by instead re-defining the inner (or outer) sequence as $Z$ when $m$ is odd and then combining terms to get $X$ and $Y$ again. In this work, we choose the former approach. 
It would be interesting to compare this with the latter approach in future work.

\subsubsection{UDD and QDD with finite-width pulses}
For real pulses with finite width $\Delta$, these formulas must be slightly augmented. First, defining $t_j$ is ambiguous since the pulse application cannot be instantaneous at time $t_j$. In our implementation, pulses start at time $t = t_j - \Delta$ so that they end when they should be applied in the ideal case. Trying two other strategies -- placing the center of the pulse at $t_j$ or the beginning of the pulse at $t_j$ -- did not result in a noticeable difference. Furthermore, $X$ and $Y$ have finite width (roughly $\Delta = 50$ ns). When $\UDDx{n}$ is applied for $n$ even, we must end on an identity, so the identity must last for a duration $\Delta$, i.e., $I = f_\tau$. A similar timing constraint detail appears for $\QDD{n}{m}$ when $m$ is odd. Here, we must apply a $Z$ pulse, but on IBM devices, $Z$ is virtual and instantaneous (see \cref{app:CvsP}). Thus, we apply $Z - f_{\Delta}$ to obtain the expected timings.

\section{Circuit \textit{vs} OpenPulse APIs}
\label{app:CvsP}

\begin{figure}[h]
    \centering
    \subfigure[Comparison of pulse and circuit implementation of $\UR{20}$ on \armonk\ with respect to Free]{\centering
    \label{subfig:ur20PulseVsCirc}
    \includegraphics[width=0.45\textwidth]{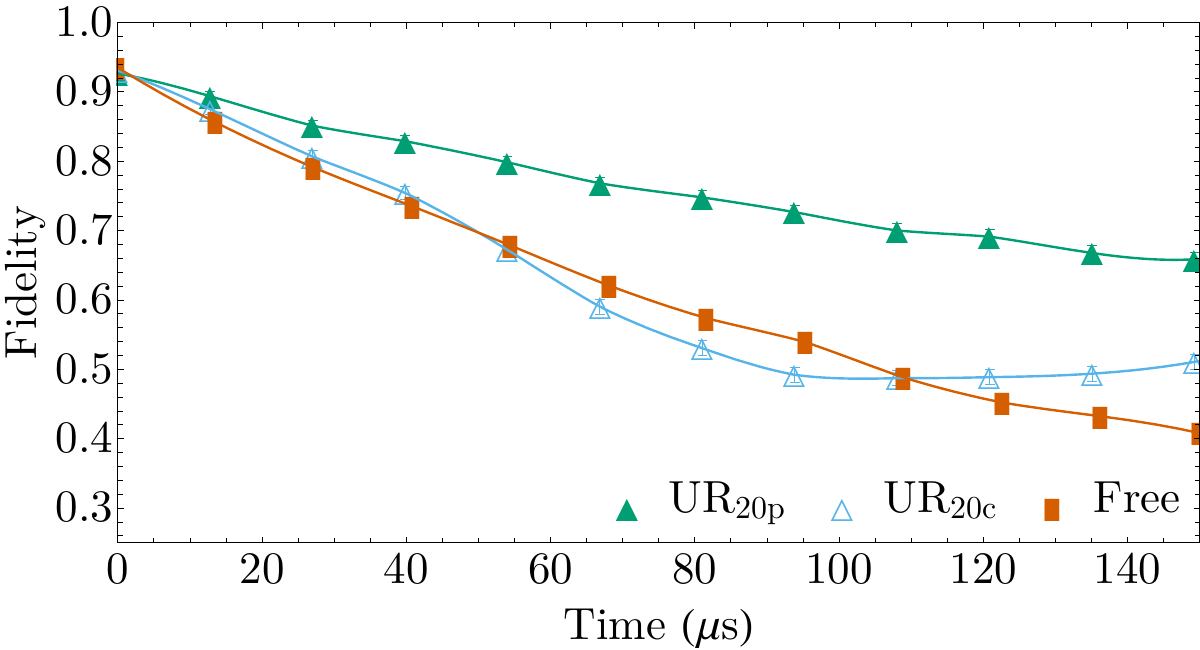}}
    \subfigure[Comparison of pulse and circuit implementation of $Y-Y$ on \armonk]{\centering
    \label{subfig:yyVsZX}
    \includegraphics[width=0.45\textwidth]{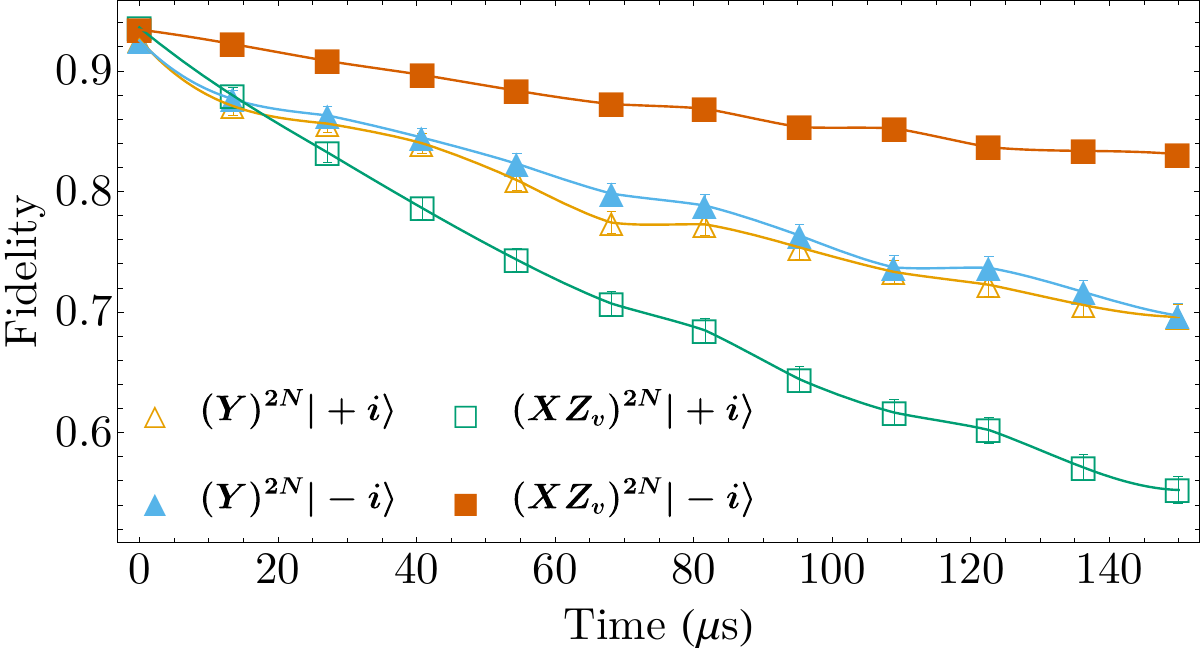}}
    \caption{Comparison of the circuit and OpenPulse API approaches to implementing (a) $\UR{20}$ and (b) $Y\freea Y\freea$. Panel (a) demonstrates that the OpenPulse version is substantially better. This is partially explained by (b). When using OpenPulse, the $Y\freea Y\freea$ sequence behaves as expected by symmetrically protecting $\ket{\pm i}$, in stark contrast to the circuit implementation, which uses virtual $Z$ gates, denoted $Z_v$.}
    \label{fig:pulseVsCirc}
\end{figure}

We first tried to use the standard Qiskit circuit API~\cite{Qiskit, mckay_qiskit_2018}. Given a DD sequence, we transpiled the circuit \cref{subfig:dd-circuit} to the respective device's native gates. However, as we illustrate in \cref{subfig:ur20PulseVsCirc}, this can lead to many advanced sequences, such as $\UR{20}$, behaving worse than expected. Specifically, this figure shows that implementing $\UR{20}$ in the standard circuit way, denoted $\UR{20\text{c}}$ (where c stands for circuit), is substantially worse than an alternative denoted $\UR{20\text{p}}$ (where p stands for pulse). 

The better alternative is to use OpenPulse~\cite{alexander_qiskit_2020}. We call this the \enquote{pulse} implementation of a DD sequence. The programming specifics are provided in Ref.~\cite{nicholas_ezzell_2023_7884641}; here, we focus on the practical difference between the two methods with the illustrative example shown in \cref{subfig:yyVsZX}. Specifically, we compare the $Y\freea Y\freea$ sequence implemented in the circuit API and the OpenPulse API.  

Under OpenPulse, the decay profiles for $\ket{+i}$ and $\ket{-i}$ are roughly identical, as expected for the $Y\freea Y\freea$ sequence. The slight discrepancy can be understood as arising from coherent errors in state preparation and the subsequent $Y$ pulses, which accumulate over many repetitions.
On the other hand, the circuit results exhibit a large asymmetry between $\ket{i}$ and $\ket{-i}$. The reason is that $Y$ is compiled into $Z_v-X$ where $Z_v$ denotes a virtual $Z$ gate~\cite{mckay_efficient_2017}. As \cref{subfig:yyVsZX} shows, $Z_v-X$ does \emph{not} behave like $Y$. The simplest explanation consistent with the results is to interpret $Z_v$ as an instantaneous $Z$.
In this case, $Z\ket{+i} = \ket{-i}$ and the subsequent $X$ rotates the state from $\ket{-i}$ to $\ket{+i}$ by rotating \emph{through the excited state.} The initial state $\ket{-i}$, on the other hand, rotates \emph{through the ground state.} Since the ground state is much more stable than the excited state on IBMQ's transmon devices, this asymmetry in trajectory on the Bloch sphere is sufficient to explain the asymmetry in fidelity.\footnote{This observation and explanation is due to Vinay Tripathi.}

Taking a step back, every gate that is not a simple rotation about the $x$ axis is compiled by the standard circuit approach into one that is a combination of $Z_v$, $X$, and $\sqrt{X}$. These gates can behave unexpectedly, as shown here. In addition, the transpiler -- unless explicitly instructed otherwise -- also sometimes combines a $Z_v$ into a global phase without implementing it right before an $X$ gate. Consequently, two circuits can be logically equivalent while implementing different DD sequences. Using OpenPulse, we can ensure the proper implementation of $(\pi)_{\phi}$. This allows the fidelity of $\UR{20p}$ to exceed that of $\UR{20c}$. 

Overall, we found (not shown) that the OpenPulse implementation was almost always better than or comparable to the equivalent circuit implementation, except for \XY\ on \armonk, where $\XY_c$ was substantially better than $\XY_p$. However,  $\XY_p$ was not the top-performing sequence. Hence, it seems reasonable to default to using OpenPulse for DD when available.

\section{Methodologies for extraction of fidelity metrics}
\label{app:fits}

In the Pauli demonstration, we compare the performance of different DD sequences on the six Pauli eigenstates for long times. More details of the method are discussed in \cref{subsec:pauliExperiment} and, in particular, \cref{subfig:dd-circuit} and \cref{fig:methods-summary}. The results are summarized in \cref{fig:pauliPlot} and in more detail in App.~\ref{app:all-pauli-data}. 

To summarize the fidelity decay profiles, we have chosen to employ an integrated metric in this work. Namely, we consider a time-averaged (normalized) fidelity, 
\begin{equation}
	\label{app-eq:norm-average-fidelity}
    F(T) \equiv \frac{1}{f_e(0)}\expval{f_e(t)}_T = \frac{1}{T} \int_0^T dt \frac{f_e(t)}{f_e(0)} .
\end{equation}
We combine this metric with an interpolation of fidelity curves, which we explain in detail in this section. We call the combined approach ``Interpolation with Time-Averaging'' (ITA).

Past work has employed a different method for comparing DD sequences, obtained by fitting the decay profiles to a modified exponential decay function. That is, to assume that $f_e(t) \sim e^{-\lambda t}$, and then perform a fit to determine $\lambda$. 
For example, in previous work~\cite{Pokharel:2018aa}, some of us chose to fit the fidelity curves with a function of the form\footnote{We have added a factor of $1 / 2$ to $\Gamma(t)$ that was unfortunately omitted in Ref.~\cite{Pokharel:2018aa}.}
\bes
    \label{app-eq:bibek-fit}
\begin{align}
    f_{P}(t) &= \frac{f_e(T_f) -f_e(0)}{\Gamma(T_f) - 1}\left[\Gamma(t) - 1\right] + f_e(0), \\ 
    \Gamma(t) &\equiv \frac{1}{2} \left(e^{-t / \lambda} \cos(t \gamma) + e^{-t / \alpha}\right),
\end{align}
\ees
where $f_e(0)$ is the empirical fidelity at $T = 0$, $f_e(T_f)$ is the empirical fidelity at the final sampled time, and $\Gamma(t)$ is the decay function that is the subject of the fitting procedure, with the three free parameters $\lambda$, $\gamma$, and $\alpha$. This worked well in the context of the small set of sequences studied in Ref.~\cite{Pokharel:2018aa}.

As we show below, in the context of our present survey of sequences, fitting to \cref{app-eq:bibek-fit} results in various technical difficulties, and the resulting fitting parameters are not straightforward to interpret and rank. We avoid these technical complications by using our integrated (or average fidelity) approach, and the interpretation is easier to understand. We devote this section primarily to explaining the justification of these statements, culminating in our preference for a methodology based on the use of \cref{app-eq:norm-average-fidelity}.

First, we describe how we bootstrap to compute the empirical fidelities (and their errors) that we use at the beginning of our fitting process.

\subsection{Point-wise fidelity estimate by bootstrapping \label{app:bootstrap-fid}}

We use a standard bootstrapping technique~\cite{efron_bootstrap_1992} to calculate the $2 \sigma$ (95\% confidence intervals) errors on the empirical Uhlmann fidelities,
\begin{equation}
    \label{eq:app-empirical-fid}
    f_e(t) = |\bra{\psi}\rho_{\text{final}}(t) \ket{\psi}|^2.
\end{equation}
To be explicit, we generate $N_s=8192$ binary samples (aka shots) from our demonstration (see \cref{subfig:dd-circuit}) for a given $\{$DD sequence, state preparation unitary $U$, total DD time $T\}$ 3-tuple. From this, we compute the empirical Uhlmann fidelity as the ratio of counted 0's normalized by $N_s$. We then generate 1000 re-samples (i.e., $N_s$ shots generated from the empirical distribution) with replacement, calculating $f_e(T)$ for each re-sample. From this set of 1000 $f_e(T)$'s, we compute the sample mean, $\expval{f_e}_T$, and sample standard deviation, $\sigma_T$, where the $T$ subscript serves as a reminder that we perform this bootstrapping for each time point. For example, the errors on the fidelities in \cref{fig:methods-summary}(a) are $2 \sigma$ errors generated from this procedure.

\subsection{A survey of empirical fidelity decay curves}

\begin{figure*}
    \centering
    \includegraphics[width=0.49\textwidth]{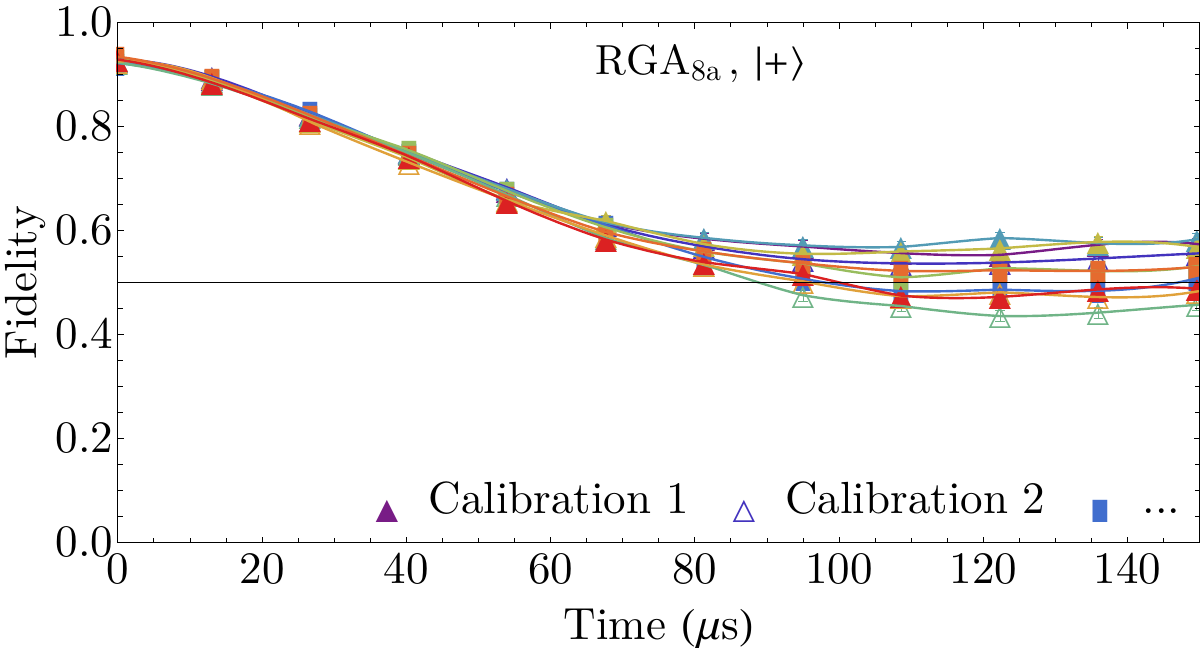}
    \includegraphics[width=0.49\textwidth]{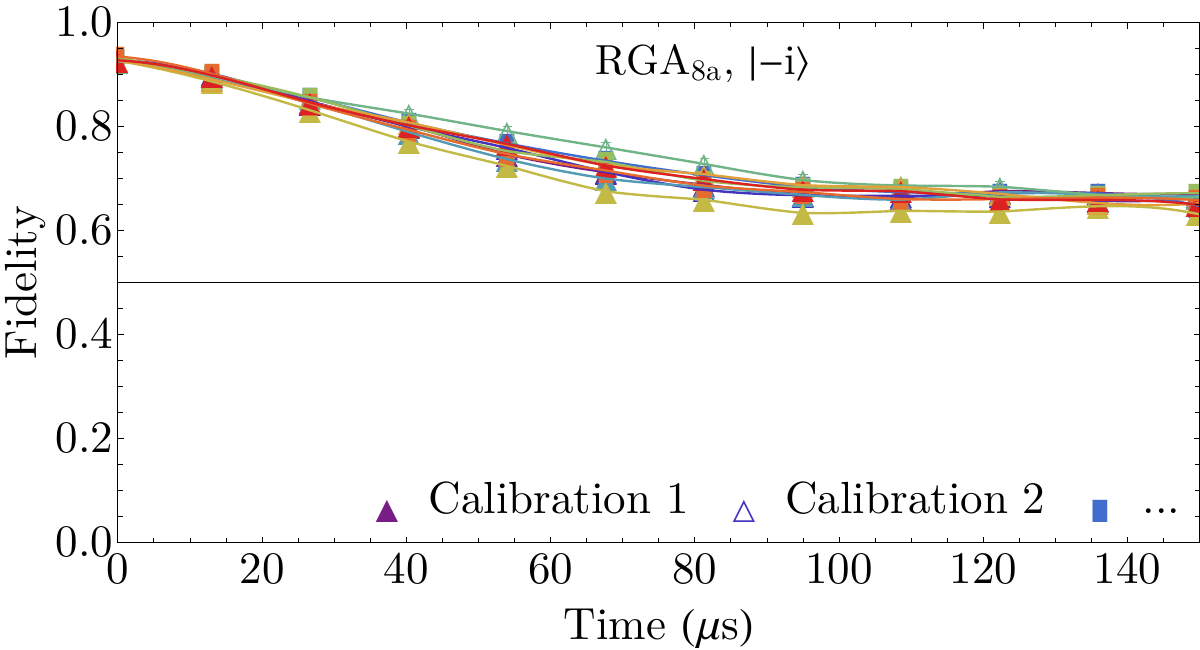}
    \includegraphics[width=0.49\textwidth]{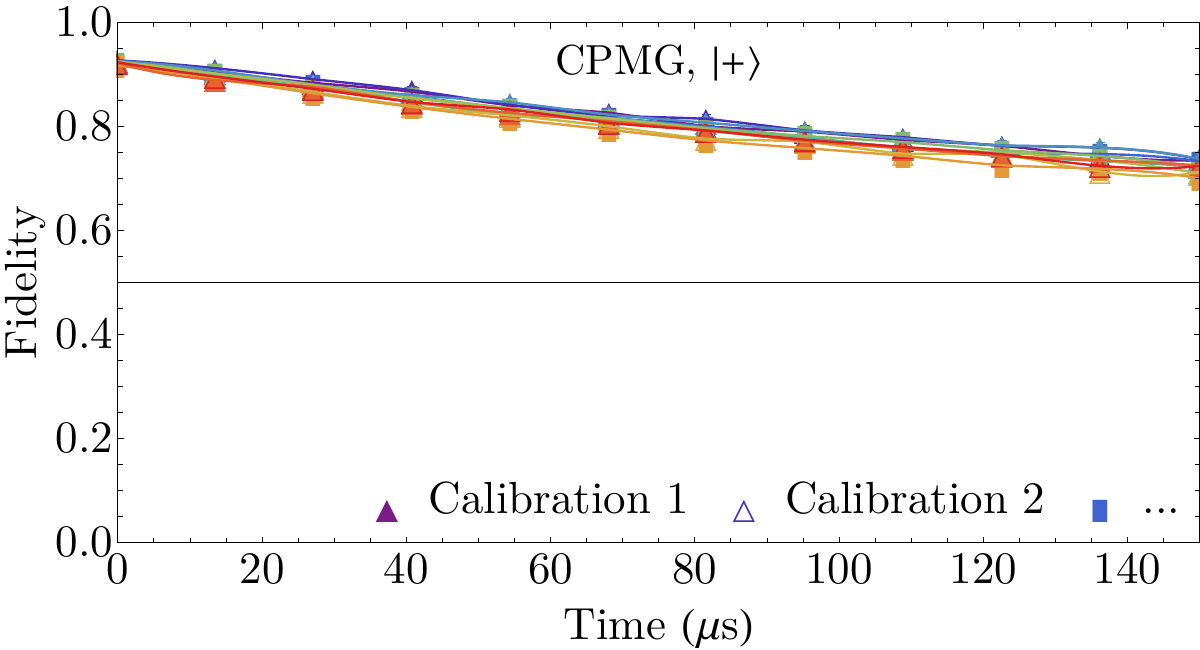}
    \includegraphics[width=0.49\textwidth]{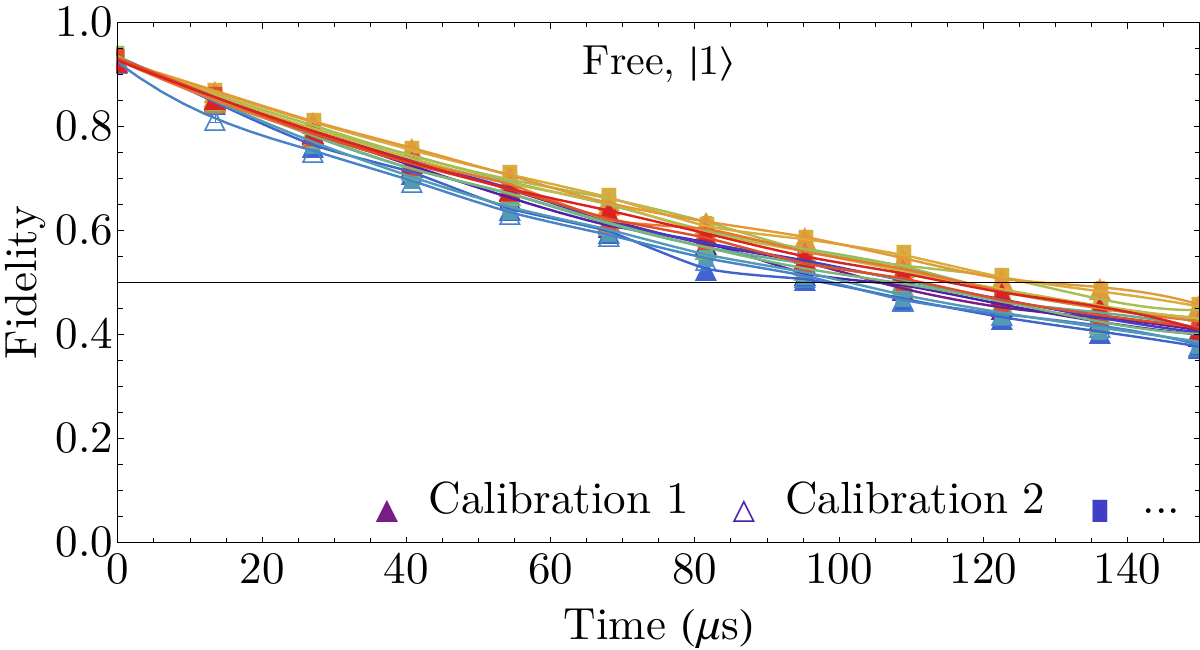}
    \includegraphics[width=0.49\textwidth]{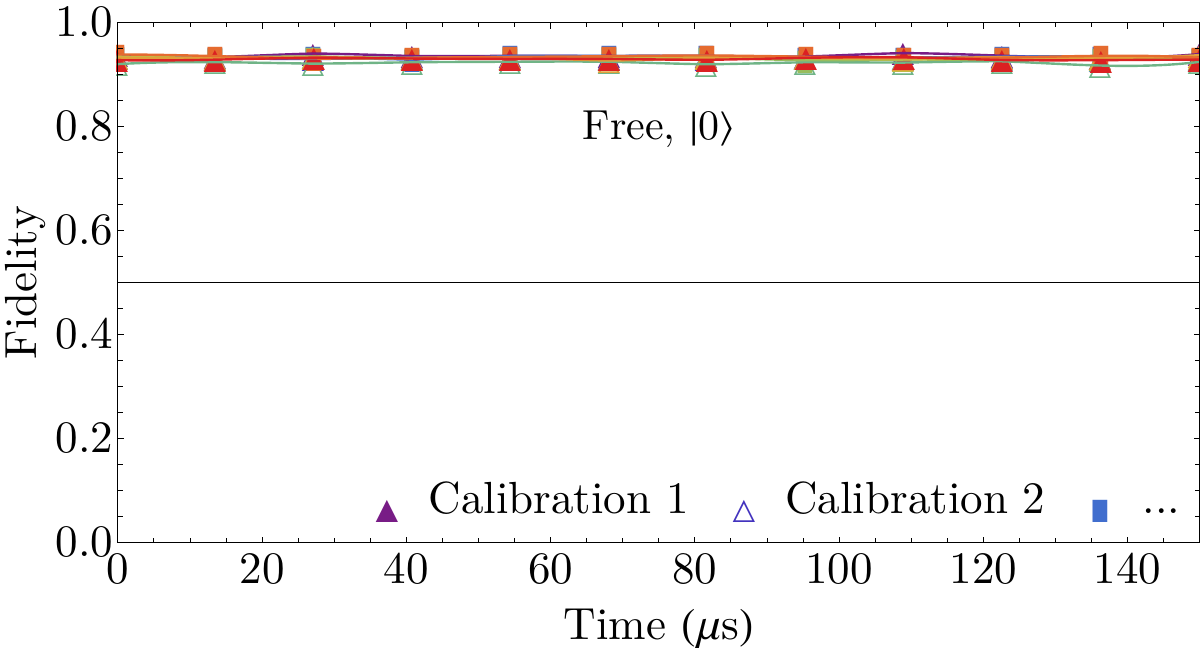}
    \includegraphics[width=0.49\textwidth]{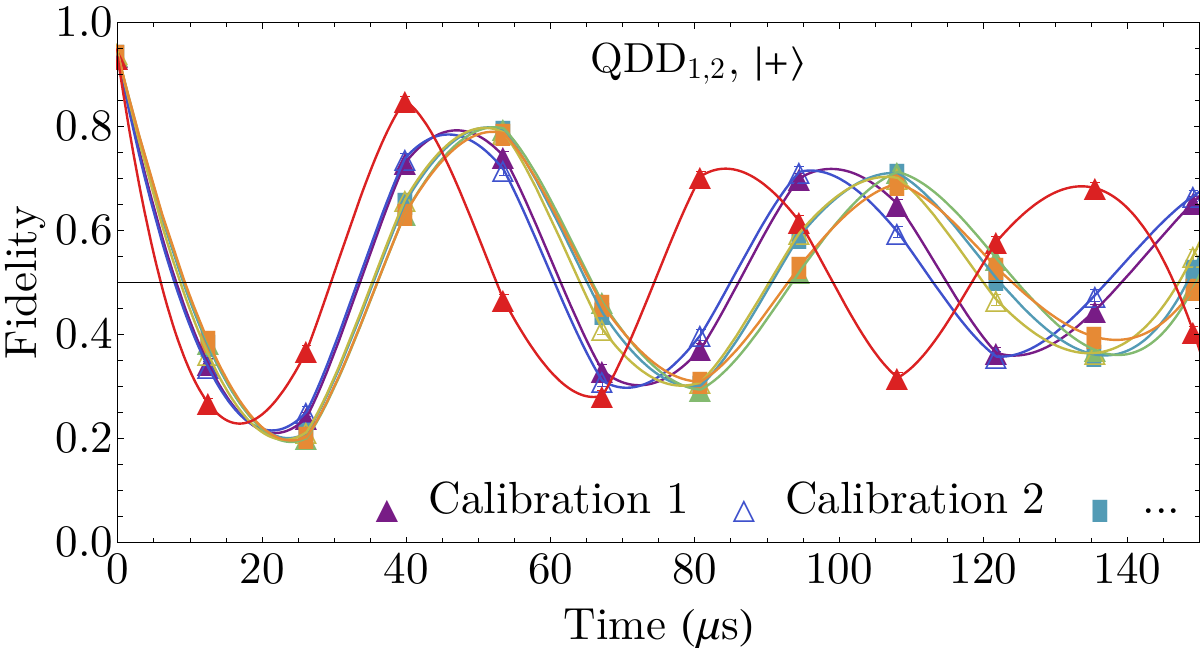}
    \includegraphics[width=0.49\textwidth]{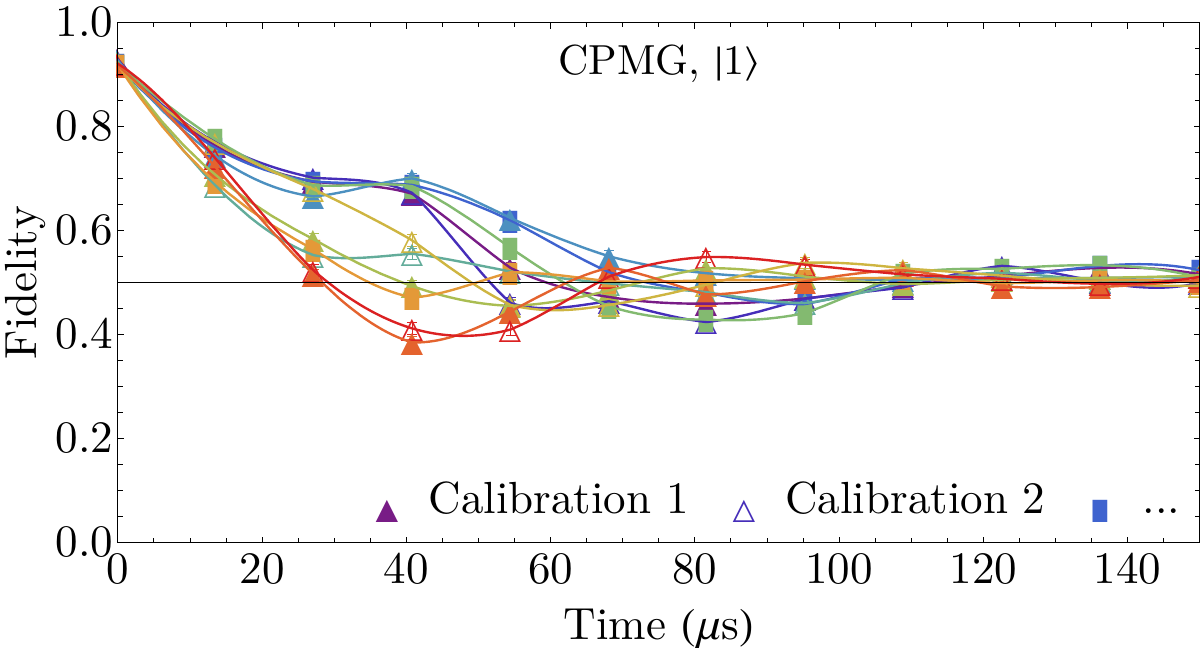}
    \includegraphics[width=0.49\textwidth]{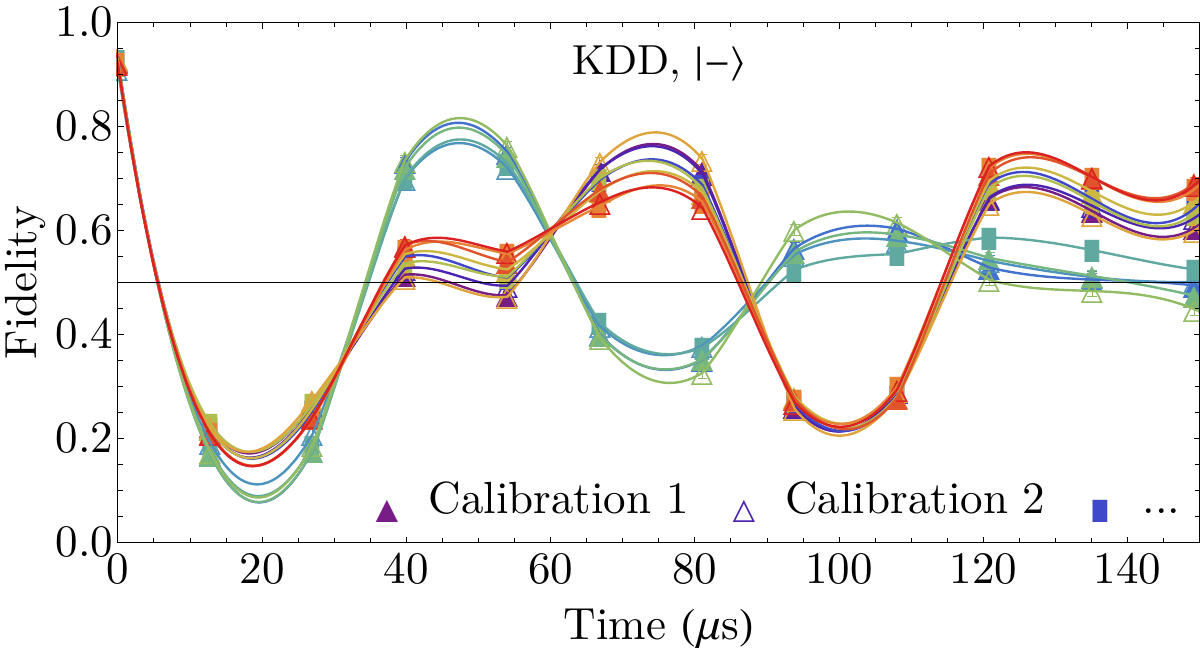}
    \caption{A sample of fidelity decay curves from \armonk\ that showcases the qualitatively different curve types. Each plot shows curves generated from ten different calibration cycles. Note that all Pauli states are sampled within the same job, so data can be compared directly without worrying about system drift across different calibrations.}
    \label{fig:sample-fid-decay-curves}
\end{figure*}

Given a systematic way to compute empirical fidelities through bootstrapping, we can now discuss the qualitatively different fidelity decay curves we encounter in our demonstrations, as illustrated in \cref{fig:sample-fid-decay-curves}. At a high level, the curves are characterized by whether they decay and whether oscillations are present. If decay is present, there is an additional nuance about what fidelity the curve decays to and whether there is evidence of saturation, i.e., reaching a steady state with constant fidelity. Finally, it matters whether an oscillation is simple, characterized by a single frequency, or more complicated. All these features can be seen in the eight examples shown in \cref{fig:sample-fid-decay-curves}, which we now discuss. 

The first four panels in \cref{fig:sample-fid-decay-curves} correspond to curves dominated by decay but that decay to different final values. For the first two $\RGA{8a}$ plots, the final value seems stable (i.e., a fixed point). For CPMG and Free, the final fidelity reached does not seem to be the projected stable fidelity. The ``Free, $\ket{0}$'' curve does not decay, consistent with expectations from the stable $\ket{0}$ state on a superconducting device. The last three plots show curves with significant oscillations. For the $\QDD{1}{2}$ plot, the oscillations are strong and only weakly damped. For CPMG, the oscillations are strongly damped. Finally, the KDD plot is a pathological case where the oscillations clearly exhibit more than one frequency and are also only weakly damped.

\subsection{Interpolation \textit{vs} curve fitting for time-series data}
 \label{app:fit-comp-high-level}
 
To obtain meaningful DD sequence performance metrics, it is essential to compress the raw time-series data into a compact form. Given a target protection time $T$, the most straightforward metric is the empirical fidelity, $f_e(T)$. Given a set of initial states relevant to some demonstration (i.e., those prepared in a specific algorithm), one can then bin the state-dependent empirical fidelities across the states in a box plot. 

The box plots we present throughout this work are generated using Mathematica's \verb^BoxWhiskerChart^ command. As a reminder, a box plot is a common method to summarize the quartiles of a finite data set. In particular, let $Q_x$ represent the $x^{\text{th}}$ quantile defined as the smallest value contained in the data set for which $x \%$ of the data lies below the value $Q_x$. With this defined, the box plot shows $Q_0$ as the bottom bar (aka the minimum), $Q_{25} - Q_{75}$ as the orange box (aka the inner-quartile range), $Q_{50}$ as the small white sliver in the middle of the inner-quartile range, and $Q_{100}$ as the top bar (aka the maximum). In our box plots, we have also included the mean as the solid black line. A normal distribution is symmetric, so samples collected from a normal distribution should give rise to a box plot that is symmetric about the median and where the mean is approximately equal to the median.

If there is no pre-defined set of states of interest, it is reasonable to sample states from the Haar distribution as we did for 
\cref{fig:haarIntervalPlot}. This is because the sample Haar mean is an unbiased estimator for the mean across the entire Bloch sphere. Indeed, for a large enough set of Haar random states (we found $25$ to be empirically sufficient), the distribution about the mean becomes approximately Gaussian. At this point, the mean and median agree, and the inner-quartile range becomes symmetric about its mean, so one may choose to report $\langle f_e(T) \rangle \pm 2 \sigma$ instead. 

When there is no target time $T$, we would like a statistic that accurately predicts performance across a broad range of relevant times. The empirical fidelity for a given state at any fixed $T$ is an unreliable predictor of performance due to oscillations in some curves (see, e.g., $\QDD{1}{2}, \CPMG,$ and KDD in the last three plots in \cref{fig:sample-fid-decay-curves}). To be concrete, consider that for $\QDD{1}{2}$ protecting $\ket{+}$, $\langle f_e(20\mu s) \rangle \approx 0.2$  whereas $\langle f_e(50\mu s) \rangle \approx 0.8$.  The standard statistic, in this case, is a decay constant obtained by a modified exponential fit [e.g., \cref{app-eq:bibek-fit}]~\cite{Pokharel:2018aa}.

\begin{figure*}
    \centering
     \includegraphics[width=0.45\textwidth]{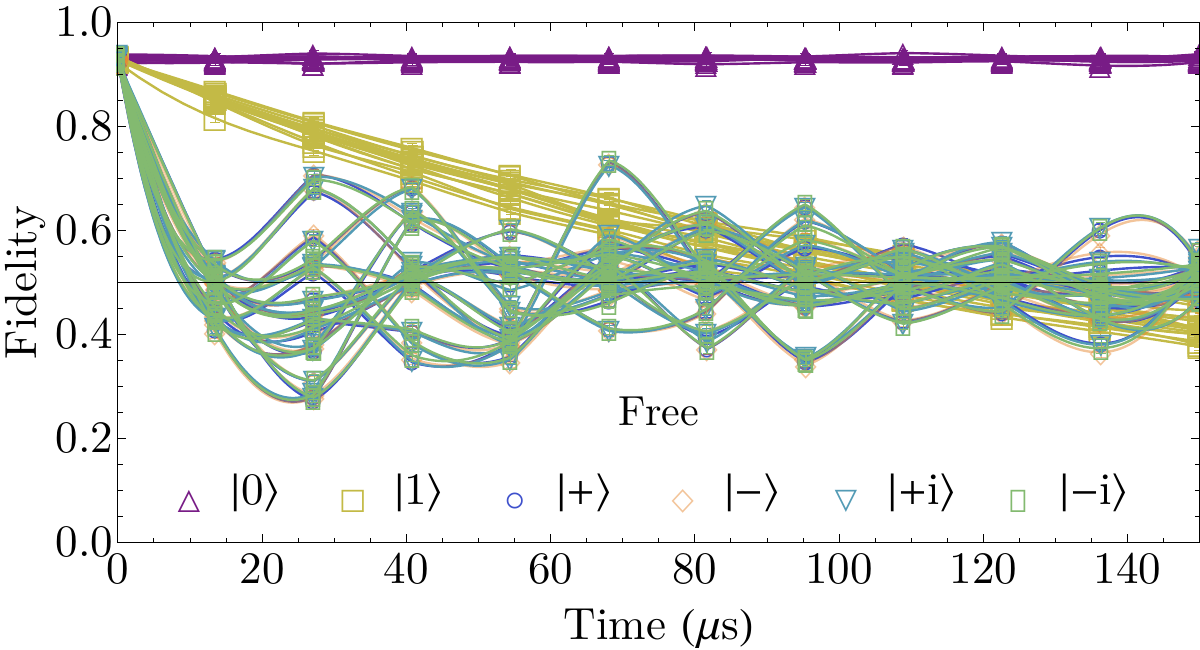}
     \includegraphics[width=0.45\textwidth]{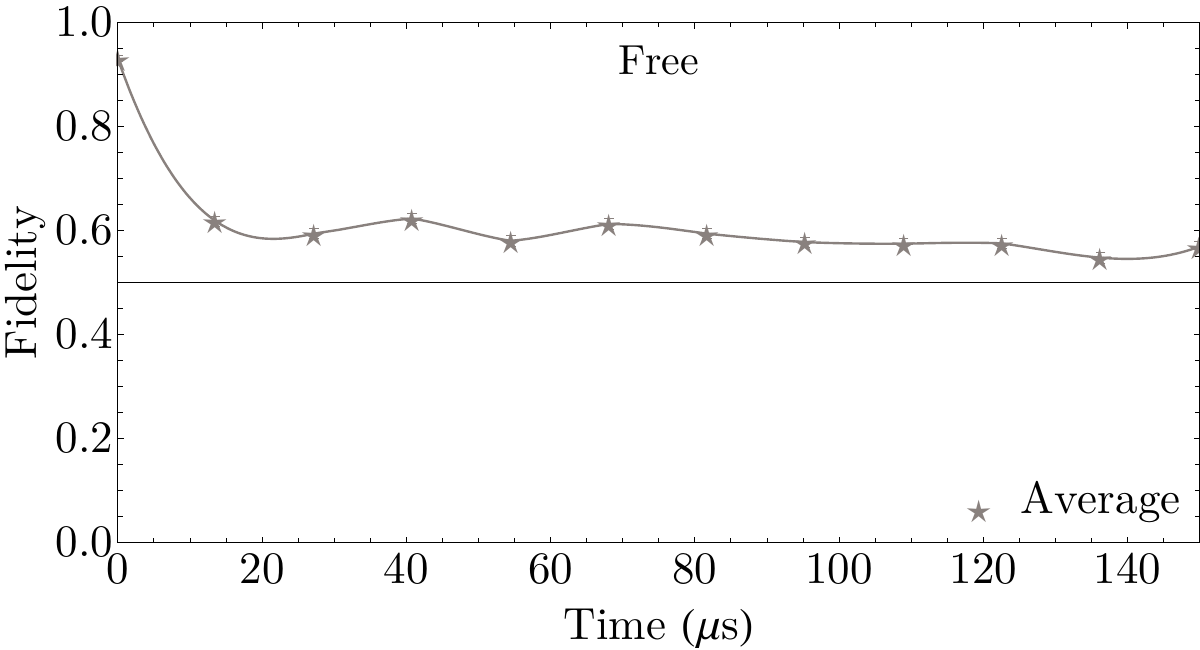}
     \includegraphics[width=0.45\textwidth]{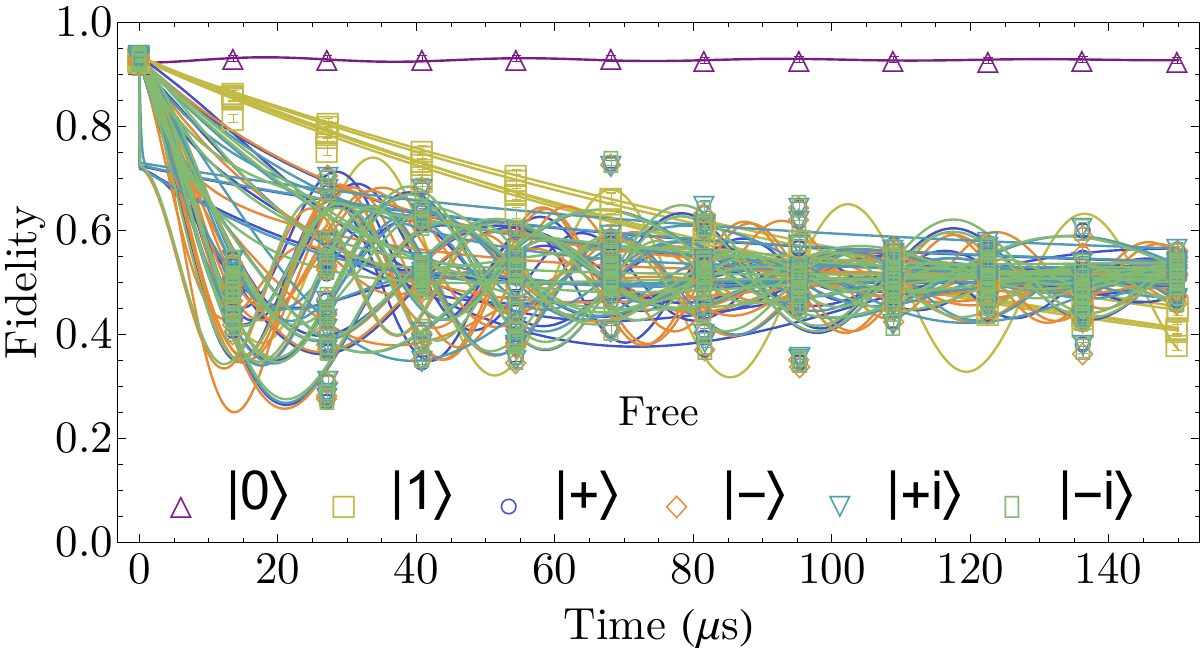}
     \includegraphics[width=0.45\textwidth]{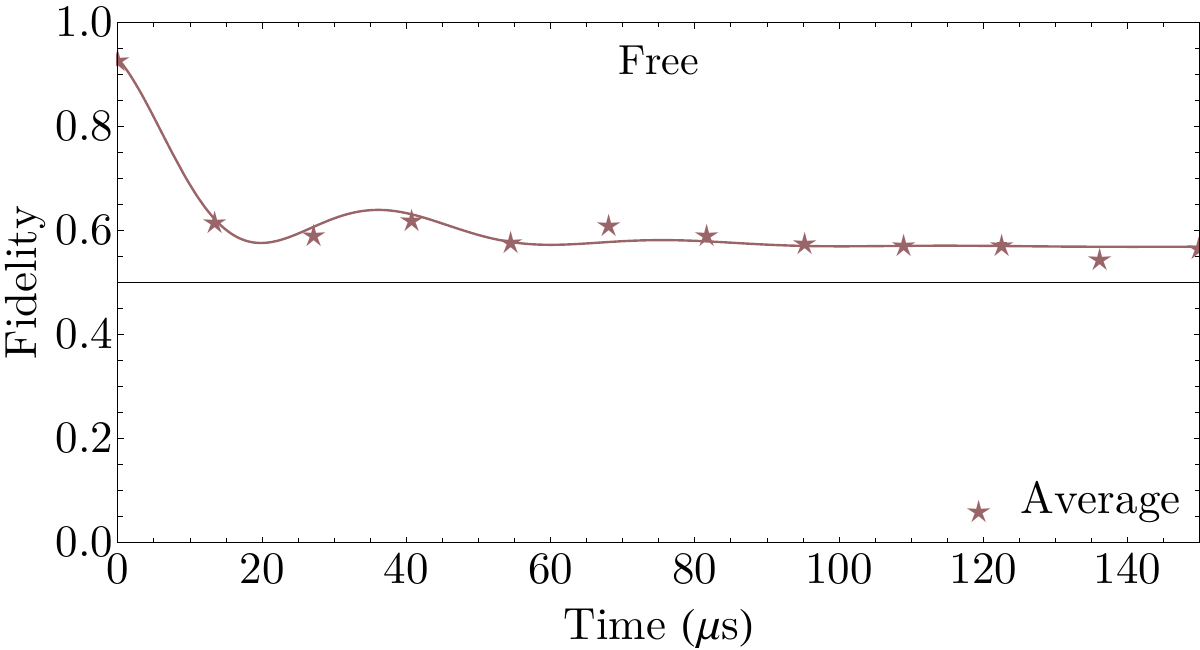}
    \caption{Illustration of the qualitative difference between different fitting procedures with free evolution (Free) data collected on \armonk. The empirical data (obtained from six different initial states and $16$ calibration cycles, for a total of $96$ time-series shown) is \emph{identical} in the top left and bottom left panels (the empty symbols). We employ two different techniques [interpolation (top) and \cref{app-eq:bibek-fit} (bottom)] to fit the data corresponding to individual time-series (left) and averaged time-series (right). Top left: curves are computed by interpolation, separately for each time-series. Bottom left: curves are computed by fitting each time-series to \cref{app-eq:bibek-fit}. The fits here must contend with three different types of behaviors: constant (no decay: $\ket{0}$), pure and slow decay ($\ket{1}$), and damped oscillations (the remaining equatorial states).  
Right: the AtF approach. We first average the fidelity across all the data for each fixed time and then fit it. Top: we utilize a piece-wise cubic spline interpolation between points. Bottom: we use the fit given by  \cref{app-eq:bibek-fit}. We remark that fitting to \cref{app-eq:bibek-fit} fails for some data sets, and this is why the bottom left panel contains fewer curves than the top left panel. This is explained more in the text and in full detail in \cref{app:fit-minutia}.}
    \label{fig:sample-fitting-approaches}
\end{figure*}

So far,  our discussion has centered around a statistic that captures the performance of individual curves in \cref{fig:sample-fid-decay-curves},  i.e., a curve for a fixed DD sequence,  state,  and calibration.  To summarize further,  we can put all found $\lambda$ values in a box plot.  For simplicity, we call any method which fits individual curves and then averages over these curves a \enquote{fit-then-average} (FtA) approach. This is the high-level strategy we advocate for and use in our work. Note that we include interpolation as a possible curve-fitting strategy as part of the FtA approach, as well as fitting to a function with a fixed number of fit-parameters. In contrast, Ref.~\cite{Pokharel:2018aa} utilized an \enquote{average-then-fit} (AtF) approach, wherein an averaging of many time-series into a single time-series was performed before fitting.\footnote{We caution that we use the term ``averaging'' to indicate both \emph{time}-averaging [as in \cref{app-eq:norm-average-fidelity}] and \emph{data}-averaging, i.e., averaging fidelity data from different calibration cycles and states.} Only fitting to a function with a fixed number of fit-parameters was used in Ref.~\cite{Pokharel:2018aa}, but not interpolation. We discuss the differences between the FtA and AtF approaches below, but first, we demonstrate what they mean in practice and show how they can give rise to different quantitative predictions. We begin our discussion by considering \cref{fig:sample-fitting-approaches}. 

The data in \cref{fig:sample-fitting-approaches} corresponds to the Pauli demonstration for free evolution (Free).  On the left,  we display the curves corresponding to each of the 96 demonstrations (six Pauli eigenstates and 16 calibration cycles in this case). The top curves are computed by a piece-wise cubic-spline interpolation between successive points, while the bottom curves are computed using \cref{app-eq:bibek-fit} (additional details regarding how the fits were performed are provided in \cref{app:fit-minutia}). For a given state,  the qualitative form of the fidelity decay curve is consistent from one calibration to the next.  For $\ket{0}$, there is no decay.  For $\ket{1}$,  there is a slow decay induced from $T_1$ relaxation back to the ground state (hence the fidelity dropping below $1/2$).  For the equatorial states,  $\ket{\pm}$ and $\ket{\pm i}$,  there is a sharp decay to the fully mixed state (with fidelity $1/2$) accompanied by oscillations with different amplitude and frequencies depending on the particular state and calibration.  The $\ket{0}$ and $\ket{1}$ profiles are entirely expected and understood, and the equatorial profiles can be interpreted as being due to $T_1$ (relaxation) and $T_2$  (dephasing) alongside coherent pulse-control errors which give rise to the oscillations.  On the right, we display the fidelity decay curves computed using the AtF approach, i.e., by averaging the fidelity from all demonstrations at each fixed time.  The two curves [top: interpolation; bottom: \cref{app-eq:bibek-fit}] display a simple decay behavior, which is qualitatively consistent with the curves in~\cite[Fig.~2]{Pokharel:2018aa}.

In \cref{sec:ITAvscurve-fitting}, we go into more detail about what the approaches in \cref{fig:sample-fitting-approaches} are in practice and, more importantly, how well they summarize the raw data. Before doing so, we comment on an important difference between the two left panels of \cref{fig:sample-fitting-approaches}. Whereas the interpolation method provides consistent and reasonable results for all fidelity decay curves, the fitting procedure can sometimes fail. A failure is not plotted, and this is why some data in the bottom right panel is missing. For example, most fits (15/16) for $\ket{0}$ fail since \cref{app-eq:bibek-fit} is not designed to handle flat \enquote{decays}. The state $\ket{1}$ also fails $11/16$ times, but interestingly, the equatorial states produce a successful fit all $16$ times. The nature of the failure is explained in detail in \cref{app:fit-minutia}, but as a prelude, a fit fails when it predicts fidelities outside the range $[0,1]$ or when it predicts a decay constant $\lambda$ with an unreasonable uncertainty. We next address the advantages of the interpolation approach from the perspective of extracting quantitative fidelity metrics.

\begin{figure}[ht]
    \centering
    \includegraphics[width=0.45\textwidth]{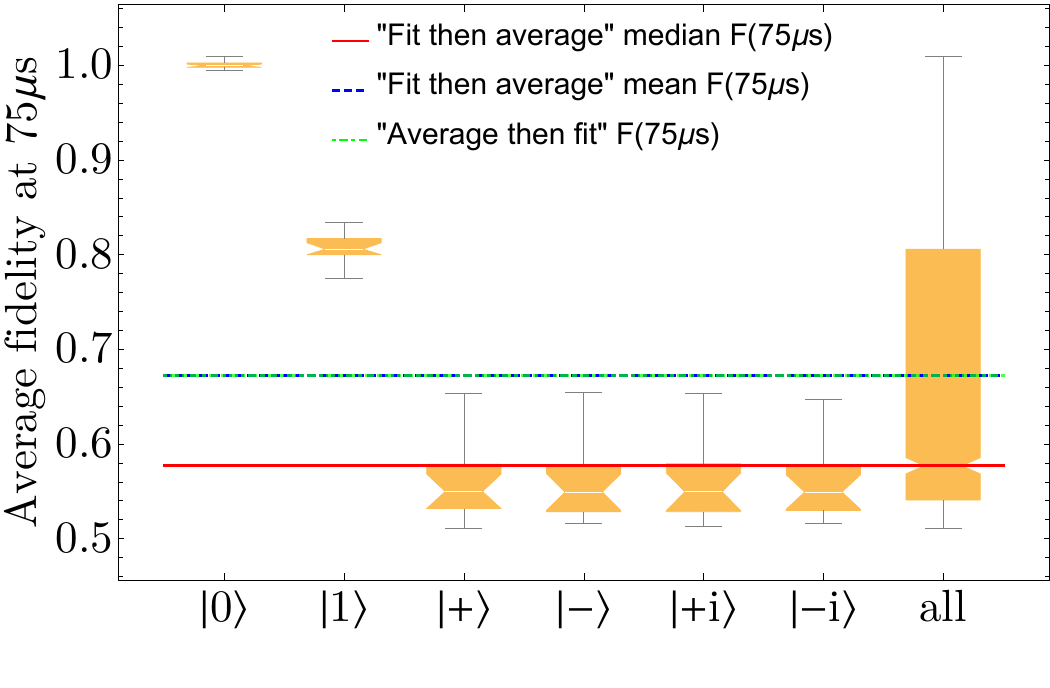}
    \includegraphics[width=0.45\textwidth]{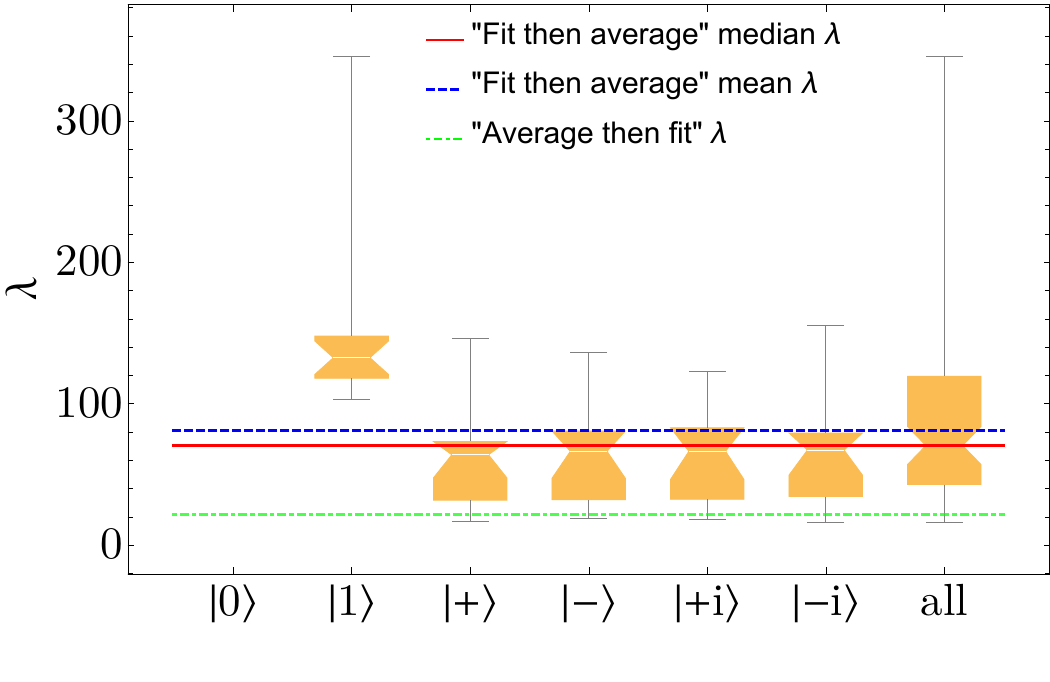}
    \caption{Summary of the integrated and fitting approaches when applied to the Free curves in \cref{fig:sample-fitting-approaches}.  For both of the resulting metrics, we display a state-by-state box plot showing the variation in the reported statistic over the $16$ calibration cycles.
Top: for each of the $6\times 16$ interpolated curves, we compute a time-averaged fidelity at $T=75\mu s$, using \cref{app-eq:norm-average-fidelity}. These average fidelities are shown in box plot format, separated by state and also combined all together. Bottom: we attempt to extract$\lambda$ for each of the $6\times 16$ ($\ket{0}$ fails 15/16 times and $\ket{1}$ fails 11/16 times) curves fitted to \cref{app-eq:bibek-fit} and shown in \cref{fig:sample-fitting-approaches}. The results are shown in box plot format, again separated by state and also combined all together. Top and bottom: also shown are the median (red line) and mean (blue dashed line) of these FtA results. The green dashed line is the AtF result (in this case, there is just a single number, hence no variation over calibration cycles or states). 
The FtA (blue) and AtF (green) curves agree in the top panel, as expected of a reasonable method, whereas they disagree significantly in the bottom panel.}
    \label{fig:sample-fitting-boxplot-results}
\end{figure}

\subsection{Interpolation with time-averaging (ITA) \textit{vs} curve fitting for fidelity metrics}
\label{sec:ITAvscurve-fitting}

To extract quantitative fidelity metrics from the fitted data, we compute the time-averaged fidelity [\cref{app-eq:norm-average-fidelity}] and the decay constant $\lambda$ [\cref{app-eq:bibek-fit}]. The results are shown in \cref{fig:sample-fitting-boxplot-results}. The box plots shown in this figure are obtained from the individual curve fits in \cref{fig:sample-fitting-approaches}. 
The top panel is the ITA approach we advocate in this work (recall \cref{app:fits}): it corresponds to the time-averaged fidelity computed from the interpolated fidelity curves in the top-left panel of \cref{fig:sample-fitting-approaches}. The bottom panel corresponds to fits computed using \cref{app-eq:bibek-fit}, i.e., the bottom-left panel of \cref{fig:sample-fitting-approaches}, from which the decay constant $\lambda$ is extracted.

Let us first discuss the bottom panel of \cref{fig:sample-fitting-boxplot-results}. First, we again comment on the effect of fit failures. Since $\ket{0}$ does not decay, \cref{app-eq:bibek-fit} is not appropriate (i.e., $\lambda\to\infty$ which is not numerically stable), and this results in only $1/16$ fits leading to a valid $\lambda$ prediction. This is insufficient data to generate a box plot; hence the absence of the $\ket{0}$ data at the bottom. Similar numerical instability issues -- though less severe -- arise for all of the fits. For example, for $\ket{1}$, only $5/16$ fits succeed. Among those that do succeed, the variation in $\lambda$ is quite large, varying in the range $[103,346]$. When compared to the raw data in \cref{fig:sample-fitting-approaches}, such a large variation should not be expected. In contrast, all the fits for the equatorial states succeed, but the variation in $\lambda$ is again relatively large, in the range $[17,146]$. To summarize, fitting \cref{app-eq:bibek-fit} to our data has instability issues which manifest as: (i) failures to fit some data sets and (ii) large variations in the reported $\lambda$ which are unphysical. The problem is not specific to \cref{app-eq:bibek-fit} and indeed would likely occur for any function which tries to reasonably model the entire set of curves found in practice as in \cref{fig:sample-fid-decay-curves}. Again, this is because (i) not all states decay as an exponential (i.e., the decay if $\ket{0}$ is flat and that of $\ket{1}$ appears roughly linear), (ii) $\lambda$ and $\gamma$ are not independent, and (iii) the notion of $\lambda$ itself depends on the final expected fidelity, which is state-dependent. 

We argue that instead, interpolating and using time-averaged fidelities, i.e., the approach shown in the top panel of \cref{fig:sample-fitting-boxplot-results}, is preferred. The results shown there demonstrate that this method gives consistent and reasonable results for any fixed state.  When significant variations are present (as in the entire data set), it is representative of the clear difference between initial states visible in \cref{fig:sample-fitting-approaches} (top-left). The key is to choose an appropriate value of $T$ to average over.  When $T$ is too small, it does not capture the difference between sequences, and when $T$ is too large, the difference between most DD sequences becomes unobservable. As a compromise, we choose $T$ long enough for oscillatory sequences to undergo at least one but up to a few oscillation periods. 

Beyond the box plot variation, it is also worth remarking that the FtA mean $F$ of $75 \mu$s in \cref{fig:sample-fitting-boxplot-results} (top; the blue dashed line) is equal to the AtF $F$ value for the interpolation approach (top; green dot-dashed line), i.e., the averaging and fitting operations commute in this sense.  Hence, the ITA approach has the nice property that we can recover the coarse-grained AtF integrated fidelity result from the granular box plot data by taking a mean.  The same is not true of fitting where the two mean values of $\lambda$ differ, as in the bottom panel of \cref{fig:sample-fitting-boxplot-results} (dashed blue and green lines).  For the interpretation issues of fitting explained in \cref{subsec:ambiguous-lambda-problem,subsec:job-to-job-fluctuations} below, alongside the practical reasons explained here, we conclude that the ITA method we use in this work is preferred to methods that attempt to directly extract the decay constant from curve fitting to functions such as $f_P(t)$. The ITA method is more versatile and yields more stable and sensible results. Notably, it is also granular and able to capture the behavior of individual (DD sequence, state) pairs. This granularity allows for finer comparisons to theory as discussed in \cref{subsec:theory-means-in-practice}) and also leads to practical benefits in our DD design as seen in \cref{app:CvsP}.

\subsection{The \enquote{ambiguous $\lambda$} problem and its resolution with ITA}
\label{subsec:ambiguous-lambda-problem}

In the previous subsection, we established that the technical difficulties encountered when using FtA as compared to AtF are resolved by interpolating and time-averaging the fidelity [ITA, \cref{app-eq:norm-average-fidelity}], instead of averaging and then fitting to a decay profile [\cref{app-eq:bibek-fit}]. In this subsection, we provide further arguments in favor of the ITA approach.

First, we take a step back and discuss what we call the \enquote{ambiguous $\lambda$} problem when attempting to fit our data using standard models such as \cref{app-eq:bibek-fit}.  Practically,  this problem makes $\lambda$ an unreliable estimator of DD sequence performance without additional information, and this additional information does not lead to a simple ranking.  As we discuss below, this problem is generic and persists whether we consider individual decay curves or averaged curves. We argue that our time-averaged fidelity metric is a reliable estimator of performance for the top sequences and resolves this issue.  After this practical ranking consideration, we consider the statistical meaning of the FtA and AtF approaches in the context of current noisy quantum devices and again argue in favor of the FtA approach. 

The crux of the issue lies in the desired interpretation of the final statistic. In the AtF approach,  the final decay constant $\lambda$ gives an estimate of how well DD does on average over an ensemble of demonstrations.  While this might be an appropriate metric for some benchmarking demonstrations,  it is not typical of any given memory demonstration or algorithm.  Here, we are interested in how well we expect a DD sequence (or free evolution) to preserve an unknown but fixed input state $\ket{\psi}$ in any given demonstration,  and as we show below, ITA is better at addressing this \enquote{typical behavior} question. 

\begin{figure}[ht]
    \centering
    \includegraphics[width=0.45\textwidth]{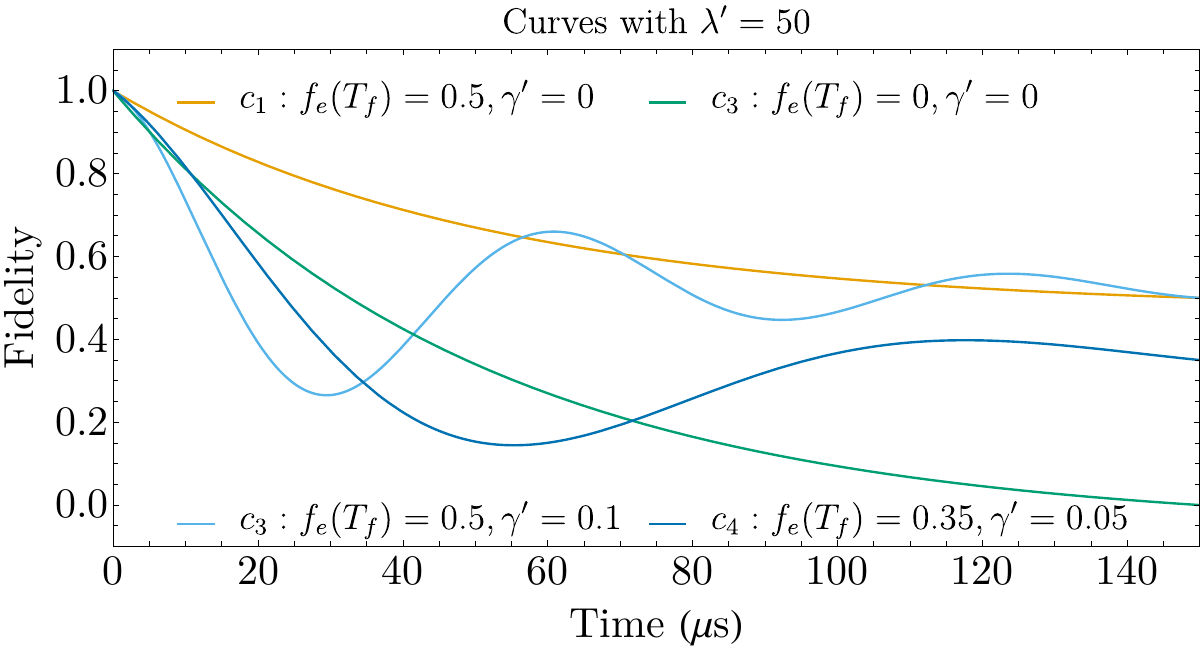}
    \includegraphics[width=0.45\textwidth]{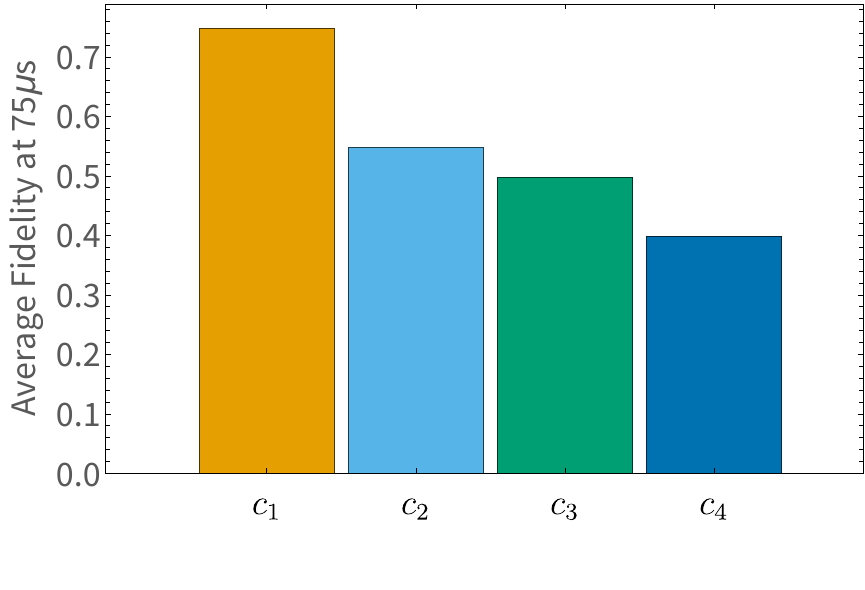}
    \includegraphics[width=0.45\textwidth]{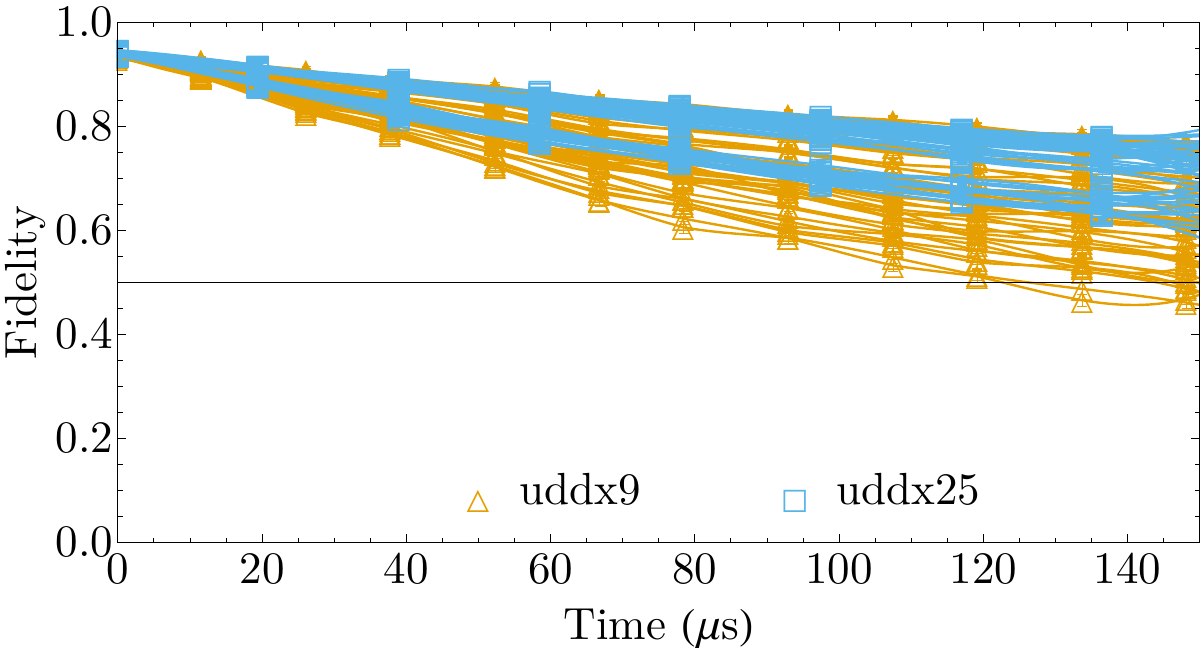}
    \caption{Examples demonstrating the problem with using $\lambda$ to rank performance. (Top) Four different artificially generated fidelity curves from \cref{app-eq:bibek-fit} with the same $\lambda$ value.  The legend labels each curve with $c_i$ and lists the value of $\gamma$ and $f_e(T_f)$ used to generate it. 
    (Middle) The time-averaged fidelity evaluated at $75\mu s$ for each curve. This metric differs for each curve. 
    (Bottom) The best sequences -- such as rank 1 and 10 in \cref{fig:pauliPlot} -- do not exhibit oscillations across the six Pauli states.}
    \label{fig:all-same-lam-prime}
\end{figure}

As we mentioned, there is an \enquote{ambiguous $\lambda$ problem} when we try to interpret the meaning of the decay constant in \cref{app-eq:bibek-fit}.
In particular,  the notion of decay in this equation is not just a function of $\lambda$ but also of $f_e(T_f) - f_e(0)$ and $\gamma$, and these fit-parameters are not mutually independent.  To make this concrete, consider the four (artificial example) curves shown in \cref{fig:all-same-lam-prime} (top panel).  These curves do not preserve fidelity equally, and yet,  by construction, they all have decay time $\lambda = 50 \mu s$.  Nevertheless, the time-averaged fidelity, $F(75\mu s)$, does distinguish between the four curves sensibly (middle panel).  First, the time-averaged fidelity metric predicts that $c_1$ is the best curve and $c_2$ is the second best, which indeed seems reasonable from the curves alone.  On the other hand, $c_3 > c_4$ is debatable and depends on the desired time scale. Had we chosen, e.g., $120 \mu s \leq T \leq 150 \mu s$, we would have found $c_4 > c_3$.  Such ambiguities are inevitable when oscillations of different frequencies are present with no preferred target time $T$.  However,  the best sequences do not have oscillations across the Pauli states, as we see by example in the bottom panel of \cref{fig:all-same-lam-prime}. Thus, this time-averaged fidelity ambiguity for oscillatory curves does not affect the ranking of the top sequences, which we present in \cref{fig:pauliPlot} as the main result of the Pauli demonstration. We remark that we choose to plot example curves with the same $\lambda$ to avoid complications with fitting noisy data, but that the real data we have already encountered faces the \enquote{ambiguous $\lambda$ problem} also due to noise.

\subsection{Job-to-job fluctuations complicate comparing data from different jobs}
\label{subsec:job-to-job-fluctuations}

We conclude our arguments in favor of the ITA approach by bringing up a subtle problem in the usage of cloud-based quantum devices.  At a high level,  users of cloud-based devices often have an implicit assumption that data taken within a fixed calibration cycle all come from roughly the same distribution.  Hence, taking data for many states, averaging, and then fitting (the AtF approach) is sensible.  In other words, the average over states is the Haar average computed on the device with the given backend properties.  However, this is not generally true for all demonstrations.  Some relevant quantum memory demonstrations (such as probing Free fidelities) violate this assumption, and data taken from one job is as if sampled from a different distribution if not appropriately handled.  In this sense, we argue against naive averaging of data not taken within the same job. Whenever possible, we prefer to only make direct comparisons within a single job unless stability is empirically observed. 

To justify this caution, we carefully test the veracity of standard \emph{a priori} assumptions on a series of identical (procedure-wise) Free demonstrations on \armonk. The Free demonstration procedure follows the standard DD protocol we established in \cref{subfig:dd-circuit}. Namely, we prepare $\ket{+}$, idle for 75$\mu s$ by applying $\approx 1000$ identity gates, unprepare $\ket{+}$, and then measure in the computational basis. To run statistical tests on the result of this demonstration, we repeat it many times. Because IBM quantum computers operate on a queuing system (similar to a high-performance computing scheduler like Slurm), there are actually several different ways to repeat an demonstration, and as we will show, which way is chosen makes a difference. This is the reason for the rather pedantic discussion that follows. 

The first notion of repeating an demonstration is to simply sample the same circuit many times by instructing the IBM \emph{job} scheduler (we define ``job'' below) to use $N_s$ shots. For example, the simplest possible job testing the above demonstration consists of sending \armonk\ the tuple $(C, N_s)$, where $C$ is the circuit encoding the above Free demonstration. Upon receiving such a tuple, \armonk\ samples $C$ for $N_s$ shots once the job reaches position one in the queue. This repetition is the same as discussed in \cref{app:bootstrap-fid}, and hence, is the basis of estimating the empirical fidelity $f_e$ [\cref{eq:app-empirical-fid}] by bootstrapping the $N_s$ bitstrings. We are interested in the stability of $f_e$ and hence the stability of repeating this entire procedure. Naively, we could repeatedly send \armonk\ the same tuple in $M$ different jobs, i.e., $[J_1 = (C, N_s), J_2 = (C, N_s), \ldots, J_M = (C, N_s)]$ and compute $M$ estimated empirical fidelities, $[f_1, f_2, \ldots, f_M]$ where we dropped the $e$ subscript for simplicity of notation. However, submitting demonstrations this way wastes substantial time waiting in the queue since $J_1$ and $J_2$ are not generally run contiguously due to other users also requesting jobs.   

More generally, a \emph{job} consists of a list of circuits that are sampled contiguously, i.e., without interruption by other users. For our purposes, each job $J_k$ shall consist of $L_k$ identical demonstration tuples, $J_k = [E_{k, 1} \equiv (C_1, N_s)_{k, 1}, E_{k, 2} \equiv (C_2, N_s)_{k, 2}, \ldots, E_{k, L_k} \equiv (C_k, N_s)_{k, L_k}]$ which allows us to compute $L_k$ empirical fidelities, $R_k = [f_{k, 1}, f_{k, 2}, \ldots, f_{k, L_k}]$ ($R_k$ standing for result $k$). Since $C_j = C\  \forall j$, the index on $C_j$ is technically redundant, but keeping it helps us make our final point on the intricacies of collecting data within a job. In particular, \armonk\ samples the $L_k \times N_s$ bitstrings by running the circuits of $J_k$ in the order $[C_1, C_2, \ldots, C_L]^{N_s}$ as opposed to $[C_1]^{N_s}\ldots [C_L]^{N_s}$, as one might naively expect. The goal of this strategy is to try and prevent device drift from biasing the results of $E_{k, 1}$ compared to, e.g., $E_{k, L_k}$. With the notation defined, we can state exactly what tests we ran.

\begin{figure}
    \centering
    \includegraphics[width=0.48\textwidth]{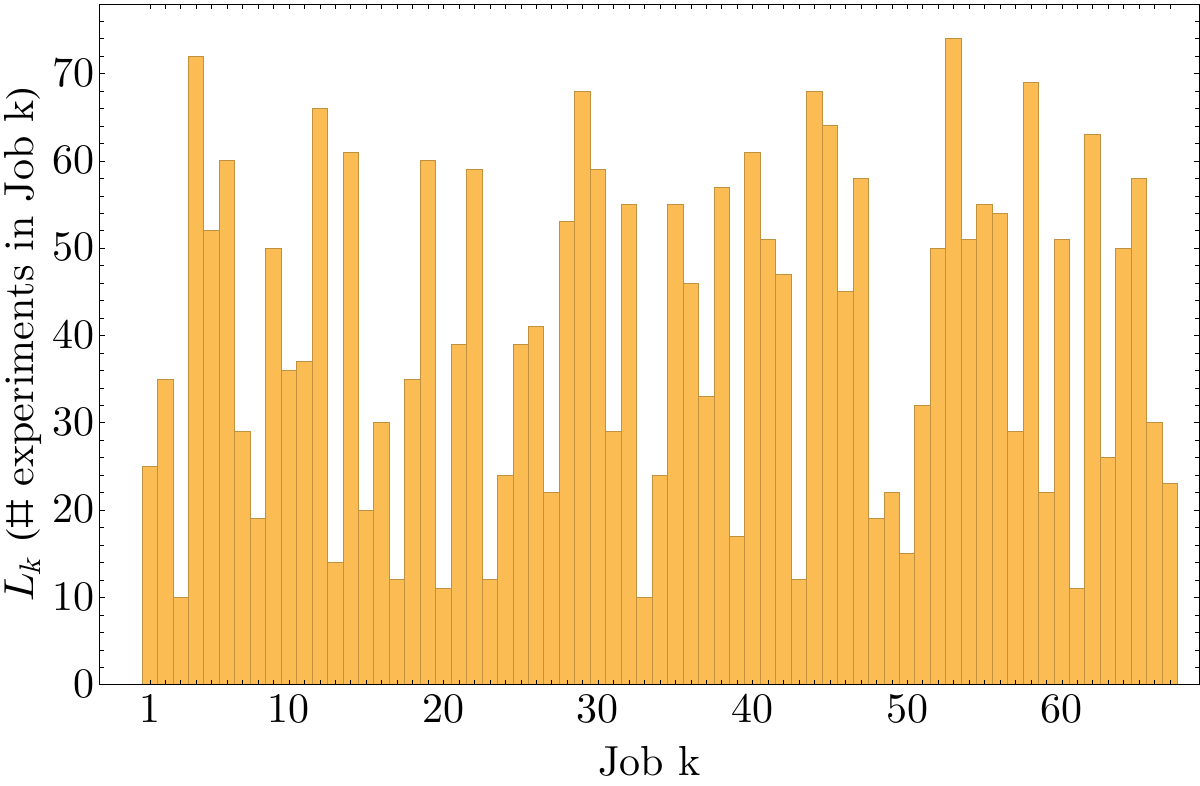}
    \includegraphics[width=0.48\textwidth]{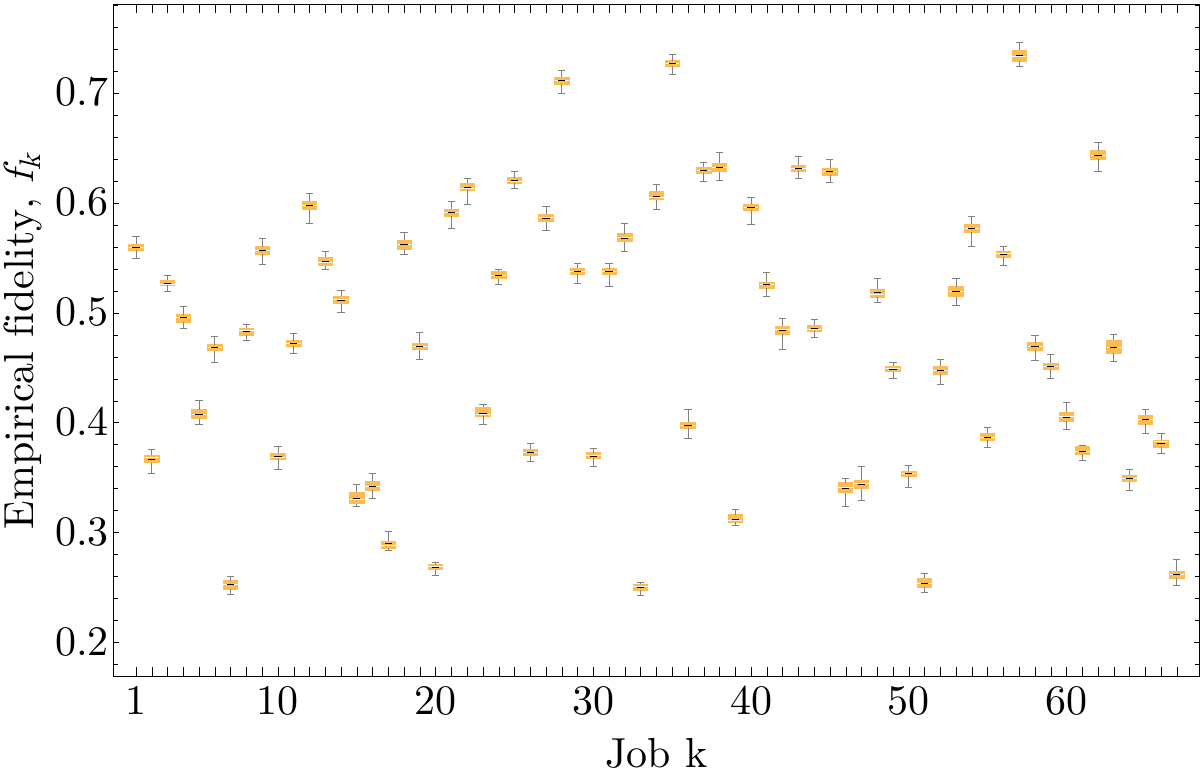}
    \caption{A coarse view of the entire \enquote{stats test data set} taken on \armonk. (Top) The random value of demonstrations, $L_k \in [10, 75]$, chosen for each job. (Bottom) The spread of fidelities for each job summarized in 67 box plots. In our earlier notation, this is a box plot over the empirical fidelities, $R_k$. All data was taken within a fixed calibration cycle over a 10 hour period, and many jobs (22 / 67) are contiguous, e.g., jobs 2 and 3 are run back-to-back with no interruption from other users' jobs. Despite this, there is a large variation in fidelity from job to job, which does not seem correlated with the job duration.}
    \label{fig:armonk-repro-tests-1}
\end{figure}

To generate our data set, we constructed 67 jobs $J_1, \ldots, J_{67}$ with a uniformly random value of $L_k \in [10, 75]$ ($75$ is the maximum demonstration allotment on \armonk) but an otherwise identical set of demonstrations $E_{k, j}$ and submitted them to the \armonk\  queue~\footnote{We actually sent $100$ jobs, but after $67$ successful completions, our program crashed due to an internet connection issue.}. The value in choosing a random $L_k$ is to check whether the duration of the job itself has any effect on results. We coarsely summarize the set of all job results in \cref{fig:armonk-repro-tests-1}, which we call the \enquote{stats test data set}. The entire data set was taken over a ten-hour period over a fixed calibration cycle. In fact, $22$ of the $67$ jobs were run contiguously (i.e., one after the other without other user's jobs being run in between).\footnote{We claim two jobs are contiguous if they began running on the device within 1 minute of each other. This is because the full circuit takes $\approx 80 \mu s$ to execute, so 8192 shots takes around $0.657s$ to execute. Hence, a job with $10$ to $75$ demonstrations takes between $6.57s$ to $49.3s$ to execute excluding any latency time to communicate between our laptop and the \armonk\ server.} Despite the jobs being localized in time, there are large jumps in average $f_k$ values from job to job, which violates our \emph{a priori} assumption of how sampling from a quantum computer should work within a fixed calibration cycle. We make this more precise in a series of assumptions and results which support or reject the given assumption.  

\begin{assumption}{1}\label{as:normal-inner-job}
Fidelities collected within a fixed job are normally distributed. Using the above notation, samples within $R_k = [f_{k, 1}, \ldots, f_{k, L}]$ are drawn from a fixed normal distribution, i.e., $f_{k, j} \sim \mathcal{N}(\mu_k, \sigma_{k})$ for some unknown mean and variance.
\end{assumption}

\begin{aresult}{1}\label{res:normal-inner-job}
Assumption~\ref{as:normal-inner-job} is well supported by hypothesis tests (Jarque-Bera, Shapiro-Wilk) and visual tests (box-plot, quantile-quantile (QQ) plot).
\end{aresult}

The Jarque-Bera test is a means to test whether a sample of points has the skewness and kurtosis matching that expected of a normal distribution~\cite{jarque-bera-test-wiki}. We implement it in Mathematica using the \verb^JarqueBeraALMTest^ command. The Shaprio-Wilk test checks whether a set of sample points has the correct order statistics matching that expected of a normal distribution~\cite{shapiro-wilk-test-wiki}. We implement it using the Mathematica command \verb^ShapiroWilkTest^. Both tests are hypothesis tests where the null hypothesis $H_0$ is that the data was drawn from a normal distribution and an alternative hypothesis $H_a$ that it was not. Like most hypothesis tests, they are rejection based methods. Namely, the tests return a $p$-value corresponding to the probability of $H_0$. If $p < 0.05$, we reject the null hypothesis with $95\%$ confidence (this is the default setting for the Mathematica command which we employ). Otherwise, we say that the data does not support rejecting $H_0$--namely, it is reasonable that a normal distribution could have produced such a sample. We summarize the results of both tests in \cref{tab:jb_and_ks_test_table}. 

\begin{table}[ht]
    \centering
    \begin{tabular}{c|c|c}
        Test used &  Job's $k$ where $H_0$ rejected & $p$-values of rejected jobs \\
        \hline
        Jaque-Bera test & $\{\}$ &  $\{\}$ \\
        \hline
        Shapiro-Wilk & $\{17, 22, 46\}$ & $\{0.028, 0.033, 0.048\}$
    \end{tabular}
    \caption{Summary of jobs where the null hypothesis that the data is sampled from a normal distribution is rejected with 95\% confidence.}
    \label{tab:jb_and_ks_test_table}
\end{table}

From \cref{tab:jb_and_ks_test_table}, we see that among the 67 jobs, only 3 give us cause for concern. Namely, jobs $\{17, 22, 46\}$ reject the null hypothesis under the Shapiro-Wilk test. From the bar chart in \cref{fig:armonk-repro-tests-1}, these jobs have $\{12, 59, 45\}$ demonstrations, respectively. Since there are several other jobs with similar lengths whose data does not reject $H_0$, there does not seem to be a correlation between job length and rejection of normality.

\begin{figure*}
    \centering
    \includegraphics[width=0.99\textwidth]{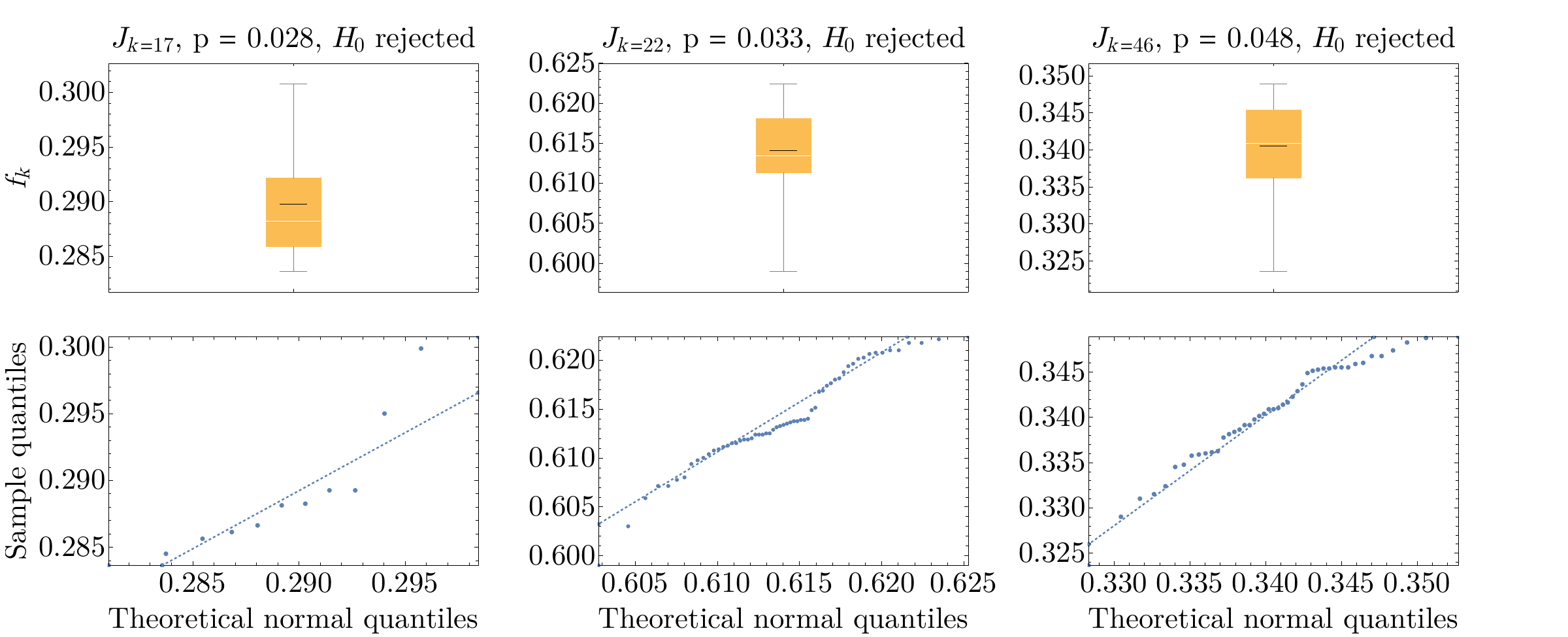}
    \includegraphics[width=0.99\textwidth]{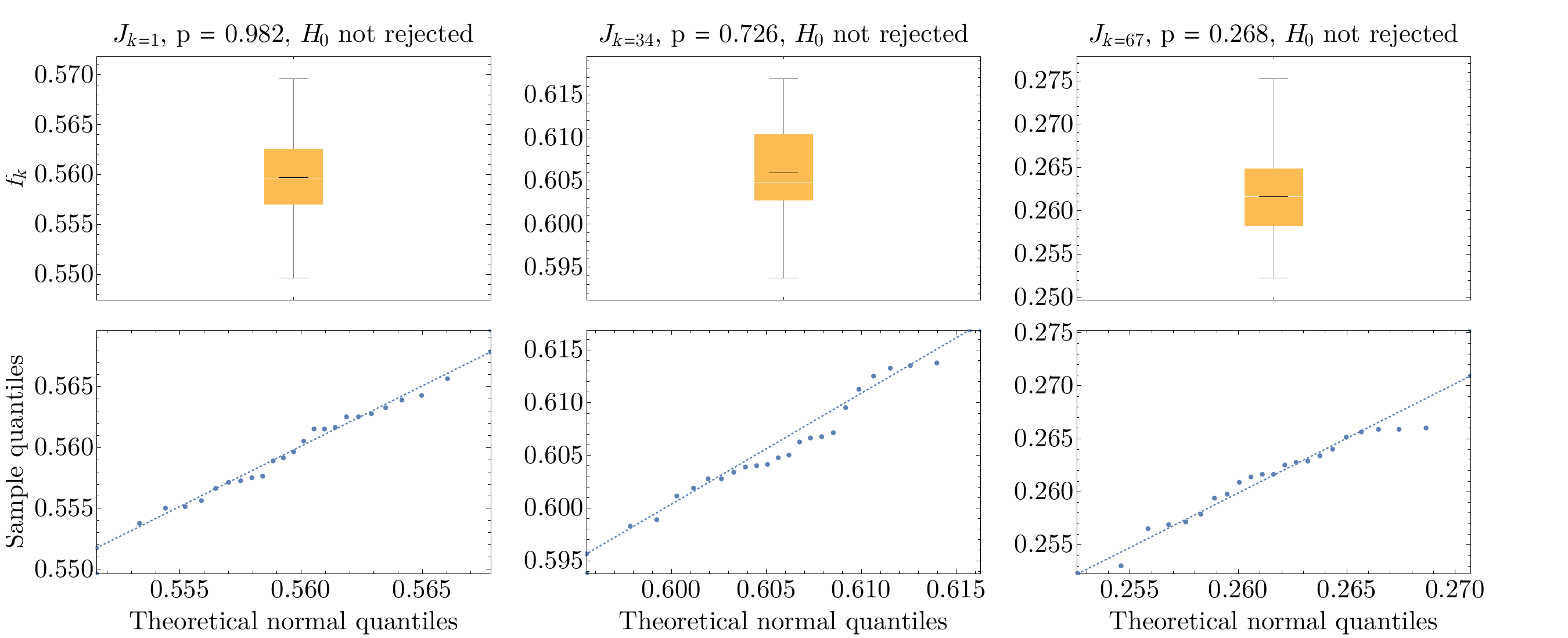}
    \caption{Box plots and quantile-quantile (QQ) plots corresponding to different job data from \cref{fig:armonk-repro-tests-1}. The top two rows show the three jobs which failed the Shapiro-Wilk test for normality. The bottom two rows show three jobs that did not fail the test and are otherwise chosen to be the first, middle, and last jobs in the set of data. Normality is characterized visually as (i) a symmetric box plot whose mean and median agree or (ii) a QQ plot whose data falls approximately along the diagonal. This is best exhibited by the $k = 1$ data in the third row. A deviation from normality is the negation of any of the above properties. For example, the $k = 17$ job in the first row departs from normality in all three respects.}
    \label{fig:armonk-repro-tests-2}
\end{figure*}

We can also check the normality of the data visually using box plots or quantile-quantile (QQ) plots as we do in \cref{fig:armonk-repro-tests-2}. Recall that $Q_x$ represent the $x^{\text{th}}$ quantile, i.e., the smallest value contained in the data set for which $x \%$ of the data lies below the value $Q_x$. 
The QQ plot is a way to compare the quantiles of two data sets. In our case, we compare the quantiles of the sample data set to that of a normal distribution with the same mean and standard deviation of the data. Namely, given samples $R_k = [f_1, \ldots, f_{L_k}]$, we compare the quantiles of $R_k$ itself to $\mathcal{N}(\overline{R}_k, s_{R_k})$, where $\overline{R}_k$ is the mean of the data and $\sigma_{R_k}$ is the uncorrected sample standard deviation, i.e., $s_{R_k} = (\frac{1}{N}\sum_i (f_i - \overline{R}_k)^2)^{1/2}.$ Letting $Q_x$ be the quantiles of our raw data and $\hat{Q}_x$ the quantiles of the corresponding normal distribution, then points on our QQ plots correspond to $(Q_x, \hat{Q}_x).$ Hence, if the two distributions are the same (or close), then the data should fall on the diagonal $Q_x = \hat{Q}_x$, which indicates normality.

We illustrate the use of these plots in \cref{fig:armonk-repro-tests-2}. In the top two rows we exhibit the three example jobs that failed the Shapiro-Wilk test for normality. For $J_{k=17}$, the corresponding box plot is (i) not symmetric and (ii) has a mean that deviates significantly from its median. The QQ plot also deviates from a diagonal line, especially for larger fidelities. As we move from $k = 17$ to $k = 22$ and then $k = 46$, the data appears increasingly more normal but still fails the Shapiro-Wilk test. The extent to which they fail the visual tests is best seen by considering a normal-looking example, which we can glean from the bottom two rows of plots, which were not rejected by either hypothesis test.

For the data shown, $J_{k=1}$ is the closest data set to normal. Here, $p = 0.982$ for Shapiro-Wilk, the box plot is symmetric, the mean/median agree, and the QQ plot data hardly deviates from the diagonal. Given this almost ideally normal data set, it is easier to see why the other data sets fail to appear as normal. For example, even those that pass hypothesis tests exhibit some non-normal features. But among those rejected, we see deviations from the diagonal line, skewness (a lack of symmetry in the box plot), and a discrepancy between the mean and the media. 

Since most jobs (64/67) have data that passes hypothesis and visual normality tests, we conclude that Assumption~\ref{as:normal-inner-job} is reasonable.

\begin{figure}
    \centering
    \includegraphics[width=0.49\textwidth]{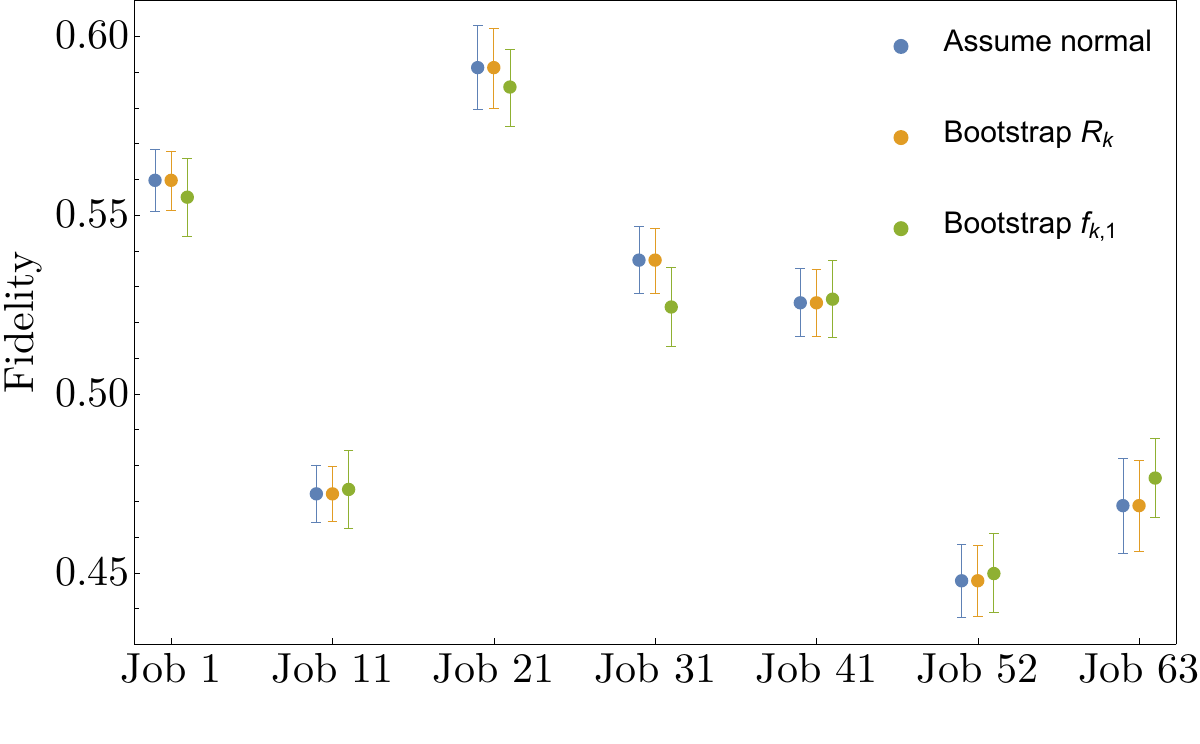}
    \caption{We compare different approaches of summarizing the fidelities from running repeated identical demonstrations within a fixed job. In other words, we compare three ways to summarize $R_k = [f_1, \ldots, f_{L_k}]$. The first method, \enquote{assume normal}, is to simply compute the sample average $\overline{f_k}$ and sample standard deviation $s_k$ and then report $\overline{f_k} \pm 2 s_k$ as the average fidelity. The second method, \enquote{bootstrap $R_k$}, is to bootstrap over the fidelities in $R_k$. To correctly estimate the population standard deviation, note that we must rescale the bootstrapped mean error by a factor of $\sqrt{L_k}$ (see text for more details). Finally, we can instead bootstrap over the counts used to compute $f_{k, 1}$ directly, which we call \enquote{bootstrap $f_{k,1}$}. This does not use the information for $f_{k, j}$ for $j > 2$ at all. Nevertheless, all three methods yield self-consistent results, as seen visually by the significant overlap in error bars between the three estimates. Note we show only a subset of the results for all jobs to ensure that the error bars are large enough to be visible. For a sense of the consistency among the approaches, note that Job 31 has the largest deviation. Hence, it is sufficient to use the bootstrap $f_{k, 1}$ method as purported in Result~\ref{res:normal-mean-and-sd}.}
    \label{fig:armonk-repro-tests-3}
\end{figure}

\begin{assumption}{2}\label{as:normal-mean-and-sd}
Any demonstration within a fixed job is representative of any other identical demonstration within the same job. Hence, it is not necessary to repeat an identical demonstration within a fixed job. In other words, $f_{k, 1} \pm \sigma_{k, 1}$ where $\sigma_{k, 1}$ is obtained by bootstrapping as in \cref{app:bootstrap-fid} is a sufficient summary of the estimates $\mu_k$ and $\sigma_k$ in $f_{k, j} \sim \mathcal{N}(\mu_k, \sigma_k)$. 
\end{assumption}
\begin{aresult}{2}\label{res:normal-mean-and-sd}
Assumption~\ref{as:normal-mean-and-sd} is justifying by a simple direct comparison of mean fidelity and standard deviation estimates. In particular, we compare (i) directly computing the sample mean and sample standard deviation and assuming normality, (ii) bootstrapping over the fidelities in $R_k$, and (iii) bootstrapping over demonstration output counts used to compute $f_{k, 1}$. For all jobs (i.e., $\forall k)$, the three methods are consistent with each other since the $2\sigma$ error bars have significant overlap in all cases. See \cref{fig:armonk-repro-tests-3}.
\end{aresult}

The basic idea is that all three methods are consistent with each other. By consistency, we mean the 
intuitive notion that the error bars have significant overlap with each other. Further rigor, in this case, is not necessary due to the degree of matching with respect to job-to-job fluctuations we discuss below. This consistency is significant because the final method shown in \cref{fig:armonk-repro-tests-3} bootstraps only over the counts used to compute $f_{k, 1}$. Hence, it is sufficient to run an demonstration only once within a single job and use the results as being representative of repeating the identical demonstration, thereby showing the claimed result.

We next detail the methods leading to the three summaries in \cref{fig:armonk-repro-tests-3}. The first is to summarize $R_k$ by the sample mean and sample standard deviation, $\overline{f_k} \pm 2 s_k$ where $\overline{f_k} = \frac{1}{L_k}\sum_{i=1}^{L_k} f_{i}$ and $s_k = \sqrt{\frac{1}{L_k - 1} \sum_{i=1}^{n} (f_i - \overline{f_k})^2}.$ By Result~\ref{res:normal-inner-job}, it is appropriate to then characterize the sample with the first two moments --- hence the name \enquote{assume normal.} 
The second and third methods, called \enquote{bootstrap $R_k$} and \enquote{bootstrap $f_{k, 1}$}, both utilize the bootstrapping procedure discussed in \cref{app:bootstrap-fid}. The difference is in what is treated as the samples. In the $R_k$ case, we bootstrap over the $f_{k, j}$ values using $n_s \equiv 10,000$, $L_k$-sized samples. From these $n_s$ estimates of $\overline{f_k}_l$, we can estimate $\langle R_k \rangle = \frac{1}{n_s}\sum_{i=1}^{n_s} \overline{f_k}_l$. We can then estimate the sample standard deviation $s_{R_k} = \sqrt{\frac{1}{n_s - 1} \sum_{i=1}^{n_s} (\overline{f_k}_l - \langle R_k \rangle )^2}.$ It is important to note that $s_{R_k}$ here has the meaning of the standard deviation of our estimate of the mean. As $L_k \rightarrow 0$, $s_{R_k} \rightarrow 0$, so $s_{R_k}$ is \emph{not} an estimate of $\sigma_k$ as in some parametric distribution $\mathcal{N}(\mu_k, \sigma_k)$. To match the same meaning as the \enquote{assume normal} estimate, we report $\langle R_k \rangle \pm 2 \sqrt{L_k} s_{R_k}$ as the \enquote{bootstrap $R_k$} estimate. In the $f_{k, 1}$ case, we bootstrap over the counts used to compute $f_{k, 1}$ in exactly the same way as discussed in \cref{app:bootstrap-fid}. Hence, this method does not utilize the demonstrations $f_{k, j}$ for $j > 2$, and the sample standard deviation is an estimate of $\sigma_k$ and is consistent with  \enquote{assume normal.}

\begin{assumption}{3}\label{as:same-normal-dist-between-jobs}
Given that $R_k$ can be modeled as samples from $\mathcal{N}(\mu_k, \sigma_k)$, data in a different job $R_p$ for $p \neq k$ but in the same calibration cycle can also be modeled as samples from $\mathcal{N}(\mu_k, \sigma_k)$.
\end{assumption}
\begin{aresult}{3}\label{res:same-normal-dist-between-jobs}
Assumption~\ref{as:same-normal-dist-between-jobs} is not supported by the data we have already seen in Figs.~\ref{fig:armonk-repro-tests-1},\ref{fig:armonk-repro-tests-2}, and \ref{fig:armonk-repro-tests-3}. It is worth reiterating that this assumption does not hold even when the calibration is fixed or even when the jobs are contiguous. 
\end{aresult}

In \cref{fig:armonk-repro-tests-1}, we see box plots scattered wildly with median fidelities ranging from $0.24$ to $0.75$ and whose ranges do not overlap. Put more precisely, recall that when \enquote{assuming normality} as in \cref{fig:armonk-repro-tests-3}, we can simplify each job's results to a single estimate $\overline{f}_k \pm 2 s_k$. Once we have done that, we find $\min_k \overline{f}_k = 0.250$ and $\max_k \overline{f}_k = 0.734$ and yet $\max_k s_k = 0.013$, so it is not possible for all the data to be consistent as described in Assumption~\ref{as:same-normal-dist-between-jobs}. This discrepancy is not resolved when considering jobs that are time-proximate or contiguous, as we have discussed before. To see this, note that all jobs were taken within a ten-hour window for a fixed calibration: the first nine jobs were taken within the first hour, the first three of which within the first ten minutes. Despite this, there are large variations that violate Assumption~\ref{as:same-normal-dist-between-jobs} even within the first three jobs. 

This result is the first major departure from expectations; it tells us that an demonstration must be repeated many times across different jobs to estimate the variance in performance from run to run. One consequence is that if we test the preservation of $\ket{\psi}$ in job 1 and $\ket{\phi}$ in job 2, it is not reliable to compare the fidelities $f(\ket{\psi}, T)$ and $f(\ket{\phi}, T)$ and draw conclusions about the results of a generic demonstration testing both $f(\ket{\psi}, T)$ and $f(\ket{\phi}, T)$. Alternatively, if we test the efficacy of two sequences $s_1$ and $s_2$ given the same state but where the data is taken from different jobs, this is not a reliable indicator of their individual relative performance either. Instead, many demonstrations must be taken where the variance can be estimated, or direct comparisons should only be made within a fixed job and then repeated to check for the stability of the result.

The next two assumptions and results address to what extent data taken across different jobs, or even calibrations, can be modeled as being normally distributed.

\begin{assumption}{4}\label{as:one-big-normal}
It is appropriate to model $f_{k, 1} \sim \mathcal{N}(\mu_C, \sigma_C)$ for $C$ some fixed calibration cycle provided all jobs $k$ are in $C$. In other words, demonstrations taken from different jobs can be viewed as sampled from a fixed normal distribution.
\end{assumption}
\begin{aresult}{4}\label{res:one-big-normal}
This assumption is justified since it is not rejected by the Jaque-Bera and Shapiro-Wilk hypothesis tests and is supported by visual indicators (box plot, QQ plot, and error bar comparison plot). 
\end{aresult}

Let $A = \{f_{k, 1}\}_{k=1}^{67}$ be the set of the first empirical fidelity from each job. Given this data set, neither the Jaque-Bera nor the Shaprio-Wilk test support rejecting the hypothesis that $f_{k, 1} \sim \mathcal{N}(\mu_C, \sigma_C)$. More precisely, the $p$-values are $0.348$ and $0.192$, respectively, so with 95\% confidence, we cannot reject the hypothesis. 

\begin{figure}
    \centering
    \includegraphics[width=0.45\textwidth]{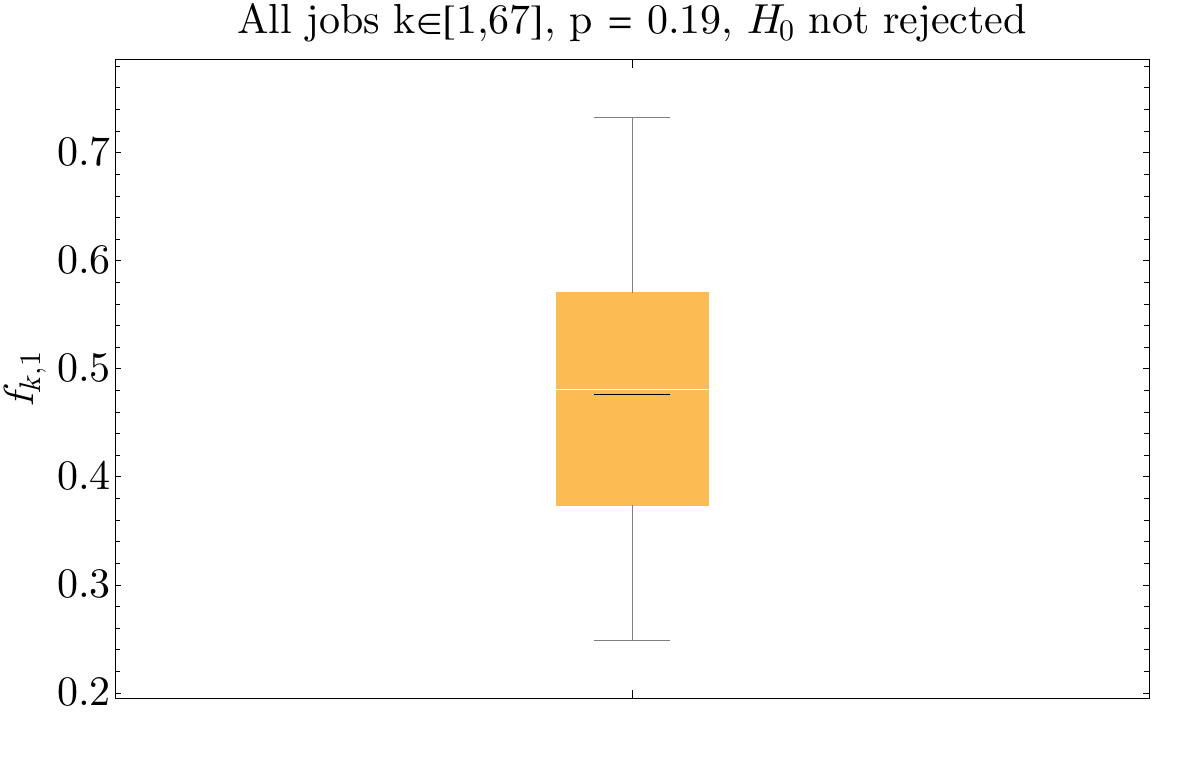}
    \includegraphics[width=0.45\textwidth]{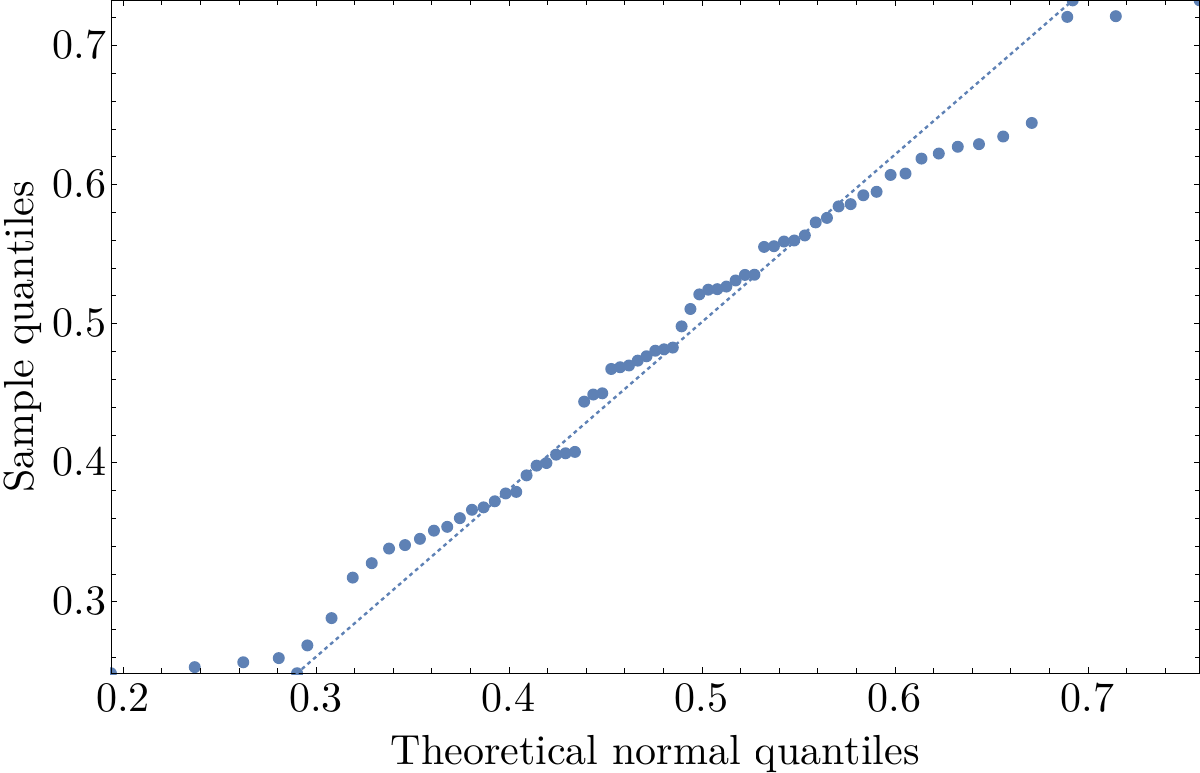}
    \includegraphics[width=0.45\textwidth]{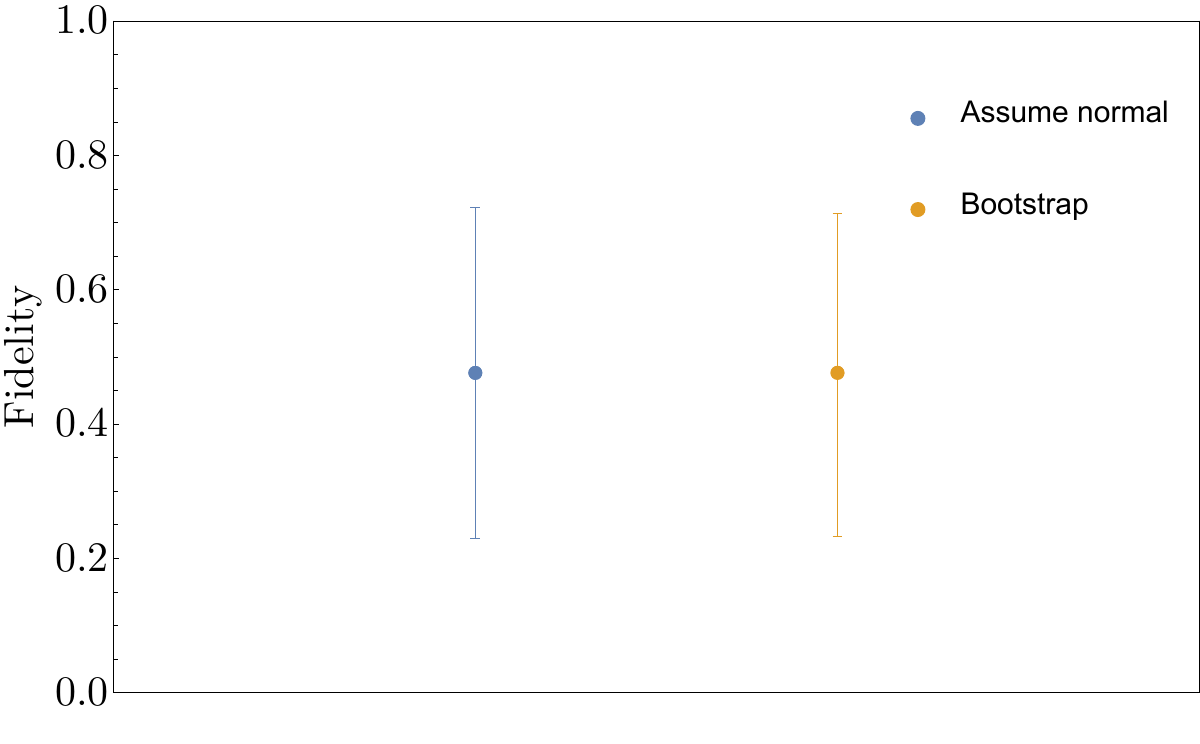}
    \caption{We graph the visual normality tests for the data set consisting of the first fidelity obtained from each job, $A = \{f_{k, 1}\}_{k=1}^{67}$. Alongside the fact that both the Jaque-Bera and Shapiro-Wilk tests do not reject normality, we can conclude that data set $A$ does indeed appear to have been sampled from a normal distribution $\mathcal{N}(\mu_C, \sigma_C)$ by the same aforementioned box-plot and QQ-plot criteria. In the bottom panel, we show the difference between the \enquote{assume normal} procedure for data set $A$ versus bootstrapping. Since both agree, it is not only reasonable to assume normality but also to use either method to make a consistent prediction for curve fitting.}
    \label{fig:armonk-repro-tests-4}
\end{figure}

In addition, we show the standard visual tests --- box plot and QQ plot --- in \cref{fig:armonk-repro-tests-4} alongside the error bar comparison plot similar to \cref{fig:armonk-repro-tests-3}. The standard visual tests also support assuming normality since the box plot is symmetric, has a mean and median that almost agree, and the QQ plot data lies mostly along the diagonal. However, perhaps the most important evidence is the error bar comparison (bottom panel). By \enquote{assume normal}, we compute the sample mean ($\overline{A}$) and standard deviation of $\sigma_{A}$ of $A$ as the estimates of $\mu_C$ and $\sigma_C$. We find $\overline{A} = 0.48$ and $\sigma_A = 0.12$. Hence, we would report $0.48 \pm 24$ as the average fidelity value across the fixed calibration. When instead bootstrapping the samples of $A$, we again find $0.48 \pm 0.24$, which remarkably has symmetric error bars. 

Hence, in all, it is reasonable to model the fidelity samples from different jobs as samples of a fixed normal distribution within a fixed calibration cycle. Note that we were careful not to try and model the entire data set as normal; instead, we held the number of samples from each job constant. Had we modeled the entirety of the data set, the Jaque-Bera and Shaprio-Wilk tests would reject normality. This is because some samples are unfairly given more weight, which is no longer normal.\footnote{By taking a different number of samples, it is as if we roll a die but write down the answer a random number of times. This sampling strategy takes longer to converge to the expected result.}

\begin{assumption}{5}\label{as:normal-across-jobs}
Let $\hat{f}_c$ be empirical fidelity samples similar to $f_{k, 1}$ from before but not necessarily sampled from a fixed calibration. Then $\hat{f}_c \sim \mathcal{N}(\mu, \sigma)$ even across different calibrations.
\end{assumption}
\begin{aresult}{5}\label{res:normal-across-jobs}
This assumption is justified for the free evolution (Free) data but is violated for some DD sequences such as KDD. Nevertheless, for average point-wise fidelity estimates of the form $\overline{f_c} \pm 2 \sigma_c$, the difference between assuming normality or bootstrapping is negligible. 
\end{aresult}

Here, we will need to consider a different data set for the first time in this series of assumptions and tests. Instead of considering data from a fixed time $T = 75\mu s$, we consider fidelity decay data as in \cref{fig:sample-fid-decay-curves,fig:sample-fitting-approaches}. Namely, we consider the preservation of $\ket{+}$ and $\ket{1}$ under Free and KDD as a function of time. In each case, we consider the fidelity at twelve roughly equidistant time steps $f^{(s)}_c(\ket{l}, T_j); T_j \approx \frac{150}{13} (j - 1) \mu s$ for $j = 1, \ldots, 12$. Here, $s$ is a placeholder for the sequence (Free or KDD), $c$ for the calibration at which the fidelity was sampled, and $l$ for the state. We test the normality for each set of fixed $(s, l, j)$, which amounts to a set such as $\hat{f}_c$ in Assumption~\ref{as:one-big-normal}.

We begin by applying both the Jarque-Bera and Shaprio-Wilk hypothesis tests. We tabulate the cases where the hypothesis of normality fails with 95\% confidence in \cref{tab:freeAndKddFailsNormal}. According to these tests, we can confidently claim that the set of KDD fidelity decay curves for $\ket{+}$ does not obey all the properties we would expect for samples of a series of normal distributions. Namely, the times $T_j$ for $j \in [3, 9]$ do not have the right order statistics since they fail the Shapiro-Wilk test. This seems reasonable given \cref{fig:sample-fid-decay-curves}, in which the KDD data appears to bi-modal for the middle times.

\begin{table}[ht]
    \centering
    \begin{tabular}{c|c|c|c}
        $s$ & $l$ & rejected $j$ by JB & rejected $j$ by SW \\
        \hline
        Free & $\ket{+}$ & \{6\} & \{6\} \\
        \hline
        Free & $\ket{1}$ & \{2\} & \{2\} \\
        \hline
        KDD & $\ket{+}$ & \{\} & \{3,4,5,6,7,8,9\} \\
        \hline 
        KDD & $\ket{1}$ & \{6\} & \{6,7,8\}
    \end{tabular}
    \caption{Summary of Free and KDD decay curve data sets that violate the null hypothesis of normality with 95\% confidence according to the Jarque-Bera (JB) or Shapiro-Wilk (SW) tests. The data set consists of a set of fidelities $\{f^{(s)}_c(\ket{l}, T_j)\}_{c = 1, \ldots, 14}$ for $s$ a DD sequence, $c$ the calibration, $\ket{l}$ the state, and $j$ the time point for $j = 1, \ldots, 12$. Note that we test normality where samples are drawn over different calibrations $c$ with the other parameters fixed.}
    \label{tab:freeAndKddFailsNormal}
\end{table}

\begin{figure}
    \centering
    \includegraphics[width=0.48\textwidth]{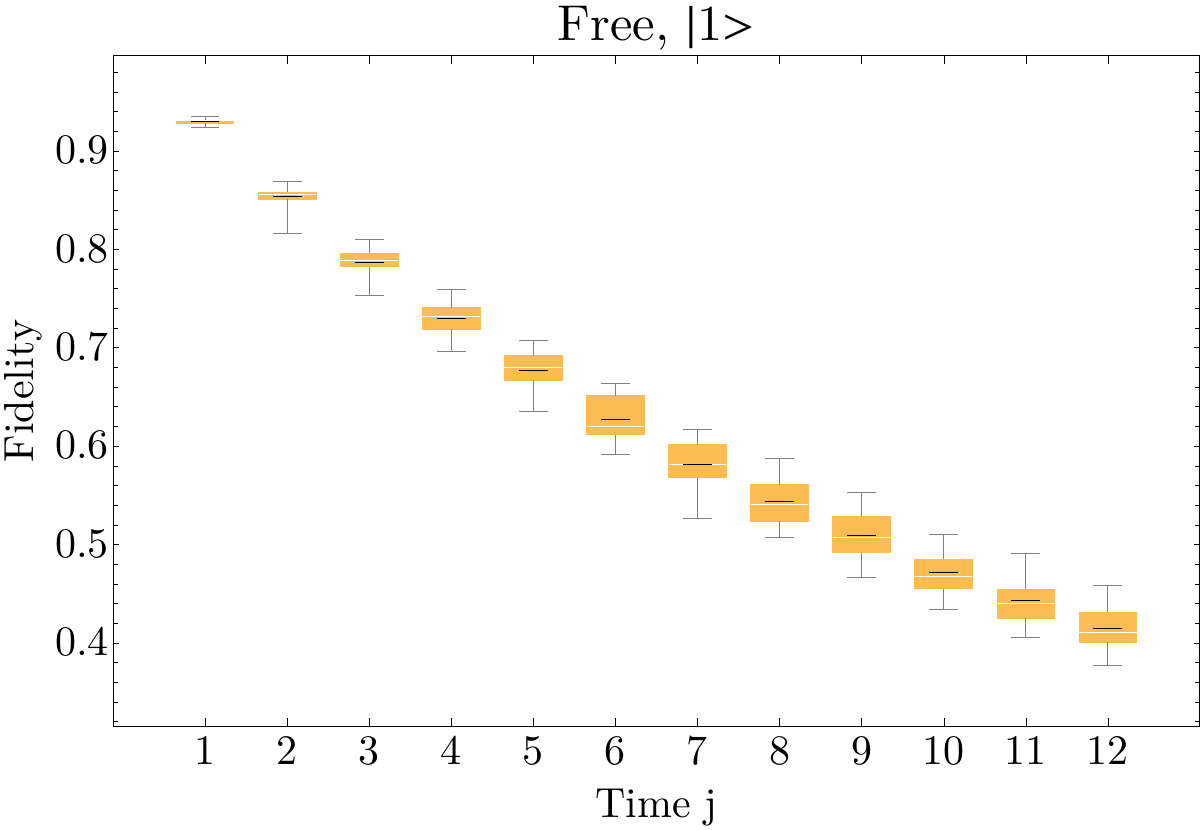}
    \includegraphics[width=0.48\textwidth]{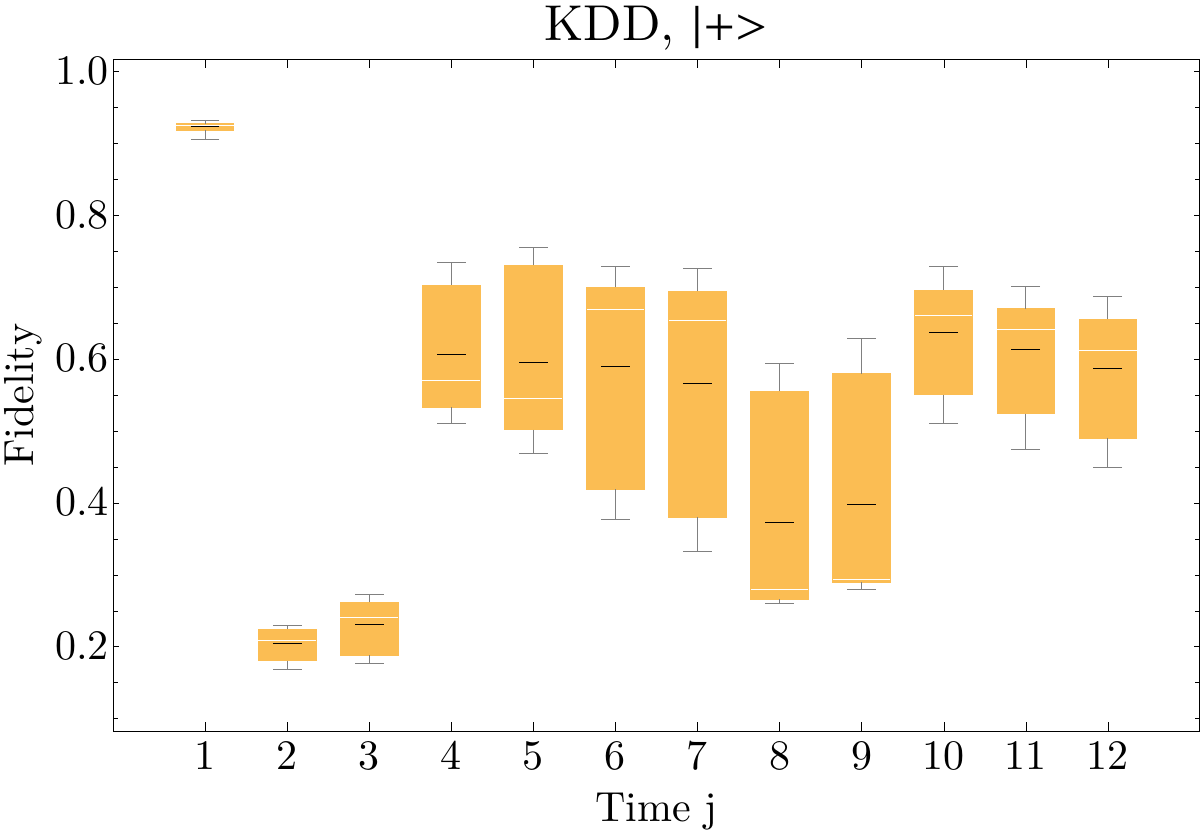}
    \caption{We plot the fidelity decay of $\ket{1}$ under Free and $\ket{+}$ under KDD. The variation in the box plots is obtained by running identical demonstrations for that time across different calibrations. The Free data appears mostly normal at each time, which is corroborated by the hypothesis tests in \cref{tab:freeAndKddFailsNormal}. The KDD data does not appear normal for the intermediate times, in agreement with the hypothesis tests.}
    \label{fig:armonkTimeSeriesNormalBox}
\end{figure}

We can also consider box plots for the data which support the claims made in \cref{tab:freeAndKddFailsNormal}. As an example,  in \cref{fig:armonkTimeSeriesNormalBox} we show the box plot for (Free, $\ket{1}$), which agrees with normality at each time point, and (KDD, $\ket{+})$, which fails to appear normal at most time points.

\begin{figure}
    \centering
    \includegraphics[width=0.48\textwidth]{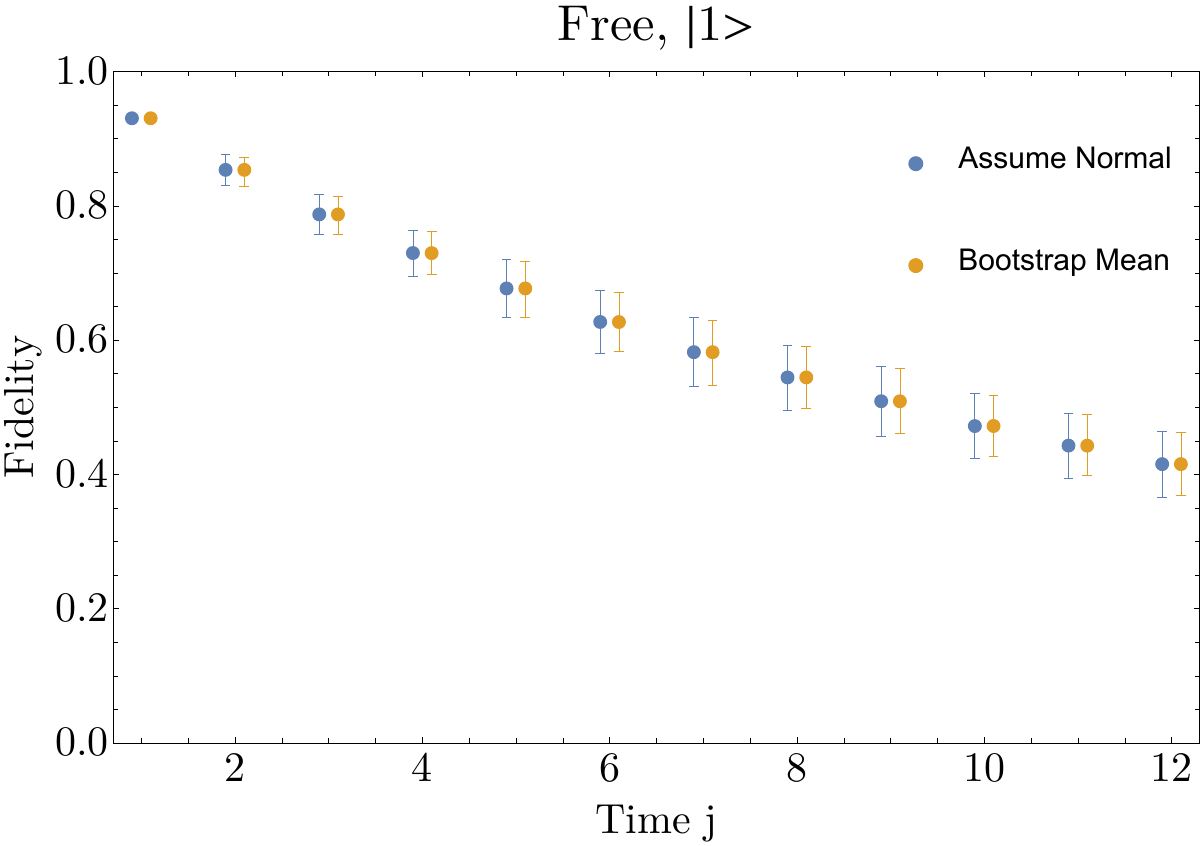}
    \includegraphics[width=0.48\textwidth]{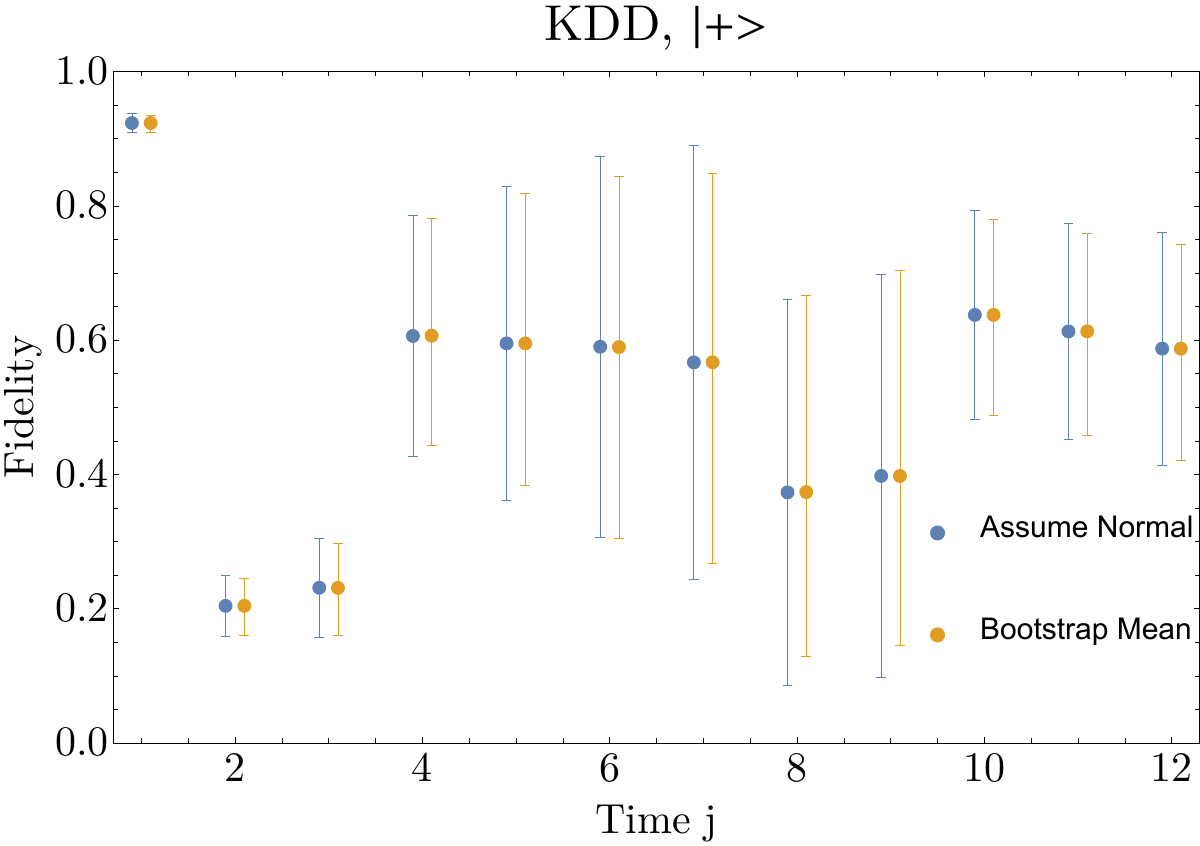}
    \caption{We show that regardless of whether the data formally appears normal, the reported mean and confidence interval is almost identical under this assumption or under bootstrapping. In particular, the top panel appears normal to various normality tests, whereas the bottom data does not. Despite this, reporting the sample mean and the standard deviation is almost identical to reporting the bootstrapped mean and confidence interval.}
    \label{fig:bootstrapVsNormalFreeKdd}
\end{figure}

Despite not passing the tests for normality, we can still formally pretend that the data is normal and compute the sample mean and variance as estimates of $\mu$ and $\sigma$. When we do this and compare the result to bootstrapping the mean and rescaling the confidence interval, the results are almost identical for both sets of data, as we can see in \cref{fig:bootstrapVsNormalFreeKdd}. By rescaling the confidence interval (CI), we mean we compute the 95\% CI of the mean estimate by bootstrapping, $(\Delta_l, \Delta_u)$, and report $(\sqrt{C} \Delta_l, \sqrt{C} \Delta_u)$ where $\sqrt{C}$ is the number of calibration cycles (i.e., the number of fidelities) we bootstrap over. The rescaling is done to obtain an estimate of the spread of the underlying distribution and not of the sample mean. In other words, we observe that averaging via the bootstrap or by the simple sample mean and standard deviation agree even for pathological data of the form we obtain for KDD.

This observation seems to suggest that we can compare DD sequences by first averaging fidelities across calibrations. This would be in line with preview work such as Refs.~\cite{Pokharel:2018aa,souza_process_2020} that did not report on the subtleties of job-to-job variation. As outlined in the main text and implied by the analysis presented in this appendix, we choose not to do this; the next section clarifies why and what we do instead.

\subsection{Effect of job-to-job fluctuations}
\begin{figure}
    \centering
    \includegraphics[width=0.48\textwidth]{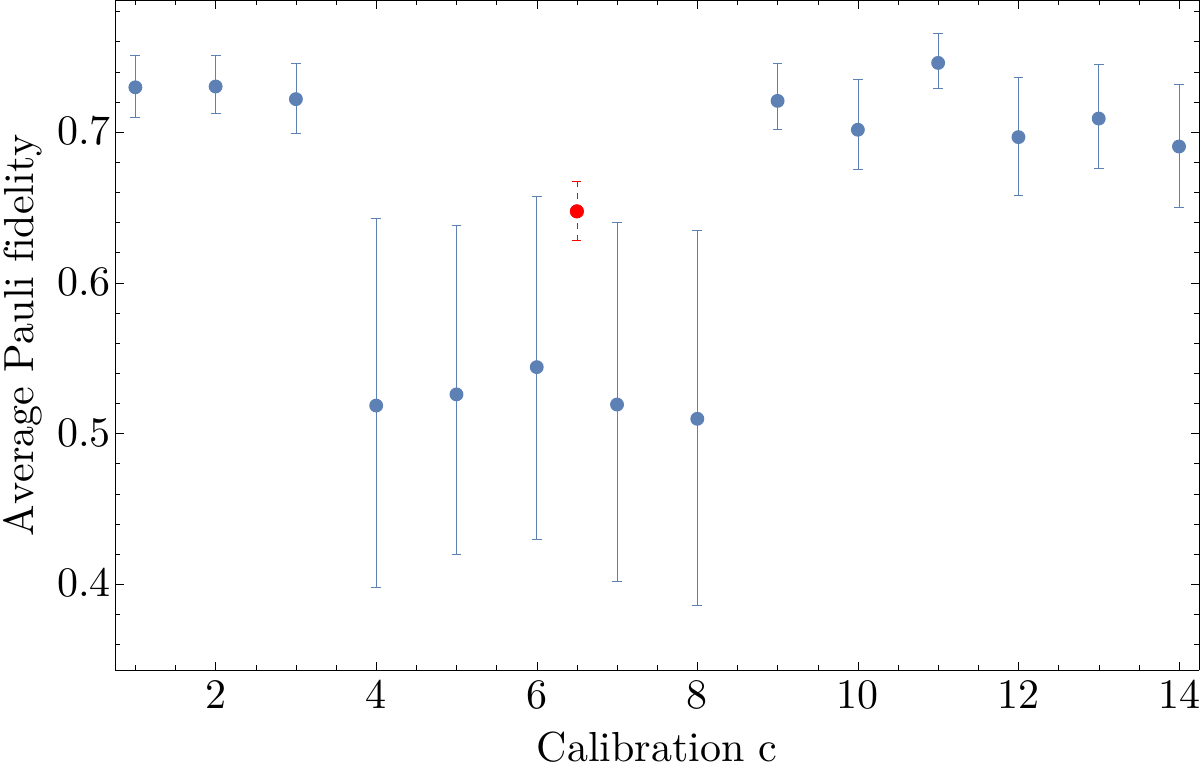}
    \caption{The average fidelity over the six Pauli eigenstates across $14$ calibrations under protection by the KDD sequence. Time is fixed to $T_6$, as defined in \cref{fig:armonkTimeSeriesNormalBox}. The red point is the mean fidelity of the entire data set. This total average is not representative of any calibration average.}
    \label{fig:avgPauliFid}
\end{figure}

Let $f^{(s)}_c$ be a fidelity sample for state $s$ and calibration $c$ for some fixed time and sequence. We saw in Result~\ref{res:normal-across-jobs} that assuming $f^{(s)}_c \sim \mathcal{N}(\mu_s, \sigma_s)$ is not formally justified for all sequences and states. In particular, the $\ket{+}$ state decay for KDD does not pass hypothesis tests for normality at intermediate times. Nevertheless, if we are only concerned with the average fidelity of a fixed state $s'$ across calibrations,
\begin{equation}
    \langle f^{(s')} \rangle_N \equiv \frac{1}{N} \sum_{c=1}^{N} f^{(s')}_c, 
\end{equation}
assuming normality or bootstrapping leads to consistent predictions. But related work---such as Refs.~\cite{Pokharel:2018aa, souza_process_2020}---was concerned with the average fidelity across states, 
\begin{equation}
    \overline{f}_{c'} \equiv \frac{1}{|S|} \sum_{s \in S} f^{(s)}_{c'},
\end{equation}
for some set of states $S$ and fixed calibration $c'$. If $f_c^{(s)} \sim \mathcal{N}(\mu_s, \sigma_s)$ and IID, then $\overline{f}_{c} \sim \mathcal{N}(\mu, \sigma)$ where $\mu = \frac{1}{|S|} \sum_{s \in S} \mu_s$ and $\sigma = \frac{1}{|S|^2}\sum_{s\in S} \sigma_s.$ Thus, we would expect the state-average fidelity to be consistent from job to job. In practice, however, this does not happen. For example, the Pauli averaged-fidelity (i.e., choosing $S = \{\ket{0}, \ket{1}, \ket{+}, \ket{-}, \ket{+i}, \ket{-i}\}$) for different calibrations for the KDD sequence at time $T_6$ is given in \cref{fig:avgPauliFid}. 

The inconsistency of the average state fidelity as exhibited in \cref{fig:avgPauliFid} has several consequences. First, data from a single calibration is generally not enough to characterize typical behavior. For example, if we only sampled a single calibration and happened to sample $c = 6$, we would be left with the wrong impression of KDD's performance. Second, data from a single calibration is also generally not representative of average behavior. Indeed, the red dot in the middle corresponds to the average over the $14$ calibration averages, and no single calibration has a mean consistent with this average. Finally, averaging over states sampled in different jobs can lead to misleading results. For example, suppose we queued up several runs of $\ket{0}$ in calibration $c=1$, of $\ket{1}$ in $c = 2$, etc\ldots. Then, according to the average fidelities in \cref{fig:avgPauliFid}, it is likely that a $\ket{+}$ run in $c=3$ would have substantially higher fidelity than a $\ket{-}$ run in $c = 4$. Yet, if we compared $\ket{+}$ to $\ket{-}$ within the same calibration, they would have roughly the same fidelity.

\begin{figure}
    \centering
    \includegraphics[width=0.48\textwidth]{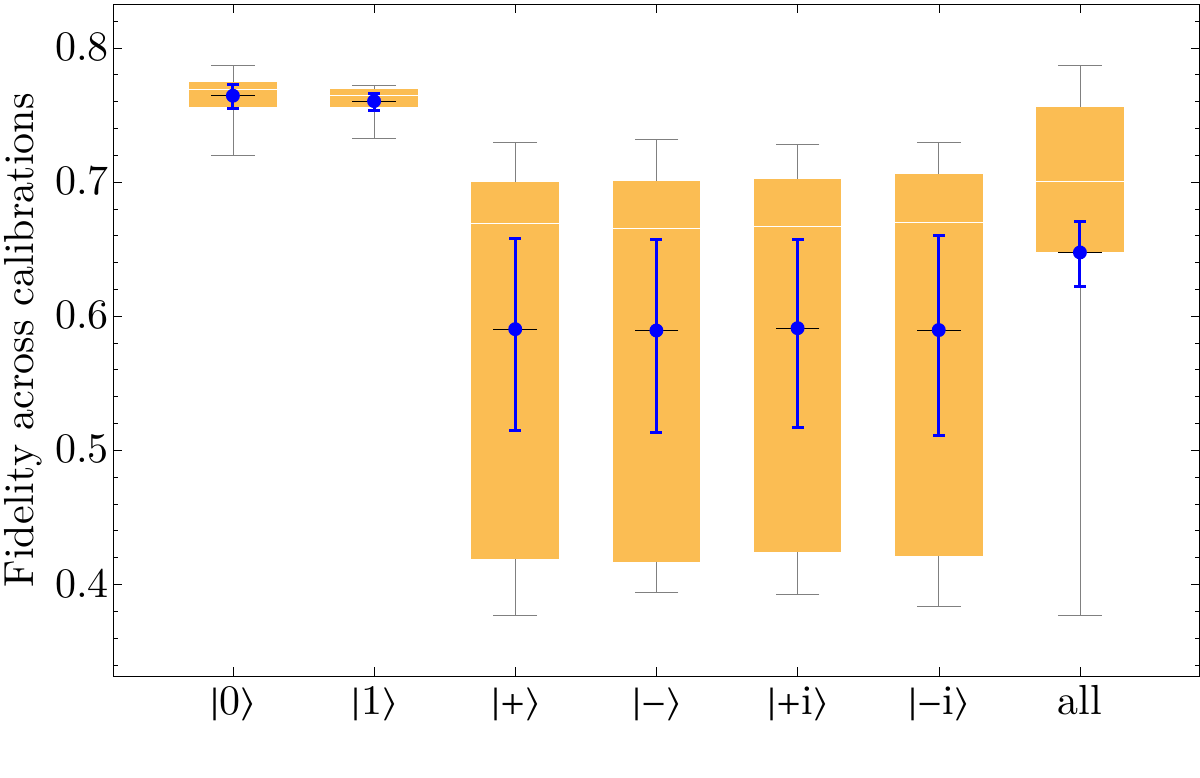}
    \caption{The variation in fidelity across $14$ calibrations for each Pauli state protected under KDD for a fixed time $T_6$. The final box compiles the data for all six Pauli states. The blue point shows the average (with 95\% bootstrapped confidence intervals) across the data in each box plot.}
    \label{fig:granulaityPlot}
\end{figure}

To clarify this averaging subtlety, we present the raw data for each state as a box plot in \cref{fig:granulaityPlot}. The variation arises from the fidelity fluctuations of each state across calibrations. Interestingly, the polar states, $\ket{0}$ and $\ket{1}$, give surprisingly consistent results, while the remaining equatorial states have large variations in performance. This suggests that in some calibrations, the equatorial states were adversely affected, while the polar states were mostly unaffected. By checking the data calibration-by-calibration, we find this to indeed be true. Namely, the equatorial states collectively have a similar performance that is worse in some calibrations and better in others. This could happen, e.g., if, in some calibrations, $T_2$ dropped while $T_1$ stayed roughly the same. Thus, to have a direct comparison of state performance, it is imperative to confine the comparison to within a single job. The same can be said about comparing the fidelities of a fixed state generated at different times. Without confining comparisons to a single job, it is unknown if the difference in fidelities is due to drift or other causes.

On the other hand, it is not possible to test all DD sequences, states, and decay times within a short enough window to avoid significant drift. Even if it were, fair-queuing enforces a maximum number of $75$ circuits per job, so this sets a hard cut-off on what can be compared reliably in a time-proximate way. As a compromise between time-proximate comparisons and the hard $75$ circuits-per-job limit, we collected data as described in \cref{sec:methods}. Namely, within a single job, we collect the fidelity decay curves for the six Pauli states. Each curve consists of $12$ time points. In total, this takes up $72$ circuits, and we pad this with an additional set of two measurement error mitigation circuits. In this sense, we guarantee that the fidelities for different states and times can be compared reliably within this data set, i.e., they are not different due to drift. We then repeat this demonstration over different calibration cycles to obtain an estimate of fidelity differences due to drift. By reporting the median across this data set, we provide an estimate of \emph{typical} performance --- the performance estimate of a hypothetical next demonstration. We emphasize that the typical performance is different from the average performance.

To further support this point, we have included average fidelities (with bootstrapped 95\% confidence intervals) superimposed on the box plots in \cref{fig:granulaityPlot}. For all data sets, the mean and median do not agree. For the polar states ($\ket{0}$ and $\ket{1})$, the difference is slight, and their respective median performance is included in the error bar of the mean. We remark that averages and medians generally agree as they do here for the best performing DD sequences (see \cref{fig:pauliPlot}). The mean is heavily biased downward for the remaining equatorial states, and the confidence intervals do not include the mean. Once averaging over the entire data set, the discrepancy is even worse. Here, the mean falls on top of the 25\% quantile, and the top of the confidence interval differs from the median by a significant amount. Hence, an average method is inappropriate for DD sequences whose performance is highly sensitive to drift effects, and a median method is more appropriate. Again, we remark that this should not affect the top performing sequences shown in \cref{fig:pauliPlot}, but it gives us a consistent way to sift through all the sequences at once to even get to a top ranking in the first place. 

To summarize, we have shown that averaging fidelities over states is a delicate issue that depends on how the data is averaged. When averaging within a fixed job, the results appear reasonable and simple averaging of the fidelities over states gives an average state fidelity that behaves as expected. When comparing state data across different jobs, a straightforward average can be biased due to a particularly bad calibration, resulting in an average sensitive to drift. A non-parametric comparison in which fidelities are assessed via a box plot, on the other hand, does not suffer from this issue. When confined to a fixed job, the median over states gives a measure of typical performance over the set of states. When fidelities are collected over both states and calibrations, the median is still a measure of typical performance, but now over both states and calibrations. Hence, a measure of typical performance is more robust than an average performance metric when considering drift. In practice, typical performance is more relevant for any given fixed demonstration. These facts, alongside the practical difficulties of averaging and fitting, are why we opted for the fit-then-average (FtA) approach.

\subsection{Summary of fitting fidelity}
The standard approach to comparing DD sequences is to generate average fidelity decay curves, fit them, and report a summary statistic like a decay constant. For this reason, we call this an ``average-then-fit'' (AtF) approach. In Ref.~\cite{Pokharel:2018aa}, for example, the authors generated fidelity decay curves for roughly 40 equally spaced times and $36$ states ($30$ Haar-random and six Pauli eigenstates). For each time, they computed a state-average fidelity constituting a single average fidelity decay curve. They then fit this curve to a modified exponential decay [\cref{app-eq:bibek-fit}] and reported the decay constant for XY4 compared to Free evolution. This was sufficient to show that, on average, applying XY4 was better than free evolution.

In this work, we aim to go beyond this approach and identify any pitfalls. To do so, we set out to first compare a large collection of DD sequences and not just \XY\ and Free. Second, we are interested in whether \XY\ retains universality for superconducting devices (see \cref{subsec:sc-physics}). This question requires us to know precisely whether, e.g., $\ket{+}$ or $\ket{1}$ is better protected under \XY. This led us to two methodological observations. (i) Using a standard fit such as \cref{app-eq:bibek-fit} to individual state decay curves results in several problems: it does not always work in practice, leading to an ambiguous interpretation of the decay constant, and it requires a complicated fitting procedure. (ii) The fidelity of a given state protected with DD for a fixed time is unstable from job to job due to device drift. This means one must be careful when comparing fidelities not collected within the same job. 

We resolved both issues by (i) focusing on typical performance instead of average and (ii) using a ``fit-than-average'' (FtA) approach where we interpolate with time-averaging (ITA). Along the way, the insistence on not first averaging results in the identification of significant and unexpected asymmetries in performance between states (see \cref{app:CvsP} and especially \cref{fig:pulseVsCirc}), which led to better DD sequence design. Ultimately, more than the issues with fitting leading to ambiguous/inconsistent results described in \cref{subsec:ambiguous-lambda-problem}, it is the core idea of comparing data within a given job that led to our eventual choice of analysis method.

\subsection{Details of fitting using Mathematica}
 \label{app:fit-minutia}

\subsubsection{Standard fit to exponential details}
At a high level, we fit the fidelity decay curves (see \cref{fig:sample-fid-decay-curves} for example curves and \cref{fig:sample-fitting-approaches} for example curves with fits) to \cref{app-eq:bibek-fit} using Mathematica's \verb^NonlinearModelFit^ (NLM)~\cite{Mathematica, reference.wolfram_2022_nonlinearmodelfit} which performs weighted least-squares regression. In more detail, NLM finds the parameters $\{ \hat{\lambda}, \hat{\gamma}, \hat{\alpha} \}$ which minimize the weighted sum of squares, 
    \begin{equation}
        \label{eq:sum-of-squares}
        S = \sum_{i=1}^N \frac{1}{\sigma_{T_i}^2} \left( \expval{f_e}_{T_i} - f_P(T_i;  \lambda, \gamma, \alpha)\right)
    \end{equation}
with the assumption that
    \begin{equation}
        \label{eq:emp-fid-to-model}
        \expval{f_e}_{T_i} = f_P(T_i;  \lambda^*, \gamma^*, \alpha^*) + \epsilon_{T_i} \ \ \ \ \  \epsilon_{T_i} \sim \mathcal{N}(0, \sigma_{T_i})
    \end{equation}
for some unknown parameters $ \lambda^*,  \gamma^*, $ and $\alpha^*$. Given a good model and fitting procedure, the NLM procedure finds $\hat{\lambda} \approx \lambda^*,  \hat{\gamma}  \approx  \gamma^*$, and $\hat{\alpha} \approx \alpha^*$~\cite{willaims-quantify-measurement}.  Furthermore,  NLM computes the covariance matrix, $C$, between the estimated parameters, 
\begin{subequations}
    \begin{align}
        \label{eq:covar-mat}
        C &= X^T W X \\
        X &= 
            \begin{bmatrix}
            \partial_\lambda F(T_1) & \partial_\gamma F(T_1) & \partial_\alpha F(T_1) \\
            \vdots & \vdots & \vdots \\
            \partial_\lambda F(T_N) & \partial_\gamma F(T_N)  & \partial_\alpha F(T_N)
            \end{bmatrix} \\
        \label{eq:weights}
        W &= \text{diag}(1 / \sigma_{T_1}^2, \ldots, 1 / \sigma_{T_N}^2),
    \end{align}
\end{subequations}
where $F(T)$ is the integrated fidelity [\cref{app-eq:norm-average-fidelity}] and this particular form of $C$ and $W$ results from setting the \verb&Weights& and \verb&VarianceEstimatorFunction& NLM settings appropriately for experimental data.\footnote{The choice of weights is exactly as stated in \cref{eq:weights}, but the variance estimator setting is more subtle. We set the variance estimator to the constant 1, which prevents the covariance matrix from being rescaled by a variance estimate of $\sigma_{T_i}$ in \cref{eq:emp-fid-to-model}. We do this since we do not need to estimate $\sigma_{T_i}$--we already did this by performing our demonstration, and the weight matrix $W$ reflects this knowledge. These settings are discussed in Mathematica's \href{https://reference.wolfram.com/language/howto/FitModelsWithMeasurementErrors.html}{Fit Models with Measurement Errors} tutorial.} As usual, we compute the parameter standard errors (aka the standard deviations of the statistics) from the covariance matrix as,
    \begin{equation}
         SE_\lambda = \sqrt{C_{11}}, \  SE_\gamma = \sqrt{C_{22}},\ SE_\alpha = \sqrt{C_{33}}.
    \end{equation}
From this, we can estimate the parameter 95\% confidence intervals by calculating a t-score. To do this, we first find the number of degrees of freedom we have, $\nu = N - \# \  \text{parameters}$. The t-score is then the $t_{0.95}^*$ such that
    \begin{equation}
        P(-t_{0.95}^* \leq T \leq t_{0.95}^*) = 0.95 \ \ \ T \sim \text{StudentTDist}(\nu).
    \end{equation}
In the end, we report our best-fit parameters with 95\% confidence using
    \begin{equation}
        \label{eq:final-fit-lam}
        \lambda = \hat{\lambda} \pm SE_{\lambda} \times t_{0.95}^* \equiv \hat{\lambda} \pm \delta_{\hat{\lambda}}
    \end{equation}
with an exactly analogous formula for $\gamma$ and $\alpha$ (i.e., same $t^*_{0.95}$ throughout).\footnote{We spell out the details for completeness, but the estimate for $\delta_{\hat{\lambda}}$ can be immediately returned from NLM using the \enquote{ParameterConfidenceIntervals} option without needing to compute the covariance matrix directly at all.}

\subsubsection{Consistency checks and selecting appropriate fitting options}
The previous section outlined the general methodology and output of Mathematica's \verb^NonlinearModelFit^ function. One detail we glossed over is the method by which \verb^NonlinearModelFit^ minimizes the weighted sum of squares, $S$. By default, $S$ is minimized using the \verb^LevenbergMarquardt^ method --- a particular implementation of the Gauss–Newton algorithm. Many of the details and optimization settings are outlined in Mathematica's \href{https://reference.wolfram.com/language/tutorial/UnconstrainedOptimizationMethodsForLocalMinimization.html#216554821}{\enquote{Unconstrained Optimization: Methods for Local Minimization}} documentation. An important feature of this method is that it is a local search method, and hence, the quality of the fit depends on the particular input settings and randomness in each run. Coupled with the fact that $\lambda$, $\gamma$, and $\alpha$ of \cref{app-eq:bibek-fit} are not independent parameters, it is important in practice to try many different settings to obtain a good fit. A full enumeration of all settings we attempted can be found in the \emph{fittingFreeData.nb} Mathematica notebook contained in our code repository~\cite{nicholas_ezzell_2023_7884641}. For example, we consider fits for which we reduce the number of parameters (e.g., setting $e^{-t / \alpha} = 1$) or where $\gamma = 0$ is either seeded or unseeded.

Upon trying many possible fits, we then need an objective way to compare their relative quality. To do so from a fit obtained using \verb^NonlinearModelFit^, one can query its associated (corrected) Akaike information criterion (AICc)~\cite{aic-wiki} via Mathematica's \verb^Fit["AICc"]^. The AIC estimates the relative quality of each model in a collection of models for a given data set. In our case, the sample size is relatively small: $12$ points for each fit. To remedy this, we employ the AICc, which corrects the tendency of AIC to favor overfitting for small sample sizes. See Ref.~\cite{aic-wiki} for a summary of the formulas for AIC and AICc. A more detailed introduction and derivations of the formulas can be found in Ref.~\cite[pp.~51-74]{konishi2008information}. 

The lower the AICc, the better the model is relative to others. However, we must also ensure that the final fit satisfies consistency conditions. Namely, a fit is considered unreasonable if it (a) predicts a fidelity less than zero or greater than one, (b) predicts a parameter whose error is larger than the value itself, or (c) violates the Nyquist-Shannon sampling theorem~\cite{nyquistshannon-wiki}, i.e., has frequency larger than the sample rate of the data itself. I.e., $\gamma$ in \cref{app-eq:bibek-fit} must satisfy $0 \leq \gamma \leq \frac{2 \pi}{2 \Delta t} \equiv B$ where $\Delta t = (150 / 12) \mu s$ is the spacing between data points in our fidelity decay curves. 
After fitting, we reject fits which violate properties (a) and (b) and correct those that violate (c). We next illustrate what this post-selection or correction looks like for our data.

\begin{figure*}
    \centering
    \includegraphics[width=0.32\textwidth]{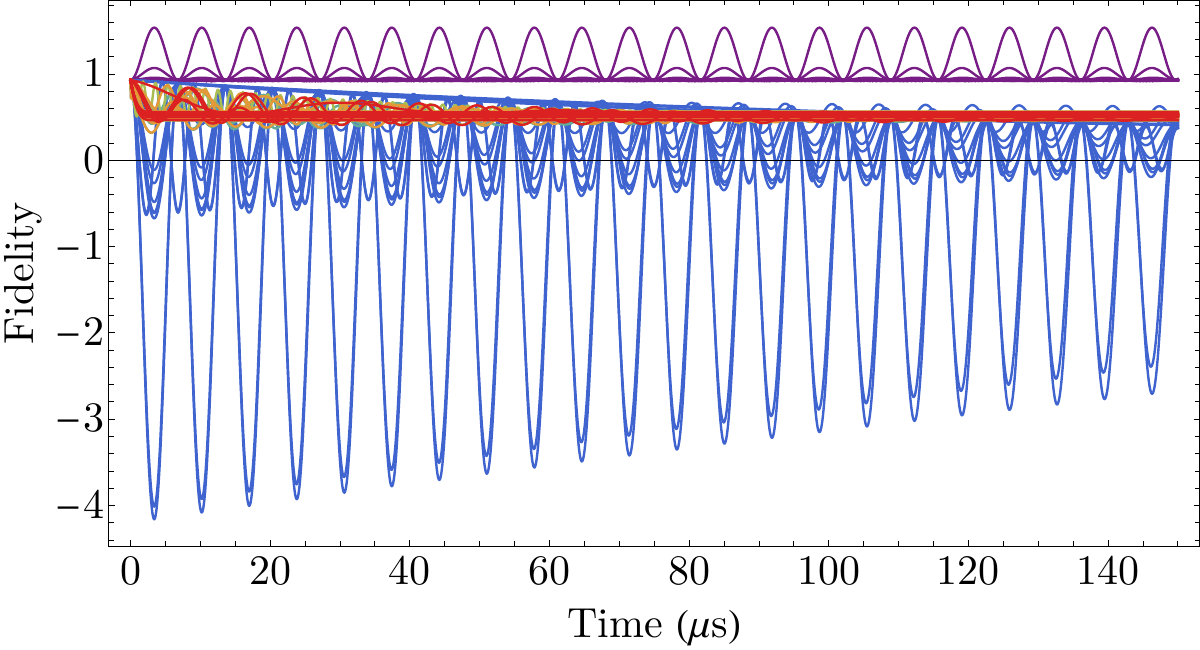}
    \hfill
    \includegraphics[width=0.32\textwidth]{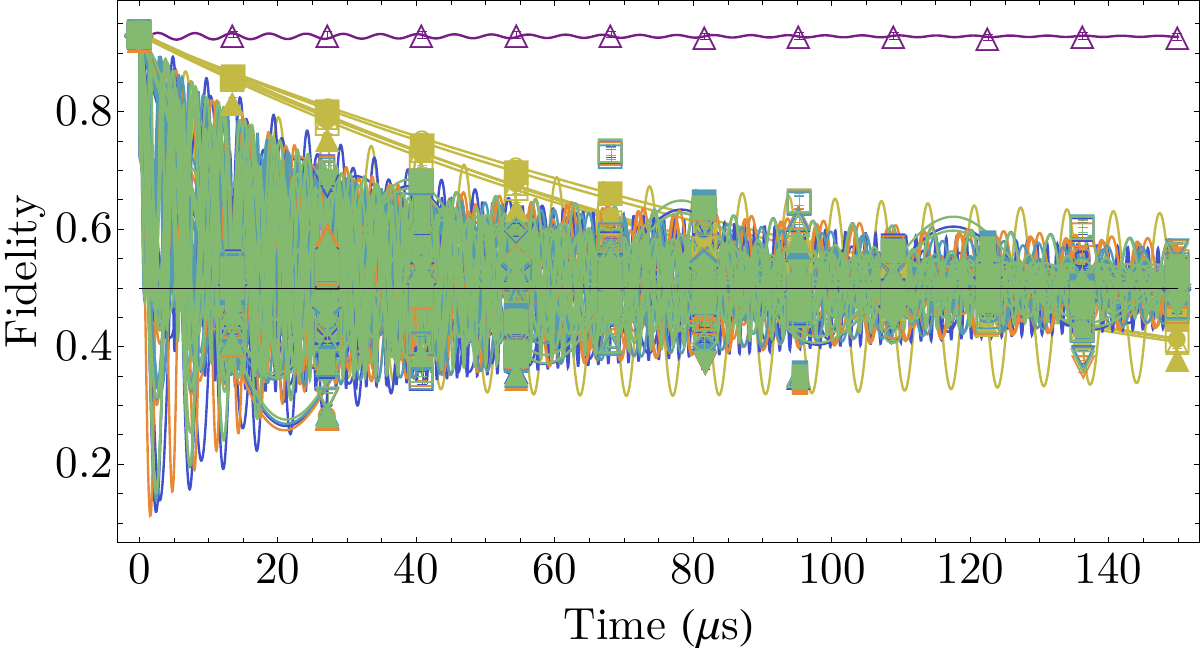}
    \hfill
    \includegraphics[width=0.32\textwidth]{newBibekFitPlot.pdf}
    \caption{A summary of the post-selection process necessary when fitting to \cref{app-eq:bibek-fit} with Mathematica's NonlinearModelFit. The data is the same Free evolution data displayed in \cref{fig:sample-fitting-approaches}. Left: for each curve, we select the fit with the lowest AICc without regard to whether the fit meets consistency conditions such as having positive fidelity. Middle: the result of first post-selecting those fits which satisfy consistency conditions and then choosing that fit with the lowest AICc. Some data sets (especially the flat $\ket{0}$ state decay) have no reasonable fits, so they are dropped. Right: finally, we rescale the frequency in accordance with the Nyquist-Shannon sampling theorem (this does not affect the predicted decay constant $\lambda$). This final plot is displayed as the result in \cref{fig:sample-fitting-approaches} and the curves for which the $\lambda$'s are reported in \cref{fig:sample-fitting-boxplot-results}. Namely, data that do not have a reasonable fit are not included in the final summary, which biases towards data that does fit \cref{app-eq:bibek-fit}.}
    \label{fig:postSelectingFits}
\end{figure*}

First, we perform many fits without regard to whether (a)-(c) are satisfied since enforcing constraints makes the nonlinear fit results unstable, and moreover, the constraints are not always enforced even when specified inside \verb^NonlinearModelFit^. We then post-select the fits that respect (a) and (b), and among these, we select the fit with the lowest AICc. If no fit respecting (a) and (b) is found, we drop that data set. Finally, we rescale the frequency via $\gamma \rightarrow \text{mod}(\gamma, B).$ This procedure --- along with the fits along the way --- is summarized in \cref{fig:postSelectingFits}. Note that we are forced to post-select fits that are well modeled by \cref{app-eq:bibek-fit} in order to report a reasonable value of $\lambda$ for the given data. For this reason, the fitting procedure thus described is biased toward data that is appropriately modeled by \cref{app-eq:bibek-fit}. 

\subsubsection{Time-averaged fidelity approach}
We construct a polynomial interpolation of the fidelity decay data using Mathematica's \verb^Interpolation^ function~\cite{Mathematica, reference.wolfram_2022_interpolation}. Then we integrate over the interpolation to compute a time-averaged fidelity according to \cref{app-eq:norm-average-fidelity}.  By default, the Interpolation uses a third-order Hermite method~\cite{cubic-hermite-spline-wiki} which we found to be sufficient to model our data adequately.  For example, the curves in \cref{fig:sample-fid-decay-curves} consist of Hermite interpolations of the triangular raw data, which appears reasonable.  To compute \cref{app-eq:norm-average-fidelity} from the interpolation, we used Mathematica's \verb^NIntegrate^. 

\begin{figure}
	\centering
	\includegraphics[width=0.45\textwidth]{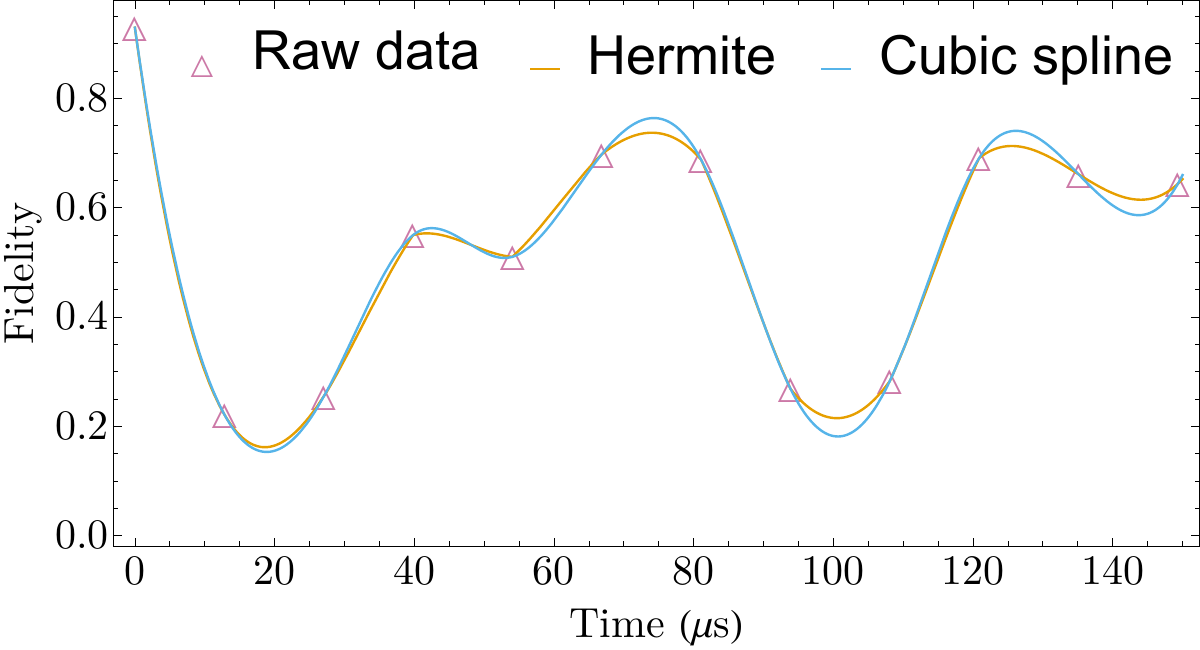}
	\includegraphics[width=0.45\textwidth]{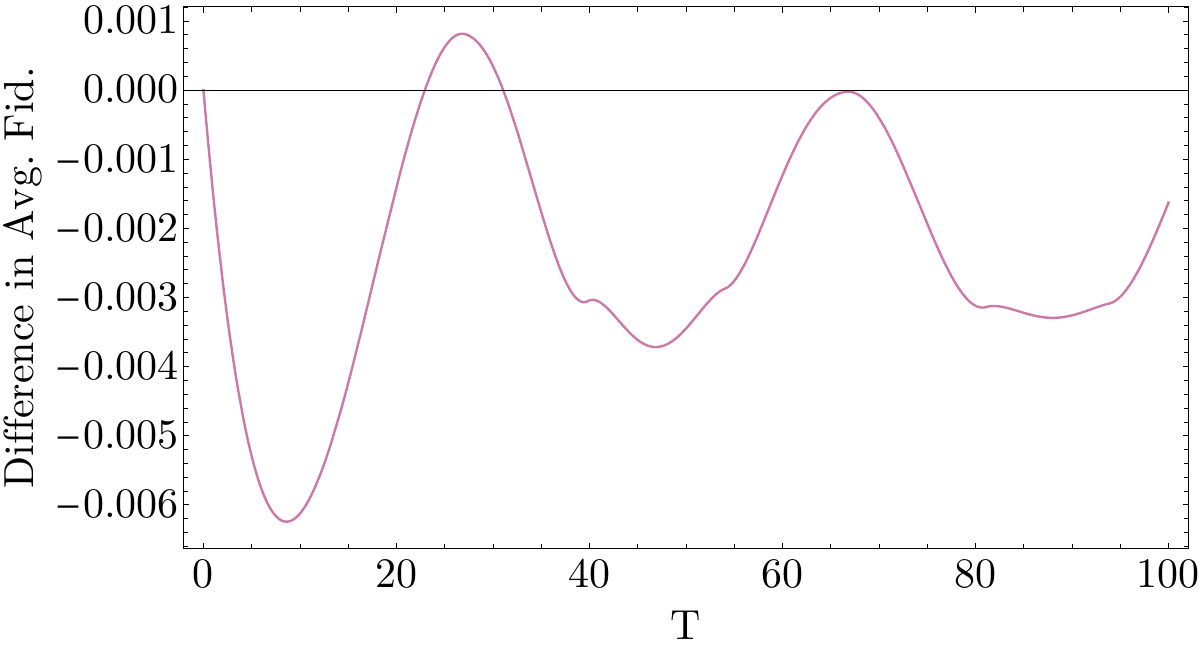}
	\caption{We compare third-order Hermite and cubic spline interpolations on the fidelity decay curve generated by KDD with the initial state $\ket{-}$.  The resulting interpolations are qualitatively similar, and the resulting average fidelities are within $6 \times 10^{-3}$ of each other for all $0 \leq T \leq 100 \mu s$.  Hence,  they are equally valid interpolations when using time-averaged fidelity as a performance metric.}
	\label{fig:hermite-vs-cubicspline}
\end{figure}

For a sense of stability,  we also tried the popular cubic spline interpolation method~\cite{cubic-spline-wiki}, but this does not affect our results outside of the interpolation error bound itself.  For example,  consider the complicated interpolation necessary for KDD using the Hermite or cubic spline methods in \cref{fig:hermite-vs-cubicspline}.  Hence,  either interpolation is equally valid when using the time-averaged fidelity metric. 

We remark that polynomial interpolation is much easier to do than fitting to a model such as \cref{app-eq:bibek-fit} since we need not (i) select reasonable fitting equations, (ii) fit them using nonlinear least squares regression, (iii) post-select fits with satisfy consistency conditions and low AICc. Additionally, the predictions are less susceptible to ambiguities as described in \cref{subsec:ambiguous-lambda-problem}. Finally, we are not forced to discard any data sets which fail the post-selection criteria, so we do not bias the final summary results in any particular way.

\section{Toggling frame}
\label{app:toggling-frame}

We assume that without any external control, the system and bath evolve under the time-independent noise Hamiltonian $H$. A DD pulse sequence is realized via a time-dependent control Hamiltonian $H_c(t)$ acting only on the system so that the system and bath evolve according to $H+H_c(t)$ (refer to \cref{sec:DD-perf} for definitions of the various Hamiltonian terms). 

For understanding the effects of the control Hamiltonian, it is convenient to use the interaction picture defined by $H_c(t)$, also known as the \emph{toggling frame} \cite{NMRBook,Viola:99,Viola:1999:4888,Viola:03,Viola:2005:060502}. The toggling-frame density operator $\tilde\rho_{SB}(t)$ is related to the Schr{\"o}dinger-picture density operator $\rho_{SB}(t)$ by
\begin{align}
\rho_{SB}(t)&= U(t,0)\rho_{SB}(0)U^\dagger (t,0)\nonumber\\
&\equiv U_c(t)\tilde\rho_{SB}(t)U_c^\dagger(t),
\end{align}
where $U(t,0)$ is the evolution operator generated by the full Hamiltonian $H+H_c(t)$. Therefore the toggling-frame state evolves according to
\begin{equation}
\tilde\rho_{SB}(t)= \tilde U(t,0) \tilde \rho_{SB}(0)\tilde U^\dagger(t,0),
\end{equation}
where the toggling-frame time evolution operator
\begin{equation}
\tilde U(t,0) \equiv U_c^\dagger(t) U(t,0)
\end{equation}
is generated by the \emph{toggling-frame Hamiltonian}
\begin{equation}
\tilde H(t)\equiv U_c^\dagger(t)HU_c(t).
\end{equation}
Since $U_c(t)$ acts nontrivially only on the system, $\tilde H(t)$ can be written as
\begin{equation}
\tilde H(t)=H_B+\tilde H_\err(t),
\label{eq:tildeH}
\end{equation}
where $\tilde H_\err(t)\equiv U_c^\dagger (t)H_\err U_c(t)$ is the toggling-frame version of $H_\err$.  Because the operator norm is unitarily-invariant, we have $\Vert\tilde H(t)\Vert=\Vert H\Vert\leq \epsilon$ and $\Vert\tilde H_\err(t)\Vert=\Vert H_\err\Vert\leq J$.

Throughout, we consider \emph{cyclic DD}, where
$U_c(t)$ returns to the identity (up to a possible irrelevant overall
phase) at the end of
a cycle taking time $t_{\rm DD}$:
\begin{equation}
U_c(\tDD)=U_c(0)=I.
\end{equation}
Therefore, at the end of
the cycle, the toggling-frame and Schr{\"o}dinger-picture states coincide.

\section{Results of all Pauli demonstrations}
\label{app:all-pauli-data}

\begin{figure*}[ht]
    \centering
    \includegraphics[width=0.75\textwidth]{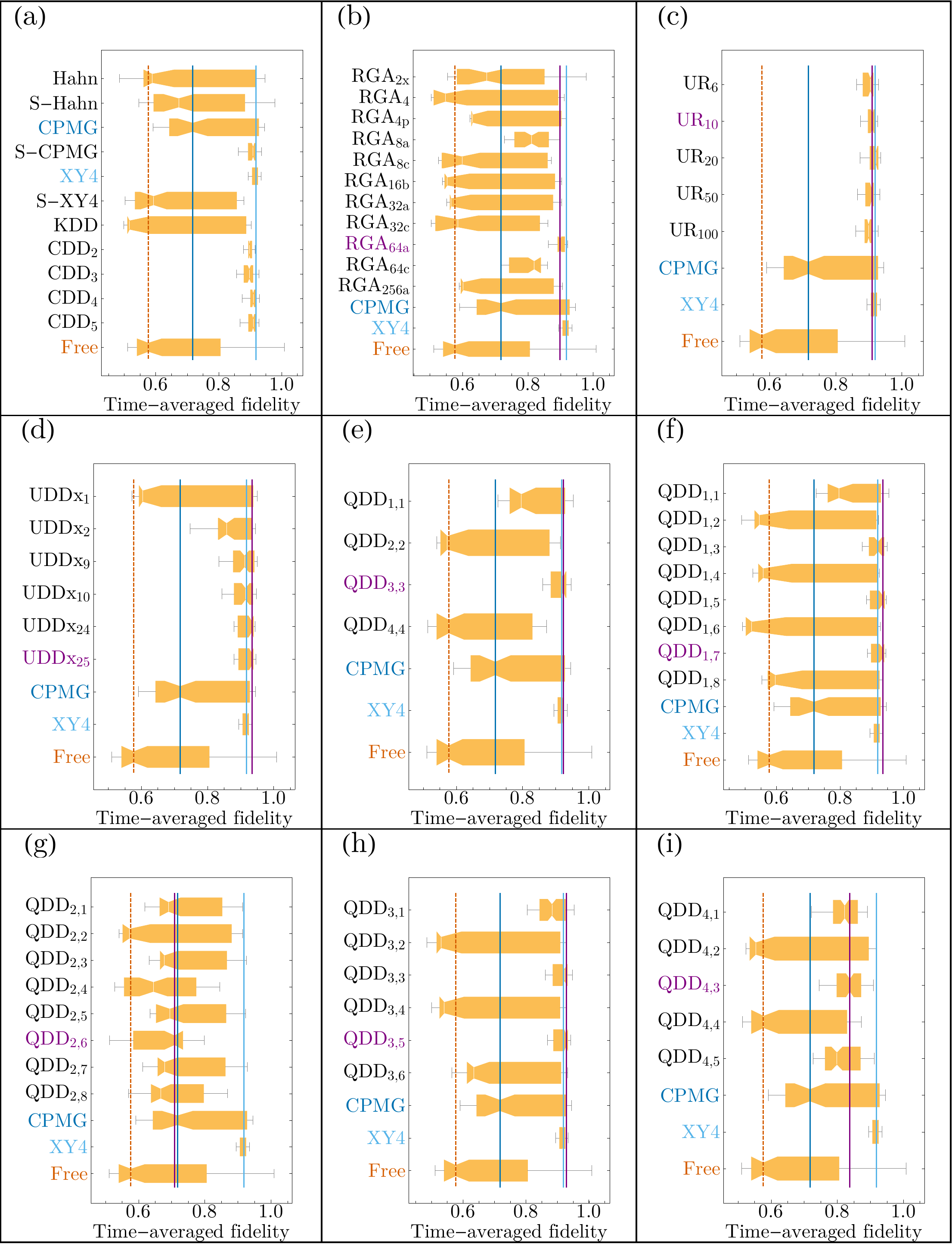}
    \caption{Collection of all Pauli-demonstration results for \armonk. The best performing sequence for each family (solid purple line) substantially outperforms Free (orange) and \CPMG\ (blue) but only marginally differs from the performance of \XY\ (cyan).   
    \newline
    \textbf{(a)} Results are not consistent with ideal-pulse theory but are sensible when considering realistic competing factors. First, S-Hahn $>$ Hahn and S-CPMG $>$ \CPMG\ are consistent with the large pulse widths on \armonk, for which $\Delta \approx 142$ ns. That \XY\ works very well is also consistent with its expected approximate universality given that $T_1 \approx \frac12 T_2$ on \armonk. Given that \XY\ already performs well, it is not surprising that its robust versions, S-XY4 and KDD, do worse. In particular, they have little room for improvement but also add extra free evolution periods that accumulate additional error. Finally, $\CDD{n}$ is roughly flat for all $n$ tested, which is consistent with an expected saturation that happens to occur at $n_{\text{opt}} = 1$.
    \newline
    \textbf{(b)} The performance of RGA does not match theoretical expectations. To see this, note that $\RGA{4}$ is itself a four-pulse sequence with error scaling $\mathcal{O}(\tau^2)$ which is the same as \XY. Hence, it should perform comparably to \XY, but it does significantly worse. Furthermore, it is also unexpected that the best RGA sequences are 8a, 64a, and 64c. Indeed, 8a and 64a are expected to work best in a flip-angle error dominated regime, but in this regime, 64c has a scaling of $\mathcal{O}(\epsilon_r^2)$, so it is not expected to do well. 
    \newline
    \textbf{(c)} The $\UR{n}$ sequence performance is consistent with theory. First note that $\UR{4} = \XY$. Thus, we expect $\UR{n}$ for $n > 4$ to improve upon or saturate at the performance of \XY, and the latter is what happens.  
    \newline
    \textbf{(d)} The $\UDD{n}$ sequence performance is consistent with theory. In particular, we expect (and observe) a consistent increase in performance with increasing $n$ until performance saturates.  
    \newline
    \textbf{(e-i)} The $\QDD{n}{m}$ results in part match theoretical expectations, since they exhibit a strong even-odd effect, as predicted in Refs.~\cite{QL:11,KL:11,Kuo:2012rf}. 
Nevertheless, the optimal choice of $n$ and $m$ has to be fine-tuned for \armonk. We note that five out of ten of the top ten sequences on \armonk\ are from the QDD family.
  }
    \label{fig:armonk-family-plot}
\end{figure*}

\begin{figure*}[ht]
    \centering
    \includegraphics[width=0.75\textwidth]{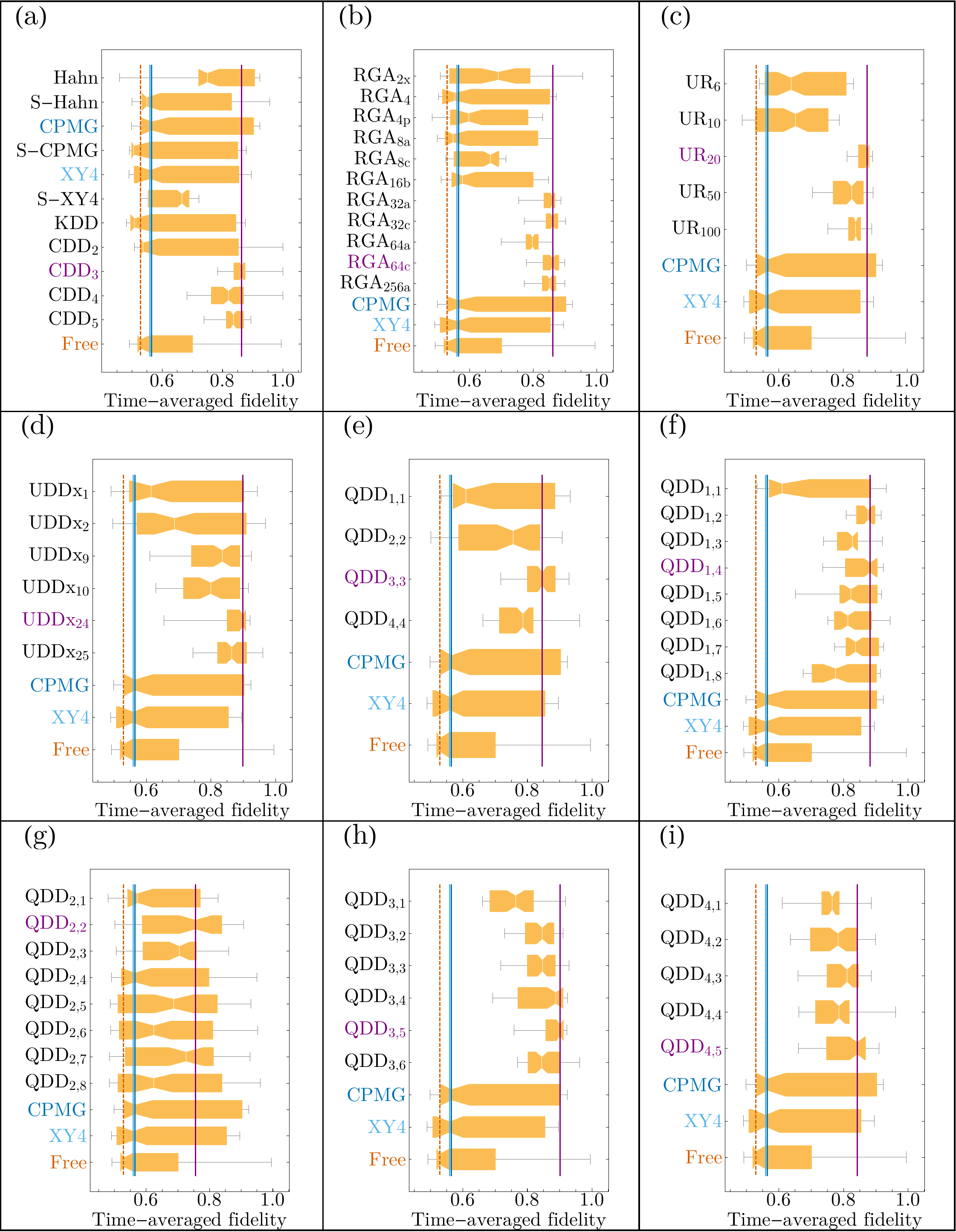}
    \caption{Collection of all Pauli-demonstration results for \bogota. The best-performing sequence for each family (solid purple line) substantially outperforms Free (orange), \CPMG\ (blue), and \XY\ (cyan).
    \newline
    \textbf{(a)} Results are not fully consistent with ideal-pulse theory. For example, it is unexpected that Hahn $>$ \CPMG\ and that Hahn $<$ S-Hahn while at the same time S-XY4 $>$ \XY. Nevertheless, some trends are expected such as $\CDD{1} < \CDD{2} < \CDD{3}$ and then saturating.  
    \newline
    \textbf{(b)} Results are somewhat consistent with ideal-pulse theory. First of all, $\RGA{4} \approx \XY$ in performance, which is sensible. The first large improvement comes from $\RGA{8c}$ and then  from numbers $32$ and greater. This trend is similar to $\CDD{n}$ increasing, which is expected since larger RGA sequences are also recursively defined (e.g., $\RGA{64c}$ is a recursive embedding of $\RGA{8c}$ into itself). However, it is again unexpected that both `a' and `c' sequences should work well at the same time. For example, $\RGA{8c} > \RGA{8a}$ theoretically means that \bogota\ has negligible flip angle error. In such a regime, the decoupling order of $\RGA{8a}$ is the same as $\RGA{64a}$ since they are designed to cancel flip-angle errors. But, we find that $\RGA{64a} > \RGA{8a}$ in practice.
    \newline
    \textbf{(c)} The $\UR{n}$ results are mostly consistent with theory. First, $\UR{6}$ is an improvement over \XY, and though $\UR{n}$ does increase with larger $n$, it is not simply monotonic as one would expect in theory. Instead, we find a more general trend with an empirical optimum at $n = 20$. 
    \newline
    \textbf{(d)} The $\UDDx{n}$ results are mostly consistent with expectations. Again, performance mostly increases with increasing $n$, but the increase is not fully monotonic. 
    \newline
    \textbf{(e-i)} The $\QDD{n}{m}$ results are fairly consistent with theory. In (e), performance of $\QDD{n}{n}$ increases with $n$ until $n = 3$. The degradation for $n = 4$ is consistent with expectations in the bandwidth-limited setting~\cite{Xia:2011uq}. In (f - i) the results are again fairly expected: aside from parity effects (odd/even $m$), for $\QDD{n}{m}$, we expect monotonic improvement with increasing $m$ until $n = m$, after which performance should saturate or even slightly improve;
this is the general empirical trend.
    }
    \label{fig:bogota-family-plot}
\end{figure*}

\begin{figure*}[ht]
    \centering
    \includegraphics[width=0.75\textwidth]{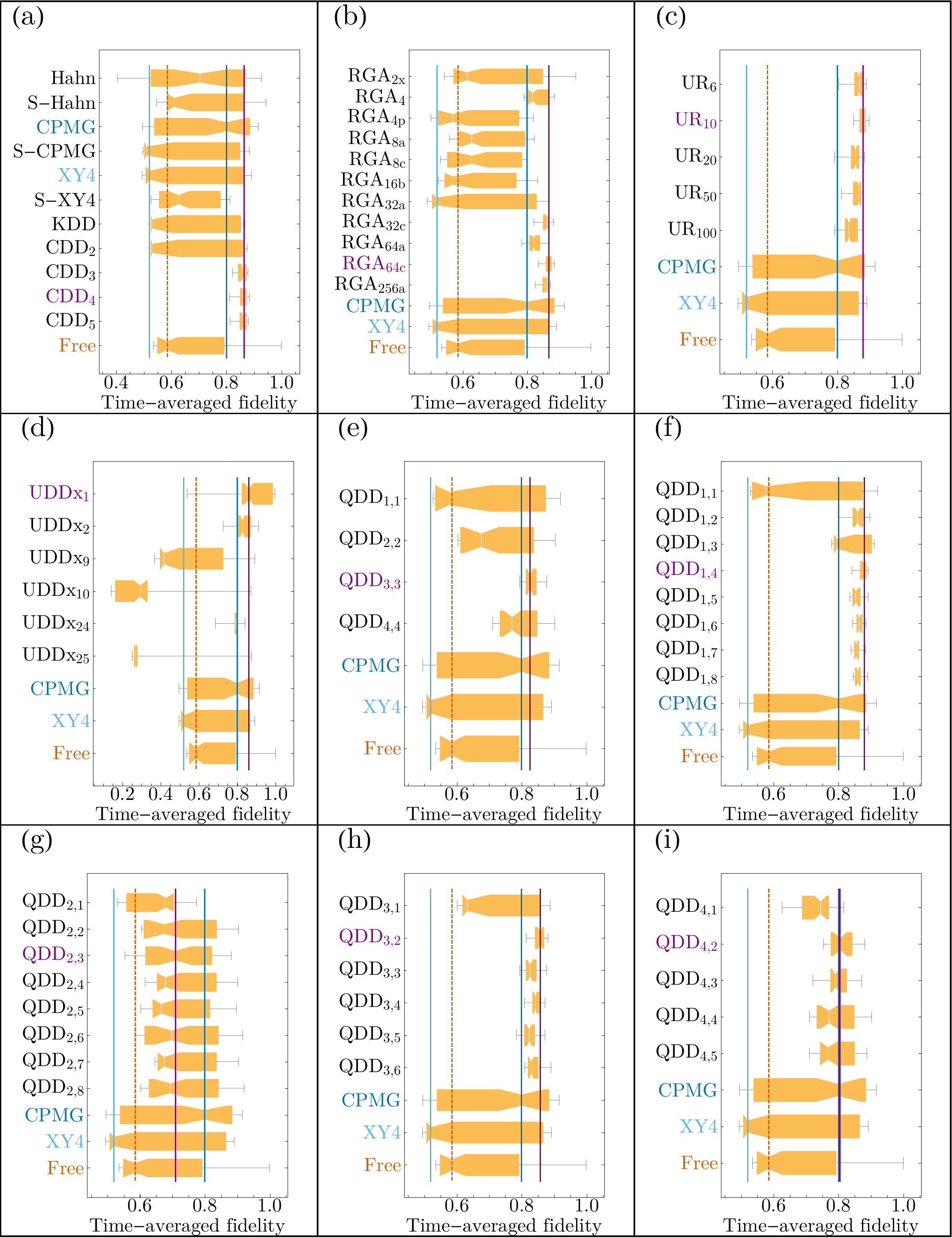}
    \caption{Collection of all Pauli-demonstration results for \jakarta. The best performing sequence for each family (solid purple line) substantially outperforms Free (orange) and \CPMG\ (blue) but only marginally differs from the performance of \XY\ (cyan). 
    \newline
    \textbf{(a)} Results are not fully consistent with ideal-pulse theory. For example, it is unexpected that Hahn $>$ \CPMG\ and that Hahn $>$ S-Hahn while at the same time S-XY4 $>$ \XY. Nevertheless, some trends are expected, such as $\CDD{1} < \CDD{2} < \CDD{3} > \CDD{4}$ and then saturating.
    \newline
    \textbf{(b)} Results are somewhat consistent with the theory. First of all, $\RGA{4} > \XY$ is sensible and implies that we are in a flip-angle error dominated regime. However, we would then  expect $\RGA{4} > \RGA{8a}$, which does not occur. Nevertheless, the recursively defined sequences (number $32$ and above) generally outperform their shorter counterparts, which is similar to $\CDD{n}$ as expected.
    \newline
    \textbf{(c)} The $\UR{n}$ results are consistent with theory. First, $\UR{6}$ is a large improvement over \XY, and from there $\UR{10} > \UR{6}$. After this, the improvement plateaus. 
    \newline
    \textbf{(d)} The performance of $\UDDx{n}$ greatly differs from the theoretical expectation of monotonic improvement with $n$. In fact, the behavior is so erratic that we suspect device calibration errors dominated the demonstration. Nevertheless, the performance of $\UDDx{1}$ was excellent in this case. 
    \newline
    \textbf{(e-i)} The $\QDD{n}{m}$ results are fairly consistent with theory, and quite similar to \cref{fig:bogota-family-plot}. The same comments as made there apply here.
    }
    \label{fig:jakarta-family-plot}
\end{figure*}

In \cref{fig:pauliPlot}, we summarized the results of the Pauli demonstration for the top $10$ sequences on each device and the baseline of CPMG, \XY, and free evolution. This appendix presents the data for all $60$ tested sequences. In particular, we split the data for each device into a $3\times 3$ grid of plots shown in Figs.~\ref{fig:armonk-family-plot}-\ref{fig:jakarta-family-plot}, separating by family as in \cref{tab:dd-seq-sum} when possible.  For convenience, \CPMG, \XY, and Free are still included in all plots along with colored reference lines matching the convention in \cref{fig:pauliPlot}. The purple reference line denotes the best sequence in the given family plot excluding the baseline sequences placed at the end. For example, in \cref{fig:armonk-family-plot}(b) $\RGA{64a}$ is the best RGA sequence even though it does not outperform \XY; it is marked with a purple reference line.

We plot the same families for each device in a $3\times 3$ grid, labeled (a)-(i). In each caption, we make a few general comments on DD performance overall and then comment on observations specific to each sequence family. 
Recall that a specific definition of each sequence is given in \cref{app:DD-summary}, and a summary of their properties along with references in \cref{tab:dd-seq-sum}. 
To avoid excessive repetition in each caption, we first provide a brief description of each of the cases (a)-(i) shown in \cref{fig:armonk-family-plot}-\ref{fig:jakarta-family-plot}: 
\begin{enumerate}
    \item[\textbf{(a)}] This \enquote{family} serves as a catch-all for the basic sequences Hahn~\cite{hahn_spin_1950}, CPMG~\cite{Carr:54, Meiboom:58}, and XY4~\cite{maudsley_modified_1986}, along with sequences born from their modifications. Namely, we also plot the Eulerian super-cycle versions~\cite{smith_degree_nodate, viola-knill-robustDD} (denoted by S), KDD~\cite{souza_robust_2012, souza_robust_2011} which is a composite pulse \XY, and $\CDD{n}$~\cite{Khodjasteh:2005xu, Khodjasteh:2007zr} which is a recursive embedding of \XY.
    \item[\textbf{(b)}] The robust genetic algorithm (RGA) family~\cite{Quiroz:2013fv}.
    \item[\textbf{(c)}] The universally robust (UR) family~\cite{Genov:2017aa}.
    \item[\textbf{(d)}] The Uhhrig dynamical decoupling (UDD) family using $X$ pulses, UDDx~\cite{Uhrig:2007qf}.
    \item[\textbf{(e--i)}] The quadratic dynamical decoupling (QDD) family with same inner and outer order, $\QDD{n}{n}$, with fixed outer order 1, $\QDD{1}{m}$ with fixed outer order 2, $\QDD{2}{m}$ with fixed outer order 3, $\QDD{3}{m}$, and with fixed outer order 4, $\QDD{4}{m}$, respectively~\cite{west_near-optimal_2010}.
\end{enumerate}

For each family, we expect the empirical hierarchy of sequence performance to be a complicated function of device-specific properties. Specifically, actual performance is a competition between (i) error cancellation order, (ii) number of free evolution periods, and (iii) systematic pulse errors due to finite width and miscalibrations, among other factors discussed in \cref{sec:DD-background}. For each family, we summarize our expectations regarding these factors
\begin{enumerate}
    \item[\textbf{(a)}] For ideal pulses, we expect $\CDD{n+1} > \CDD{n} \geq \text{KDD} =  \text{S-}\XY = \XY >  \text{S-CPMG} =  \text{CPMG} =  \text{S-Hahn} >  \text{Hahn}$. With a finite bandwidth constraint, we expect $\CDD{n+1} > \CDD{n}$ to only hold up until some optimal concatenation level $n_{\text{opt}}$ after which performance saturates. Using finite width pulses with systematic errors, we expect $\text{S-Hahn} >  \text{Hahn}$ (and similarly for other S sequences) and $ \text{KDD} > \XY$ provided the additional robustness is helpful. I.e., if the pulses are extremely well calibrated and errors are dominated by latent bath-induced errors, then we should instead see $ \text{Hahn} >  \text{S-Hahn}$.
    \item[\textbf{(b)}] The expected performance hierarchy for RGA is rather complicated, as indicated by the labeling, and is best summarized in depth using Table~II of Ref.~\cite{Quiroz:2013fv}. A quick summary is that if we have strong pulses dominated by miscalibration errors ($\epsilon \Delta \ll \epsilon_r$), then we expect $\RGA{8a}$ and $\RGA{64a}$ to do well. In the opposite limit, we expect $\RGA{4}, \RGA{8c}, \RGA{16b}, \RGA{64c}$ to do well. The increasing number indicates the number of pulses, and as this increases, the decoupling order $\mathcal{O}(\tau^n)$ increases, and the same competition between order-cancellation and free evolution periods as in $\CDD{n}$ also applies. 
    \item[\textbf{(c)}] The $\UR{n}$ sequence provides $\mathcal{O}(\epsilon_r^{n/2})$ suppression of flip angle errors at the expense of using $n$ free evolution periods. The relationship of $n$ to $\mathcal{O}(\tau)$ decoupling is not well established in Ref.~\cite{Genov:2017aa}, but by construction seems to be $\mathcal{O}(\tau^2)$ for all $n$. Thus our expectation is that $\UR{n}$ improves with increasing $n$ until performance saturates and the $\mathcal{O}(\tau^2)$ contribution dominates the $\mathcal{O}(\epsilon_r^{n/2})$ contribution. To see this, note that for a fixed time, the number of free evolution periods will be roughly the same regardless of $n$. 
    \item[\textbf{(d)}] Ideally, for a fixed demonstration duration $T$, the performance of $\UDDx{n}$ should scale as $\mathcal{O}(\tau^n)$, and hence improves monotonically with increasing $n$. In practice, this performance should saturate once the finite-pulse width error $\mathcal{O}(\Delta)$ is the dominant noise contribution.
    \item[\textbf{(e--i)}] An extensive numerical study of $\QDD{n}{m}$ performance is discussed in Ref.~\cite{QL:11} with corresponding rigorous proofs in Ref.~\cite{KL:11}. For ideal pules, the decoupling order is expected to be at least $\mathcal{O}(t_s^{\min\{m, n\}})$ where $t_s$ is the total evolution time of implementing a single repetition. Since we are instead interested in a fixed total time $T$ consisting of multiple sequence repetitions with a minimal pulse interval, the interplay of competing factors is quite complicated. Further, we are forced to apply rotations about $X$, $Y$, and $Z$ to implement $\QDD{n}{m}$ when $m$ is odd, but as noted in \cref{app:CvsP}, $Z$ is virtual without OpenPulse. In summary, the naive theoretical expectation is that $\QDD{n}{m}$ should improve with increasing $\min\{n, m\}$, eventually saturating for the same reasons as $\UDDx{n}$. However, we expect the fixed $T$ and virtual-$Z$ set-up to complicate the actual results. 
\end{enumerate}

\clearpage

\bibliography{refs.bib, new_bib_no_repeat_for_main2.bib}

\end{document}